%% file: main.tex
\documentclass[11pt,dvipsnames,twoside,openright]{memoir}
\usepackage[dvipsnames,table]{xcolor}
\usepackage[utf8]{inputenc}
\usepackage[T1]{fontenc}
\usepackage[american]{babel}
\usepackage{amsmath, amssymb, amsthm}
\usepackage{todonotes}
\usepackage{graphicx}
\usepackage{tikz}
\usepackage{hhline}
\usepackage{physics}
\usepackage{mathtools}
\usepackage{tikz}
\usepackage{bm}
\usepackage[linesnumbered,ruled]{algorithm2e}
\usepackage{listings}
\usepackage{float}
\usepackage[url=false,eprint=true,style=phys,biblabel=brackets,maxbibnames=10,backend=bibtex,sorting=none]{biblatex}
\usepackage{subcaption}
\usepackage[font={small}]{caption}
\usepackage{xfrac}
\usepackage{multirow}
\usepackage{makecell}
\usepackage[chapter,draft]{minted}
\usepackage[colorlinks=true, urlcolor=RoyalBlue, linkcolor=blue]{hyperref}

\checkandfixthelayout
\DeclareCaptionLabelFormat{subfigure}{\textbf{#2.}}
\colorlet{theader}{ProcessBlue!30}
\colorlet{tsubheader}{ProcessBlue!20}

\theoremstyle{definition}
\newtheorem{example}{Example}[chapter]

\newtheorem{theorem}{Theorem}

\DeclareRobustCommand\tikzdot{\tikz[overlay,yshift=0.5ex] \fill[RoyalBlue,thick] (0.ex,0.ex) circle (0.3ex);}
\DeclareRobustCommand\tikzcircle{\tikz[overlay,yshift=0.5ex] \draw[Tan,thick] (0.ex,0.ex) circle (0.4ex);}
\DeclareRobustCommand\tikzquad{\tikz[overlay,yshift=0.ex,xshift=0.1ex] \draw[LimeGreen,thick] (0.ex,0.ex) rectangle (1ex,1ex);}

\title{Validation and benchmarking of quantum annealing technology}
\author{Konrad Jałowiecki}
\date{January 2022}

\include{definitions}
\colorlet{lstbg}{black!80}

\AtBeginBibliography{\small}

\DeclareFieldFormat[article]{title}{{\itshape #1}}
\DeclareFieldFormat[book]{title}{\href{\thefield{url}}{\itshape #1}}
\DeclareFieldFormat[thesis]{title}{\href{\thefield{url}}{\itshape #1}}
\allowdisplaybreaks[4]

\makeatletter
\def\PYG@reset{\let\PYG@it=\relax \let\PYG@bf=\relax%
    \let\PYG@ul=\relax \let\PYG@tc=\relax%
    \let\PYG@bc=\relax \let\PYG@ff=\relax}
\def\PYG@tok#1{\csname PYG@tok@#1\endcsname}
\def\PYG@toks#1+{\ifx\relax#1\empty\else%
    \PYG@tok{#1}\expandafter\PYG@toks\fi}
\def\PYG@do#1{\PYG@bc{\PYG@tc{\PYG@ul{%
    \PYG@it{\PYG@bf{\PYG@ff{#1}}}}}}}
\def\PYG#1#2{\PYG@reset\PYG@toks#1+\relax+\PYG@do{#2}}

\@namedef{PYG@tok@w}{\def\PYG@tc##1{\textcolor[rgb]{0.73,0.73,0.73}{##1}}}
\@namedef{PYG@tok@c}{\let\PYG@it=\textit\def\PYG@tc##1{\textcolor[rgb]{0.24,0.48,0.48}{##1}}}
\@namedef{PYG@tok@cp}{\def\PYG@tc##1{\textcolor[rgb]{0.61,0.40,0.00}{##1}}}
\@namedef{PYG@tok@k}{\let\PYG@bf=\textbf\def\PYG@tc##1{\textcolor[rgb]{0.00,0.50,0.00}{##1}}}
\@namedef{PYG@tok@kp}{\def\PYG@tc##1{\textcolor[rgb]{0.00,0.50,0.00}{##1}}}
\@namedef{PYG@tok@kt}{\def\PYG@tc##1{\textcolor[rgb]{0.69,0.00,0.25}{##1}}}
\@namedef{PYG@tok@o}{\def\PYG@tc##1{\textcolor[rgb]{0.40,0.40,0.40}{##1}}}
\@namedef{PYG@tok@ow}{\let\PYG@bf=\textbf\def\PYG@tc##1{\textcolor[rgb]{0.67,0.13,1.00}{##1}}}
\@namedef{PYG@tok@nb}{\def\PYG@tc##1{\textcolor[rgb]{0.00,0.50,0.00}{##1}}}
\@namedef{PYG@tok@nf}{\def\PYG@tc##1{\textcolor[rgb]{0.00,0.00,1.00}{##1}}}
\@namedef{PYG@tok@nc}{\let\PYG@bf=\textbf\def\PYG@tc##1{\textcolor[rgb]{0.00,0.00,1.00}{##1}}}
\@namedef{PYG@tok@nn}{\let\PYG@bf=\textbf\def\PYG@tc##1{\textcolor[rgb]{0.00,0.00,1.00}{##1}}}
\@namedef{PYG@tok@ne}{\let\PYG@bf=\textbf\def\PYG@tc##1{\textcolor[rgb]{0.80,0.25,0.22}{##1}}}
\@namedef{PYG@tok@nv}{\def\PYG@tc##1{\textcolor[rgb]{0.10,0.09,0.49}{##1}}}
\@namedef{PYG@tok@no}{\def\PYG@tc##1{\textcolor[rgb]{0.53,0.00,0.00}{##1}}}
\@namedef{PYG@tok@nl}{\def\PYG@tc##1{\textcolor[rgb]{0.46,0.46,0.00}{##1}}}
\@namedef{PYG@tok@ni}{\let\PYG@bf=\textbf\def\PYG@tc##1{\textcolor[rgb]{0.44,0.44,0.44}{##1}}}
\@namedef{PYG@tok@na}{\def\PYG@tc##1{\textcolor[rgb]{0.41,0.47,0.13}{##1}}}
\@namedef{PYG@tok@nt}{\let\PYG@bf=\textbf\def\PYG@tc##1{\textcolor[rgb]{0.00,0.50,0.00}{##1}}}
\@namedef{PYG@tok@nd}{\def\PYG@tc##1{\textcolor[rgb]{0.67,0.13,1.00}{##1}}}
\@namedef{PYG@tok@s}{\def\PYG@tc##1{\textcolor[rgb]{0.73,0.13,0.13}{##1}}}
\@namedef{PYG@tok@sd}{\let\PYG@it=\textit\def\PYG@tc##1{\textcolor[rgb]{0.73,0.13,0.13}{##1}}}
\@namedef{PYG@tok@si}{\let\PYG@bf=\textbf\def\PYG@tc##1{\textcolor[rgb]{0.64,0.35,0.47}{##1}}}
\@namedef{PYG@tok@se}{\let\PYG@bf=\textbf\def\PYG@tc##1{\textcolor[rgb]{0.67,0.36,0.12}{##1}}}
\@namedef{PYG@tok@sr}{\def\PYG@tc##1{\textcolor[rgb]{0.64,0.35,0.47}{##1}}}
\@namedef{PYG@tok@ss}{\def\PYG@tc##1{\textcolor[rgb]{0.10,0.09,0.49}{##1}}}
\@namedef{PYG@tok@sx}{\def\PYG@tc##1{\textcolor[rgb]{0.00,0.50,0.00}{##1}}}
\@namedef{PYG@tok@m}{\def\PYG@tc##1{\textcolor[rgb]{0.40,0.40,0.40}{##1}}}
\@namedef{PYG@tok@gh}{\let\PYG@bf=\textbf\def\PYG@tc##1{\textcolor[rgb]{0.00,0.00,0.50}{##1}}}
\@namedef{PYG@tok@gu}{\let\PYG@bf=\textbf\def\PYG@tc##1{\textcolor[rgb]{0.50,0.00,0.50}{##1}}}
\@namedef{PYG@tok@gd}{\def\PYG@tc##1{\textcolor[rgb]{0.63,0.00,0.00}{##1}}}
\@namedef{PYG@tok@gi}{\def\PYG@tc##1{\textcolor[rgb]{0.00,0.52,0.00}{##1}}}
\@namedef{PYG@tok@gr}{\def\PYG@tc##1{\textcolor[rgb]{0.89,0.00,0.00}{##1}}}
\@namedef{PYG@tok@ge}{\let\PYG@it=\textit}
\@namedef{PYG@tok@gs}{\let\PYG@bf=\textbf}
\@namedef{PYG@tok@gp}{\let\PYG@bf=\textbf\def\PYG@tc##1{\textcolor[rgb]{0.00,0.00,0.50}{##1}}}
\@namedef{PYG@tok@go}{\def\PYG@tc##1{\textcolor[rgb]{0.44,0.44,0.44}{##1}}}
\@namedef{PYG@tok@gt}{\def\PYG@tc##1{\textcolor[rgb]{0.00,0.27,0.87}{##1}}}
\@namedef{PYG@tok@err}{\def\PYG@bc##1{{\setlength{\fboxsep}{\string -\fboxrule}\fcolorbox[rgb]{1.00,0.00,0.00}{1,1,1}{\strut ##1}}}}
\@namedef{PYG@tok@kc}{\let\PYG@bf=\textbf\def\PYG@tc##1{\textcolor[rgb]{0.00,0.50,0.00}{##1}}}
\@namedef{PYG@tok@kd}{\let\PYG@bf=\textbf\def\PYG@tc##1{\textcolor[rgb]{0.00,0.50,0.00}{##1}}}
\@namedef{PYG@tok@kn}{\let\PYG@bf=\textbf\def\PYG@tc##1{\textcolor[rgb]{0.00,0.50,0.00}{##1}}}
\@namedef{PYG@tok@kr}{\let\PYG@bf=\textbf\def\PYG@tc##1{\textcolor[rgb]{0.00,0.50,0.00}{##1}}}
\@namedef{PYG@tok@bp}{\def\PYG@tc##1{\textcolor[rgb]{0.00,0.50,0.00}{##1}}}
\@namedef{PYG@tok@fm}{\def\PYG@tc##1{\textcolor[rgb]{0.00,0.00,1.00}{##1}}}
\@namedef{PYG@tok@vc}{\def\PYG@tc##1{\textcolor[rgb]{0.10,0.09,0.49}{##1}}}
\@namedef{PYG@tok@vg}{\def\PYG@tc##1{\textcolor[rgb]{0.10,0.09,0.49}{##1}}}
\@namedef{PYG@tok@vi}{\def\PYG@tc##1{\textcolor[rgb]{0.10,0.09,0.49}{##1}}}
\@namedef{PYG@tok@vm}{\def\PYG@tc##1{\textcolor[rgb]{0.10,0.09,0.49}{##1}}}
\@namedef{PYG@tok@sa}{\def\PYG@tc##1{\textcolor[rgb]{0.73,0.13,0.13}{##1}}}
\@namedef{PYG@tok@sb}{\def\PYG@tc##1{\textcolor[rgb]{0.73,0.13,0.13}{##1}}}
\@namedef{PYG@tok@sc}{\def\PYG@tc##1{\textcolor[rgb]{0.73,0.13,0.13}{##1}}}
\@namedef{PYG@tok@dl}{\def\PYG@tc##1{\textcolor[rgb]{0.73,0.13,0.13}{##1}}}
\@namedef{PYG@tok@s2}{\def\PYG@tc##1{\textcolor[rgb]{0.73,0.13,0.13}{##1}}}
\@namedef{PYG@tok@sh}{\def\PYG@tc##1{\textcolor[rgb]{0.73,0.13,0.13}{##1}}}
\@namedef{PYG@tok@s1}{\def\PYG@tc##1{\textcolor[rgb]{0.73,0.13,0.13}{##1}}}
\@namedef{PYG@tok@mb}{\def\PYG@tc##1{\textcolor[rgb]{0.40,0.40,0.40}{##1}}}
\@namedef{PYG@tok@mf}{\def\PYG@tc##1{\textcolor[rgb]{0.40,0.40,0.40}{##1}}}
\@namedef{PYG@tok@mh}{\def\PYG@tc##1{\textcolor[rgb]{0.40,0.40,0.40}{##1}}}
\@namedef{PYG@tok@mi}{\def\PYG@tc##1{\textcolor[rgb]{0.40,0.40,0.40}{##1}}}
\@namedef{PYG@tok@il}{\def\PYG@tc##1{\textcolor[rgb]{0.40,0.40,0.40}{##1}}}
\@namedef{PYG@tok@mo}{\def\PYG@tc##1{\textcolor[rgb]{0.40,0.40,0.40}{##1}}}
\@namedef{PYG@tok@ch}{\let\PYG@it=\textit\def\PYG@tc##1{\textcolor[rgb]{0.24,0.48,0.48}{##1}}}
\@namedef{PYG@tok@cm}{\let\PYG@it=\textit\def\PYG@tc##1{\textcolor[rgb]{0.24,0.48,0.48}{##1}}}
\@namedef{PYG@tok@cpf}{\let\PYG@it=\textit\def\PYG@tc##1{\textcolor[rgb]{0.24,0.48,0.48}{##1}}}
\@namedef{PYG@tok@c1}{\let\PYG@it=\textit\def\PYG@tc##1{\textcolor[rgb]{0.24,0.48,0.48}{##1}}}
\@namedef{PYG@tok@cs}{\let\PYG@it=\textit\def\PYG@tc##1{\textcolor[rgb]{0.24,0.48,0.48}{##1}}}

\def\PYGZus{\char`\_}

\def\PYGZca{\char`\^}

\def\PYGZlt{\char`\<}
\def\PYGZgt{\char`\>}
\def\PYGZsh{\char`\#}
\def\PYGZpc{\char`\%}

\def\PYGZhy{\char`\-}
\def\PYGZsq{\char`\'}

\makeatother

\begin{document}

\begin{titlingpage}
  \begin{center}
    \includegraphics[width=0.4\textwidth]{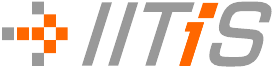}\\
    \vspace{0.5em}
    \textsc{\large Institute of Theoretical and Applied Informatics, Polish Academy of Sciences}
    \vspace*{1in}
    \hrule
    \vspace*{0.5em}
    \textsc{\huge Validation and benchmarking of quantum annealing technology}
    \vspace*{0.5em}
    \hrule
    \vspace*{1em}
    \textsc{\large Doctoral dissertation}
    \par
    \vspace{1.5in}
    {\large mgr Konrad \textsc{Jałowiecki}}\\
    \vspace{0.25in}
    Supervisor:\\ dr hab. Bartłomiej Gardas\\
    \vspace{0.25in}
    Co-supervisor:\\ dr hab. inż. Łukasz Pawela\\
    \vfill
    {Katowice, November 24, 2023}
  \end{center}
\end{titlingpage}

\begin{titlingpage}
  \begin{center}
    \includegraphics[width=0.4\textwidth]{iitis_logo.pdf}\\
    \vspace{0.5em}
    \textsc{\large Instytut Informatyki Teoretycznej i Stosowanej Polskiej Akademii Nauk}
    \vspace*{1in}
    \hrule
    \vspace*{0.5em}
    \textsc{\huge Walidacja i testowanie porównawcze technologii kwantowego wyżarzania}
    \vspace*{0.5em}
    \hrule
    \vspace*{1em}
    \textsc{\large Rozprawa doktorska}
    \par
    \vspace{1.5in}
    {\large mgr Konrad \textsc{Jałowiecki}}\\
    \vspace{0.25in}
    Promotor:\\ dr hab. Bartłomiej Gardas\\
    \vspace{0.25in}
    Promotor pomocniczy:\\ dr hab. inż. Łukasz Pawela\\
    \vfill
    {Katowice, 24 listopada 2023}
  \end{center}
\end{titlingpage}

\frontmatter

\tableofcontents*
\newpage
\input{acknowledgements.tex}
\input{published_work.tex}
\input{abstract_en.tex}
\input{abstract_pl.tex}

\mainmatter

\input{introduction.tex}
\input{ising.tex}
\input{near-term-technologies.tex}
\input{simulating-dynamics-with-dwave.tex}
\input{tensor_networks}
\input{bruteforce.tex}
\input{trains.tex}
\input{summary.tex}

{
  \small
  \printbibliography
}
\appendix
\input{appendix.tex}

\end{document}

%% file: definitions.tex
\def\NN{\mathbb{N}}
\def\RR{\mathbb{R}}
\def\CC{\mathbb{C}}
\def\HH{\mathcal{H}}

\DeclareMathOperator*{\argmin}{arg\,min}

\def\clockop{\mathcal{C}}
\def\coefmatrix{\mathcal{A}}
\def\JJ{\mathcal{J}}
\def\tout{t_\text{out}}
\def\tin{t_\text{in}}
\def\ttout{t_\text{out}^\text{timetable}}
\def\ttin{t_\text{in}^\text{timetable}}
\def\pt{p^\text{timetable}}
\def\pmin{p^\text{min}}
\def\tauu{\tau_{(1)}}
\def\tauuu{\tau_{(2)}}

\def\ssigma{\hat{\sigma}}

\newcommand{\q}[1]{q^{(#1)}}
\newcommand{\bq}[1]{\mathbf{q}^{(#1)}}
\DeclareMathOperator*{\flipop}{flip}
\newcommand{\flip}[2]{\flipop(#1,#2)}
\DeclareMathOperator{\diff}{diff}
\newcommand{\ppair}{p_{\text{pair}}}
\newcommand{\psum}{p_{\text{sum}}}
\newcommand{\psucc}{p_{\text{succ}}}
\newcommand{\ptarget}{p_{\text{target}}}
\newcommand{\Ppair}{\mathcal{P}_{\text{pair}}}
\newcommand{\Psum}{\mathcal{P}_{\text{sum}}}
\newcommand{\Sj}{\mathcal{S}_j}
\newcommand{\Bj}{\mathcal{B}_j}
\newcommand{\Sjs}{\mathcal{S}_{j}^{*}}
\DeclareMathOperator*{\eend}{end}

%% file: acknowledgements.tex
\chapter{Acknowledgements}

I am deeply grateful to my supervisor, dr hab. Bartłomiej Gardas, whose unwavering support and insightful guidance have been instrumental in shaping this doctoral thesis. His expertise, encouragement, and mentorship have been invaluable, and I am truly fortunate to have had the opportunity to work under his supervision.

I would also like to express my sincere appreciation to my co-supervisor, dr hab. inż. Łukasz Pawela, for his constructive feedback, especially in the field of software engineering. His expertise and willingness to share knowledge have significantly enriched the quality of this thesis.

I would also like to express my gratitude to all my colleagues from the Institute who contributed to this thesis via many fruitful conversations I had with them. In particular, I would like to thank Krzysztof Domino for sharing his knowledge and expertise in the railway dispatching field.

Additionally, I extend my deepest gratitude to my friends, Alexander Juda, Michał Stęchły, and Paweł Grzybek, for taking the time to read parts of this thesis. Their valuable feedback contributed greatly to improving the readability and overall quality of this work.

This project was partially supported by the National Science Center (NCN), Poland, under Projects: Sonata Bis 10, No. 2020/38/E/ST3/00269
and the National Centre for Research and Development (NCBR), Poland, under Project No. POIR.01.01.01-00-0061/2. I would also like to thank The Quantum Data Center Corporation for providing me with access to several GPUs used for benchmarks presented in this thesis.


%% file: published_work.tex
\chapter{Published work}

\section*{Publications relevant for this dissertation}

\nocite{parallelintime,tn,bruteforce,railwaydispatching,omnisolver,pyqbench}
\printbibliography[heading=none,keyword=my-used]

\section*{Other publications}
\printbibliography[heading=none,keyword=my-unused]


%% file: abstract_en.tex
\chapter{Abstract}

In this thesis, we focus on the problem of validating and benchmarking quantum annealers in a
practical context. To this end, we propose two algorithms for solving real-world problems and test how
well they perform on the current generation of quantum annealers. The first algorithm allows for
solving the dynamics of quantum systems (or, in fact, any dynamical systems). The second of the proposed
algorithms is suitable for solving a particular family of railway dispatching problems: the delay
and conflict management on single-track railway lines. We assess the performance of those
algorithms on the current generation of D-Wave quantum annealers with the assistance of two novel,
classical strategies for solving an Ising model also presented in the thesis. The first, tensor
network-based approach is a heuristic algorithm specifically tailored for solving instances defined
on Chimera-like graphs, thus making it ideal for providing a baseline with which the results from
physical annealers can be compared. The other presented approach is a massively parallel
implementation of the exhaustive search through the whole solution space, also known as the
brute-force approach. Although the brute-force approach is limited to moderate instance sizes, it
has the advantage of being able to compute the low energy spectrum and certify the solutions. Thus,
it can be used to obtain additional insight into the solution space structure. The results obtained
in our experiments suggest that already present-day quantum annealers are capable of solving a
subset of the aforementioned optimization problems. In particular, we show that the D-Wave annealers
are capable of capturing the dynamics of a simple two-level quantum system in a specific regime of
parameters, and can be used to obtain good-quality solutions for instances of railway conflict
management problems. Finally, our findings make it clear that the current generation of the D-Wave
annealers is far from perfect. We discuss problem instances for which the annealers failed to find
a good or even feasible solution. We also provide, where possible, a plausible explanation of why
some of the presented problems might be hard for the annealers.

%% file: abstract_pl.tex
\chapter{Streszczenie}

\begin{otherlanguage}{polish}
  W niniejszej pracy skupiamy się na problemie walidowania i benchmarkowania
  wyżaraczy kwantowych w praktycznym kontekście. W tym celu, przedstawiamy dwa
  algorytmy służące do rozwiązywania rzeczywistych problemów, oraz sprawdzamy,
  jak dobrze sprawdzają się na obecnej generacji wyżaraczy kwantowych. Pierwszy z
  algorytmów pozwala na rozwiązywanie dynamiki kwantowych układów (lub, w gruncie
  rzeczy, dowolnych układów dynamicznych). Drugi z przedstawianych algorytmów
  może z kolei zostać użyty do rozwiązywania pewnego podzbioru kolejowych
  problemów dyspozytorski: zarządania opóźnieniami i konfliktami w sieciach
  kolejowych o jednej linii. Oceny działania obu w.w. algorytmów na bieżącej
  generacji wyżaraczy D-Wave dokonujemy z pomocą dwóch, nowatorskich, klasycznych
  strategii rozwiązywania szkieł spinowych Isinga, które również prezentujemy w
  niniejszej rozprawie. Pierwszym z nich jest opierający się na sieciach tensorowych
  heurystyczny algorytm stworzony specjalnie do rozwiązywania szkieł spinowych
  zdefiniowanych na grafach przypominających topologię Chimera, co sprawia, że
  idealnie nadaje się do wyznaczania referencyjnych rozwiązań, do których można
  porównać wyniki z fizycznych wyżarzaczy. Drugim z prezentowanych podejść jest
  masywnie równoległa implementacja wyczerpującego przeszukiwania całej
  przestrzeni rozwiązań, tzw. brute-force. Mimo, że użycie algorytmu brute-force
  jest ograniczone do instancji o niewielkich rozmiarach, posiada on tę zaletę,
  że może wyznaczać niskoenergetyczne spektrum, oraz certyfikować rozwiązania. W
  związku z~tym, algorytm przeszukiwania wyczerpującego może slużyć do uzyskania
  dodatkowego wglądu w strukturę przestrzeni rozwiązań. Wyniki otrzymane w
  naszych eksperymentach sugerują, że już współczesne wyżarzacze są w stanie
  uchwycić dynamikę prostych, dwupoziomowych układów kwantowych w specyficznym
  reżimie parametrów, oraz mogą znaleźć dobrej jakości rozwiązania instancji
  kolejowych problemów dyspozytorskich. Wreszcie, nasze eksperymenty pokazują jasno, że obecna
  generacja wyżaraczy D-Wave nie jest idealna. Wymieniamy instancje problemów,
  dla których wyżarzanie nie potrafily znaleźść wysokojakościowych, lub nawet
  dopuszczalnych rozwiązań. Tam gdzie to możliwe, omawiamy również możliwe
  wyjaśnienie dlaczego niektóre z prezentowanych instancji mogą być dla wyżaraczy
  wymagające.
\end{otherlanguage}


%% file: introduction.tex
\chapter{Introduction}
The previous century has witnessed what is now called the digital revolution.
The introduction of digital computers dramatically altered multiple aspects of
our lives. In particular, almost every area of science benefitted hugely from
the increasingly available computational power \cite{winsberg}. Physics was no
exception, and numerical simulations now commonly assist experiments.

Simulating quantum systems -- a holy grail of modern computational physics --
is a highly challenging task for classical computers \cite{feynman.82}. The
difficulties can be blamed on the enormous number of possible configurations of
such systems. Direct, naive simulations would require solving systems of
differential equations with the number of variables exponential in the number
of particles. But what about using more sophisticated algorithms? Surprisingly,
it is commonly believed that a sufficiently efficient classical algorithm for
simulating quantum systems does not exist \cite{feynman.82, poplavskii}.
Matters seem even worse when one considers that the increase in the classical
devices' computational power cannot accelerate infinitely. Moore's law
\cite{mack}, which so far well predicted this growth, is expected to slow down
in the years to come \cite{waldrop, kumar}.

If classical computers cannot simulate quantum physics efficiently, what device
can? In the 1980s, Richard Feynman and Paul Benioff put forward the idea that
quantum devices can be used to carry simulations of quantum systems
\cite{feynman.82,benioff.80}. This idea led to the development of several
quantum computation models. In 1985 David Deutsch described a universal,
gate-based quantum computer \cite{deutsch}, a device capable of simulating any
other quantum computer with at most polynomial slowdown. The 1990s and the
early 2000s saw the emergence of another model of quantum computation,
Adiabatic Quantum Computing (AQC) \cite{kadowaki,farhi}. Interestingly, AQC was
later proven to be equivalent to the standard gate-based model \cite{aharonov}.

Just like a classical computer, a quantum computer needs software to run, and
software is based on algorithms, describing how the computation should be
performed. It is not a surprise that quantum computers operate in a very
different way than classical ones, and require different, specialized
algorithms. What is surprising is that several notable quantum algorithms were
developed even before the first quantum computers were constructed. In 1994
Peter Shor published his, now famous, algorithm for integer factorization
\cite{shor}. Shor's algorithm demonstrated that quantum computers are (in
principle) capable of solving problems intractable by the classical ones
\cite{kleinjung}. It was also shown that quantum computers could offer a
significant performance boost for easier problems. For instance, in 1996 Grover
presented a quantum algorithm for unstructured database search \cite{grover},
offering a quadratic speed-up over classical algorithms solving the same problem.

The invention of specialized quantum algorithms further fuelled interest in the
field. In recent years, we observed the development of hardware that brings us
closer to the quantum revolution. Several implementations of gate--based
quantum computers \cite{ionq, bohnet} and quantum annealers \cite{johnson,
  dattani} were constructed and made publicly available. This allowed scientists
to benchmark them and further research their possible applications.

However promising, current quantum computers are far from perfect
\cite{pitfalls,preskill}. Can those noisy devices already be used to solve some
real-world problems? And how does one approach validating if this is the case?
In this thesis, we try to answer these questions, focusing solely on a specific
type of quantum computer -- namely, the D-Wave quantum annealers.

\subsection*{Layout of the thesis}

We begin the thesis with an introduction to Ising and QUBO models (collectively known as
Binary Quadratic Models) in Chapter \ref{chapter:ising}. This chapter's purpose
is to lay the necessary foundations for understanding optimization problems
that can be, at least in principle, solved using quantum annealers.

In Chapter \ref{chapter:near-term} we introduce technologies and devices used
for conducting research presented in this thesis. Quite naturally, the first of
those devices are quantum annealers. We briefly describe the principle of
operation of these devices and then move on to discuss currently available
models. We also describe NVIDIA CUDA, another technology that we used for
implementing the brute-force algorithm presented in Chapter
\ref{chapter:bruteforce}.

It is widely believed that a noiseless universal quantum computer would be
capable of simulating quantum systems. But what about the near-term quantum
devices? In Chapter \ref{chapter:simulating} we explore the idea of simulating
the evolution of dynamical (not necessarily quantum) systems using quantum
annealers. We describe how to represent the task of simulating the dynamics as
a static optimization problem and then present experimental results obtained
from the D-Wave annealer. We find out that for small systems, the annealer is
able to faithfully capture the dynamics. We also discuss possible sources of
errors for the problem instances that the annealers failed to solve. While our
algorithm is only a proof of concept, it exemplifies possible directions of
future research.

A key component in assessing the performance of current quantum annealers is
comparing them to the classical algorithms solving the same problems. While
there exists a plethora of general heuristic methods for finding a ground state
of Ising spin-glass, one can ask if it is possible to construct a better
algorithm tailored for problems defined on the same graph as the physical
device. In Chapter \ref{chapter:tn}, we present a recent, heuristic algorithm
for finding the low-energy spectrum of an Ising spin-glass based on tensor
networks, specifically suited for problems defined on Chimera-like graphs.

Chapter \ref{chapter:bruteforce} describes a fast, parallel approach to
exhaustively searching for a low-energy spectrum of Ising spin-glass problems.
Our method is suitable for solving small (less than 54 spins), but otherwise
arbitrary instances. The presented approach can be used for benchmarking other
algorithms that cannot certify their solution. Moreover, the possibility of
finding a low-energy spectrum (instead of a single solution) is extremely
useful for analyzing the structure of the energy landscape of the problem. We
exemplify the usage of our algorithm by conducting benchmarks of a recent
MPS-based algorithm on a set of random spin-glass problems. Compared to the
original algorithm presented in \cite{bruteforce}, the algorithm described in
Chapter \ref{chapter:bruteforce} contains several new, non-trivial
optimizations further increasing the problem sizes that it can tackle. To the
best of our knowledge, those optimizations make our implementation the fastest
brute-force solver for Ising problems available on the market.

Lastly, in Chapter \ref{chapter:trains}, we present the application of quantum
annealing to solving certain railway dispatching problems. We discuss how such
problems can be converted to QUBO problems suitable for running on the
annealer. We then report the performance of the current generation of quantum
annealers on a set of dispatching problems constructed for real Polish railway
networks. Presented benchmarks extend results presented in
\cite{railwaydispatching} to the newer generation of quantum annealers.
Compared to \cite{railwaydispatching}, we also include a more detailed
discussion on the influence of penalty terms on the quality of results.

%% file: ising.tex
\chapter{Ising model and QUBO problem}
\label{chapter:ising}

Quantum annealers are fundamentally different from classical computers. For
one, they don't execute programs written as a sequence of instructions in their
memory. Instead, they are single-purpose devices capable (in principle) of
solving a specific optimization problem. Namely, annealers are designed to find
the lowest energy configuration (called \emph{ground state}) of instances of
the Ising spin--glass model, which we introduce in this chapter.

The potential usefulness of quantum annealers stems from the fact that the
optimization problem they are supposed to solve is hard for classical
computers. But what does it formally mean for a problem to be hard? To answer
this question, we will need a brief recap of complexity theory, which is a
second point of this chapter.

Finding a ground state of the Ising spin--glass model may be hard for classical
computers, but there exists a plethora of heuristic, classical algorithms
capable of finding solutions that are at least ``good enough''. As the next
point in this chapter, we provide a brief overview of the most popular
classical algorithms for solving the Ising spin-glass problems. These
algorithms will serve as a baseline for comparison with quantum annealing and a
recent tensor network-based approach discussed later in the thesis.

As the last point in the chapter, we define the Quadratic Unconstrained Binary
Optimization (QUBO) problem, which is equivalent to the problem of finding the
ground state of the Ising model. We will use the QUBO formulation on several
occasions in the thesis, as it oftentimes results in a more natural phrasing of
the problem, or leads to a surprising performance improvement when implementing
software solvers.

\section{Ising model}

The Ising spin--glass model was introduced in 1920 by Wilhelm Lenz \cite{lenz}
as a description of ferromagnetism in solids but is named after his student
Ernst Ising, who studied and solved it in the one-dimensional case
\cite{ising}. For purposes of this thesis, however, we will forget about the
physical interpretation of the model, treating it merely as a description of a
particular optimization problem.

\begin{figure}[H]
  \centering
  \includegraphics{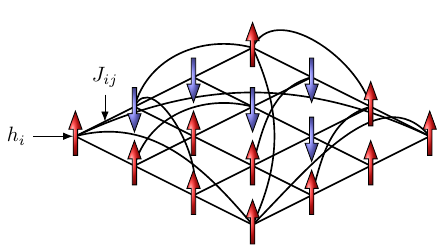}
  \caption{Symbolic representation of Ising spin--glass defined on the graph with $N=16$
    nodes. Here, $h_i$ is a real number associated with $i$-th node, and $J_{ij}$
    denotes coupling strength associated with an edge between $i$-th and $j$-th
    node. The configuration of each spin is marked by a red arrow pointing upwards
    (+1) or a blue arrow pointing downwards (-1).} \label{fig:my_label}
  \label{fig:ising}
\end{figure}

Consider a simple\footnote{That is, one that does not contain duplicate edges
  or loops.}, undirected graph $G = (V, E)$ with $N$ nodes labeled by consecutive
natural numbers. With each node $i \in V$ we associate a spin variable $s_i \in
  \{-1, 1\}$. To each edge $\{i, j\} \in E$ we assign an interaction strength
$J_{ij}$ and to each node $i \in V$ we assign a local magnetic field $h_i$.
Here, all $J_{ij}$ and $h_i$ are real numbers. For such a system, one can
define the following energy function (Hamiltonian):
\begin{equation}
  \label{eq:ising-hamiltonian}
  H(\mathbf{s}) = \sum_{\langle i, j \rangle} J_{ij} s_i s_j +  \sum_{i=1}^N h_i s_i,
\end{equation}
where $\mathbf{s} = (s_i, \ldots, s_N)$ and the first sum runs over all edges
exactly once\footnote{ In the literature, the Ising Hamiltonian
  \eqref{eq:ising-hamiltonian} is often negated. However, the definition provided
  here is consistent with the one used by D-Wave, and thus more suitable for use
  in this thesis.}. In this thesis, we will call the instances of the Ising model
\emph{Ising spin-glasses}. An illustrative representation of a spin-glass is depicted in
Fig. \ref{fig:ising}.

For fixed model coefficients, one is typically interested in finding its
\emph{ground state}, a configuration $\mathbf{s}$ that minimizes $H$. More
generally, it might be desirable to search for $k \ll 2^N$ configurations with
the lowest energy, a so-called \emph{low-energy spectrum of size $k$}.

\begin{example}
  Consider an Ising model instance with 3 spins given by the Hamiltonian $H$:
  \begin{equation}
    \label{eq:isingexample}
    H(s_1, s_2, s_3) = s_1 - s_2 +2s_3 - 2s_2s_3 + 3s_1s_2
  \end{equation}
  This instance has 8 possible states:

  \begin{table}[h]
    \begin{center}
      \begin{tabular}{|c|c||c|c|}
        \hline
        \rowcolor{theader}
        $\mathbf{s}=(s_1, s_2, s_3)$ &
        $H(\mathbf{s})$              &
        $\mathbf{s}=(s_1, s_2, s_3)$ &
        $H(\mathbf{s})$                                      \\\hline
        (-1, -1, -1)                 & -1 & (1, -1, -1) & -5 \\ \hline
        (-1, -1, 1)                  & 7  & (1, -1, 1)  & 3  \\ \hline
        (-1, 1, -1)                  & -5 & (1, 1, -1)  & 3  \\ \hline
        (-1, 1, 1)                   & -5 & (1, 1, 1)   & 3  \\ \hline
      \end{tabular}
    \end{center}
    \caption{All possible configurations for the Ising Hamiltonian in the equation
      \eqref{eq:isingexample}.} \label{tab:isingexample}
  \end{table}
  Observe that the lowest attainable energy is -5 and there are 3 states with
  exactly this energy. Hence, all the configurations $(-1, 1, -1)$, $(-1, 1, 1)$,
  $(1, -1, -1)$ are ground states. This situation, i.e. when two or more states
  share the same energy, is called \emph{degeneracy} and the states in question
  are called \emph{degenerate}. For this instance, a low energy spectrum of size
  $k=5$ comprises all ground states, the $(-1, -1, -1)$ state with $H(-1, -1, -1)
    = -1$ and any of the states with $H(\mathbf{s})=3$.
\end{example}

Despite the simple formulation, the problem of finding a ground state of Ising
spin--glass is computationally hard \cite{barahoma}. Before expanding on this
idea, let us first introduce the hierarchy of complexity classes.

\section{Algorithms and complexity}

Solving the computational problem requires a suitable \emph{algorithm}, a
description of steps to be performed by a computer to obtain a solution. It is
hardly surprising that some problems might be solved in more than one way, i.e.
there might exist different algorithms performing essentially the same task.
Different algorithms solving the same problems might vastly differ in their
demand on various resources, like memory or time needed to execute them. In
practice, the execution time (and usage of other resources) of a given
algorithm might also vary between its implementations, depending on factors
like programming language or libraries used and the hardware it is executed on.
Moreover, measuring execution time can only tell us how the given
implementation performs on a specific problem. But if we increase the problem
size tenfold, will the execution time be 10 times slower? Or maybe 100 times
slower? Or maybe it will remain unchanged? Clearly, measuring execution times
is useful, but cannot be used for comparing algorithms (instead of their
implementations). Instead, it is more informative to characterize algorithms
based on how their execution time scales asymptotically with increasing problem
size \cite{arora}. For instance, given an algorithm with execution time roughly
proportional to the input size $N$, one might suspect that for problem
instances large enough, it will perform better than the one with execution time
proportional to $N^{2}$. This characteristic, known as computational
complexity\footnote{Note that here we focus only on \emph{time complexity}, but
  other notions like memory complexity can be defined similarly}, can be
formalized by a big-$O$ notation (see appendix for a more detailed
description). Using this notation, the algorithms from the above example would
be classified as $O(N)$ and $O(N^{2})$ respectively.

\section{Complexity classes}
Although there might exist more than one algorithm for solving a particular
problem, one might consider the smallest time complexity needed to do so.
Consequently, one might group computational problems based on their demand on
resources. In this view, sets of similar problems are called \emph{complexity
  classes} \cite{arora}. The definition of some complexity classes might also be
restricted to specific types of problems. For instance, one might consider only
decision problems, i.e. problems to which the answer is yes or no \cite{arora}.
Finally, to define any complexity class one has to assume some model of
computation. In many cases, this model is assumed to be a Turing Machine
\cite{arora}, a theoretical device manipulating symbols on a tape using some
table of rules, or its non-deterministic variant.

\begin{figure}
  \includegraphics[width=\textwidth]{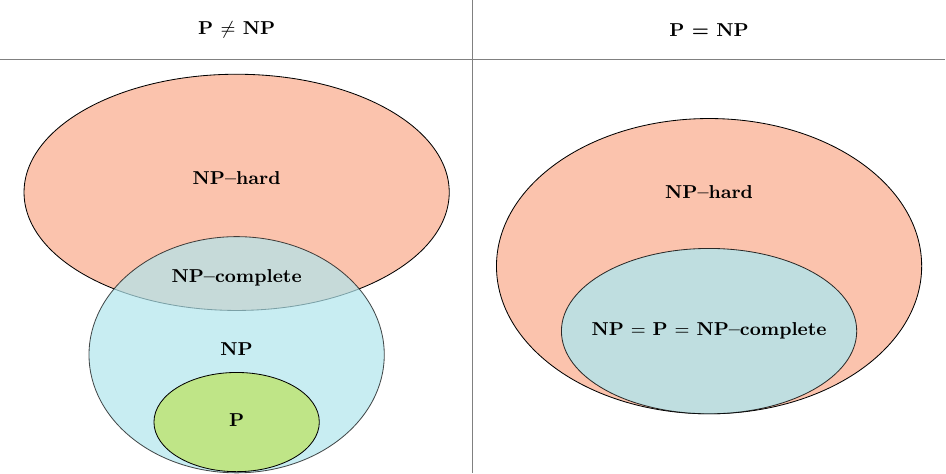}
  \caption{Hierarchy of basic complexity classes. Under the assumption of $\textbf{P} \ne
      \textbf{NP}$ (left), the hierarchy is richer and there exist problems in
    \textbf{NP} that are not \textbf{NP}--complete. Under the opposite assumption
    (right), the hierarchy collapses. Notice that in both cases there exist
    \textbf{NP}--hard problems that are not in \textbf{NP} } \label{fig:complexity}
\end{figure}

One of the fundamental complexity classes is \textbf{P}, a class of decision
problems solvable in polynomial time on a deterministic Turing Machine
\cite{arora}. Another class, \textbf{NP}, comprises all decision problems whose
solution can be verified in polynomial time using a deterministic Turing
Machine \cite{arora}. Immediately, one can see that \textbf{P} $\subset$
\textbf{NP}. Indeed, if a problem is solvable in polynomial time, then it is
also trivially verifiable in polynomial time. However, it is not immediately
obvious if the inclusion is strict, and the question of whether \textbf{P}
$\ne$ \textbf{NP} is one of the most important, yet unsolved problems in
theoretical computer science \cite{fortnow}. The class of \textbf{NP--hard}
problems comprises all the problems that are at least as hard as every problem
in \textbf{NP}. More formally, a decision problem $S$ is \textbf{NP--hard} if
and only if solving every problem in \textbf{NP} can be reduced to solving $S$
a polynomial number of times \cite{arora}. A particular subclass of
\textbf{NP--hard} problems, \textbf{NP--complete}, is an intersection of
\textbf{NP} and \textbf{NP--hard} \cite{arora}. Figure \ref{fig:complexity}
shows the relationship between the discussed complexity classes, both under
assumptions \textbf{P} = \textbf{NP} and \textbf{P} $\ne$ \textbf{NP}.

Problems in the complexity class \textbf{P} are often considered tractable, or
efficiently solvable, whereas problems not in \textbf{P} are perceived as hard
and computationally demanding, a statement known as the Cobham's thesis
\cite{cobham, arora}. At first, one might find it strange and unintuitive - a
decision problem for which the best-known algorithm runs in $O(N^{10^5})$ time
is definitely in \textbf{P}, but can hardly be called efficiently solvable.
However, such large polynomial complexities are rarely encountered in practice.
Furthermore, even in such cases, it is not uncommon that a better algorithm is
found shortly after the original one is discovered \cite{arora}.

\section{Ising model and complexity}

Thus far, we only discussed classes of decision problems. How do they relate to
the problem of finding a ground state of the Ising model? Suppose we are given
an Ising model instance with hamiltonian $H$ and let $x \in \RR$ be some fixed
number. Consider the problem of deciding whether there exists $\mathbf{s}$ such
that $H(\mathbf{s}) \le x$. We will call this problem a \emph{decision version
  of the Ising problem}.

If we can minimize $H$, we can also solve the decision problem by simply
finding a ground state and checking if its energy exceeds threshold $x$. On the
other hand, the sole capability of solving the decision version of a problem
does not give us an algorithm for solving an original optimization problem.
Therefore, one can see that the optimization problem is at least as hard as the
corresponding decision problem. Of course, the same reasoning applies for other
optimization problems. Hence, if the decision version of an optimization
problem is \textbf{NP--hard}, the optimization problem is sometimes also called
\textbf{NP--hard}, even if it slightly abuses the terminology. For simplifying
the vocabulary, in what follows we will use this slightly imprecise but more
concise convention.

It was shown that finding a ground state of the Ising spin--glass in the case
of three-dimensional lattices, as well as for some planar graphs, is
\textbf{NP--hard} \cite{barahoma}. The decision version of the problem is
\textbf{NP--complete}. Multiple known \textbf{NP--hard} problems, such as
Travelling Salesman Problem, Hamiltonian Cycles Problem or Set Cover Problem
are reducible to finding the ground state of Ising spin-glass \cite{lucas}.

As a side note, one might be tempted to think that the \textbf{NP--hard}ness of
finding Ising model's ground state is trivial, because its enormous state space
comprises $2^{N}$ states. However, it is important to remember that the size of
the solution space itself is not enough to reason about the problem's hardness.
For instance, the number of possible spanning trees in the complete graph of
$N$ vertices is $N^{(N-2)}$, yet the minimum spanning tree problem is solvable
in polynomial time via several algorithms \cite{clrs}.

\section{Algorithms for solving Ising model}

As is the case with many \textbf{NP--hard} optimization problems, there are
many heuristic approaches for solving the Ising model. One family of such
algorithms relies on the Metropolis-Hastings \cite{beichl} algorithm for
sampling configurations from the underlying Boltzmann distribution. In
\emph{simulated annealing} \cite{cook, isakov}, one samples states from the
system while lowering the temperature over time. Thus, the chance of accepting
a locally worse solution is greater at the start of the algorithm, which helps
avoid getting stuck in a local minimum, and decreases with each iteration. In
another approach from the same family, \emph{parallel tempering}, one simulates
several replicas of the system, each of them in a different temperature.
Neighboring replicas are allowed to exchange states, with exchange probability
depending on their energy and temperature difference \cite{swendsen}. Replicas
with higher temperatures explore state space rapidly (thus reseeding the
algorithm), while ones with lower temperatures refine the best solutions found
so far. Various modifications of the aforementioned algorithms exist. For
instance, one could employ isoenergetic cluster moves \cite{zhu} or adaptive
choosing of the number of sweeps performed between replica exchanges
\cite{bittner}. Population annealing is yet another Monte Carlo method, sharing
similarities with simulated annealing and parallel tempering \cite{wang}. Other
approaches for solving Ising spin--glasses include methods involving
branch--and--bound framework \cite{rendl}, its chordal extensions
\cite{baccari} or methods based on simulating dynamical systems \cite{sheldon}.

\begin{figure}[bh]
  \centering
  \includegraphics[width=0.91\textwidth]{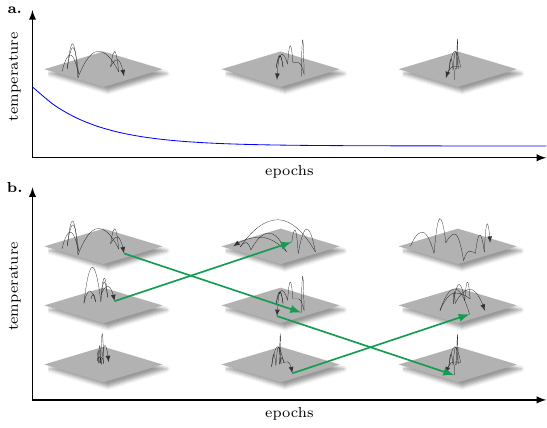}
  \caption{\textbf{a.} Schematic representation of the simulated annealing algorithm.
    \textbf{b.} Schematic representation of the Parallel tempering algorithms. In simulated annealing, a single copy of the
    system is simulated. The temperature of the system is decreased with each
    epoch, thus reducing movement through the state space. In parallel tempering,
    several copies (replicas) of the system are simulated, each with a fixed
    temperature. Hotter replicas move through the state space rapidly and less
    predictably, while colder replicas move conservatively. Between epochs,
    replicas can exchange states, which helps avoid being stuck at local minima.
    Exchanging replicas can also be viewed as reseeding of the colder replicas by
    randomized solutions provided by hotter replicas.} \label{fig:sa}
\end{figure}

\section{Quadratic Unconstrained Binary Optimization}

Let us now shift our attention to \emph{QUBO} -- the Quadratic Unconstrained Binary
Optimization problem. QUBO is essentially the same as the problem of finding
the ground state of Ising spin--glass, except that in QUBO one uses binary
variables $q_{i} \in \{0, 1\}$ instead of $\pm 1$ spin variables. To
distinguish between the two problems, we will use symbols $a_{ij}$ and $b_{i}$
to denote respectively quadratic and linear coefficients in QUBO, so the energy
function to be minimized can be written as:
\begin{equation}
  \label{eq:qubo}
  F(q_1, \ldots, q_N) =  \sum_{\langle i, j \rangle} a_{ij} q_i q_j + \sum_{i=1}^N b_iq_i,
\end{equation}
where, as in the Ising model, the first sum runs through all the edges of the
graph on which the problem is defined.

The QUBO and Ising formulations are essentially equivalent. Indeed, it is
always possible to transform the Ising Hamiltonian \eqref{eq:ising-hamiltonian}
into the QUBO cost function by a linear substitution of variables $s_i \mapsto
  2q_i-1$. Then, one obtains function $F$ like in the equation
\eqref{eq:qubo}, with the following values for $a_{ij}$ and $b_{i}$:
\begin{equation}
  \label{eq:toQUBO}
  a_{ij}= 4J_{ij},
  \quad
  b_i= 2h_i - 2 \sum_{\langle i, j \rangle} J_{ij},
\end{equation}
where the last sum runs over all neighbors of node $i$. The obtained function
$F$ differs from the original $H$ by the constant offset:
\begin{equation}
  F(\mathbf{q}) - H(\mathbf{s}) =\sum_{i=1}^N h_i - \sum_{\langle i, j \rangle} J_{ij},
\end{equation}
which is irrelevant to the optimization process.

\begin{example}
  Let us go back to the previous example and convert the Ising Hamiltonian from the equation \eqref{eq:isingexample}
  to an equivalent QUBO. We compute the coefficients using formulas from the equation \eqref{eq:toQUBO} to obtain:
  \begin{equation}
    \begin{alignedat}{6}
      b_{1} &= 2h_{1} - 2 J_{12} = -4            &\quad & a_{12} &= 4J_{12}=12\\
      b_{2} &= 2h_{2} - 2(J_{12} + J_{23}) = -4 &\quad & a_{23} &= 4J_{23}=-8\\
      b_{3} &= 2h_{3} - 2J_{23} = 8.             & & &
    \end{alignedat}
  \end{equation}
  This gives the following energy function:
  \begin{equation}
    \label{eq:quboexample}
    F(q_{1}, q_{2}, q_{3}) = -4q_{1} -4q_{2}+8q_{3}+12q_{1}q_{2}-8q_{2}q_{3}.
  \end{equation}
  The possible system configurations and their energies are listed in the table
  \ref{eq:quboexample} below. Observe that the difference between QUBO and Ising
  energies for corresponding configurations is always $1$, which is exactly what
  we get if we computed the offset explicitly:
  \begin{equation}
    \mbox{offset} = h_{1} + h_{2} + h_{3} - J_{12} - J_{23} = 1.
  \end{equation}

  \begin{table}[ht!]
    \begin{center}
      \begin{tabular}{|c|c||c|c|}
        \hline
        \rowcolor{theader}
        $\mathbf{q}=(q_1, q_2, q_3)$ &
        $F(\mathbf{q})$              &
        $\mathbf{q}=(q_1, q_2, q_3)$ &
        $F(\mathbf{q})$                                    \\\hline
        (0, 0, 0)                    & 0  & (1, 0, 0) & -4 \\ \hline
        (0, 0, 1)                    & 8  & (1, 0, 1) & 4  \\ \hline
        (0, 1, 0)                    & -4 & (1, 1, 0) & 4  \\ \hline
        (0, 1, 1)                    & -4 & (1, 1, 1) & 4  \\ \hline
      \end{tabular}
    \end{center}
    \caption{All configurations for example QUBO from the equation \eqref{eq:quboexample}.}
    \label{tab:quboexample}
  \end{table}

\end{example}

We will conclude this chapter by discussing alternative notation for QUBO
problems when the problem is defined on a complete graph. In such a case, the
first sum in the equation \eqref{eq:qubo} runs through all the possible pairs
${i, j}$, and thus $F$ can be written as:
\begin{equation}
  \label{eq:qubocomplete}
  F(q_1, \ldots, q_N) =  \sum_{i=1}^N \sum_{j=i+1}^N a_{ij} q_i q_j + \sum_{i=1}^N b_iq_i,
\end{equation}
One can now define a $n \times n$ real symmetric matrix $Q$ with coefficients:
\begin{equation}
  Q_{ij} = \begin{cases}
    b_{i}  & i = j \\
    a_{ij} & i < j \\
    a_{ji} & j < i
  \end{cases}
\end{equation}
Having in mind that squaring a binary variable does not change its value, we
can again rewrite $F$ as:
\begin{equation}
  \label{eq:qubomatrix}
  F(q_{i}, \ldots, q_{N}) = \sum_{i \le j} Q_{ij} q_{i}q_{j} = \sum_{j \le i} Q_{ij} q_{i}q_{j}.
\end{equation}
Moreover, since it is always possible to view any given QUBO as a one defined
on a complete graph (by introducing artificial edges with weights 0), the
equation \eqref{eq:qubomatrix} provides a one--to--one correspondence between
real symmetric matrices and QUBO problems. We will see the benefits of
this correspondence when discussing the brute-force algorithm in Chapter \ref{chapter:bruteforce}.


%% file: near-term-technologies.tex
\chapter{Quantum annealing and GPU computing}
\label{chapter:near-term}

After introducing the Ising model, our next task is to present the technologies
used in the research conducted for this thesis. Since the main point of this
thesis is benchmarking quantum annealers, it is only natural that we start by
introducing the reader to the concepts of adiabatic quantum computations and
quantum annealing. The second part of the chapter is devoted to Nvidia CUDA, a
technology allowing massively parallel computations on general-purpose graphics
processing units (GPUs).

\section{Adiabatic quantum computation and quantum annealing}
\sectionmark{AQC and quantum annealing}

\subsection{Adiabatic Quantum Computation}
One of the possible models of quantum computing is Adiabatic Quantum
Computation (AQC) \cite{farhi}. AQC ties closely with quantum annealing, and
hence we will shortly discuss how it works in general. Before we describe how
the computations are performed in this model, we will take a closer look at the
underlying adiabatic theorem, which can be stated as follows \cite{farhi,
  born}:

\begin{theorem}[Adiabatic theorem]
  Suppose we are give a time-dependent Hamiltonian $\tilde{H}(t)$ with
  eigenenergies $E_1(t) \le E_2(t) \le \ldots \le E_i(t) \le \ldots$ and
  corresponding eigenstates $\ket{\psi_i(t)}$. Further, suppose we are given a
  physical system $\mathcal{S}$ evolving according to $H(t) = \tilde{H}(t/T)$ and
  let $\ket{\psi(t)}$ denote the state of $\mathcal{S}$ at time $t$. If
  $\ket{\psi(0)} = \ket{\psi_n(0)}$, then also $\ket{\psi(t)} = \ket{\psi_n(t)}$
  for all time $t$, provided that $T$ is large enough and for all $t$ there
  exists a non-zero difference between $E_n(t)$ and the rest of the $H(t)$'s
  spectrum.
\end{theorem}

One conclusion to the adiabatic theorem is of particular importance to quantum
computation. If the system is prepared in a ground state, has a non-zero gap
between its ground energy and the energy of the first excited state, and is
evolved slowly enough, it will stay in the ground state during the whole
evolution. Knowing this we can finally discuss how AQC works. First, an
optimization problem to be solved is encoded as a ground state of some
Hamiltonian $\HH_{\text{target}}$. Then, a physical system is prepared in a ground
state of some simpler Hamiltonian, $\HH_{\text{initial}}$. After that, the system is
driven slowly from $\HH_{\text{initial}}$ to $\HH_{\text{target}}$. By adiabatic theorem, the
system ends up in a ground state of $\HH_{\text{target}}$, and after the measurement
is performed the solution to the original problem can be decoded.

In Quantum Annealing (QA), one follows essentially the same procedure as in
Adiabatic Quantum Computing. What is different, is that the evolution of the
system in QA does not have to be adiabatic \cite{Vinci2017}. We will describe
in more detail how Quantum Annealing works on a concrete example later in this
chapter when we discuss D-Wave annealers.

\subsection{D-Wave quantum annealers}

The first commercially available quantum annealer was D-Wave One, which was
introduced by D-Wave company in $2011$ \cite{johnson}, featuring 128 qubits.
Since then, multiple improved generations of D-Wave annealers have been
released. At the time of writing, the newest series of D-Wave annealers is
called the Advantage system. Devices in this series utilize a chip with at
least $5000$ qubits. Table \ref{tab:dwave} summarizes the release history of
D-Wave annealers and highlights the differences between their generations.

As already mentioned in the introduction, D-Wave annealers are built to find
the ground states of the classical Ising spin--glasses. In these devices, the
spin variables correspond to physical two-level systems, called qubits, which
are implemented using Josephson junctions \cite{chad,bauch2006quantum}. At the end of the
annealing process, the (quantum) Hamiltonian of the annealer has to correspond
to the classical Ising Hamiltonian of the spin-glass instance being solved,
i.e.:
\begin{equation}
  \HH_{\text{target}} = \sum_{i=1}^N h_i \ssigma^{(i)}_z + \sum_{<i, j>} J_{ij} \ssigma^{(i)}_z \ssigma^{(j)}_z,
\end{equation}
where $N$, as previously, is the number of spins, $\ssigma^{(i)}_x$,
$\ssigma^{(i)}_z$ denote Pauli operators acting on $i$-th qubit, and $h_i,
  J_{ij} \in \mathbb{R}$ are coefficients of the instance. Note that finding the
ground state of such Hamiltonian is equivalent to finding the ground state of
its classical counterpart (c.f. eq. \eqref{eq:ising-hamiltonian}). For small
system sizes, this can be accomplished by listing all possible configurations
and sorting them with respect to their energies.

\begin{figure}
  \centering
  \includegraphics[width=0.7\textwidth]{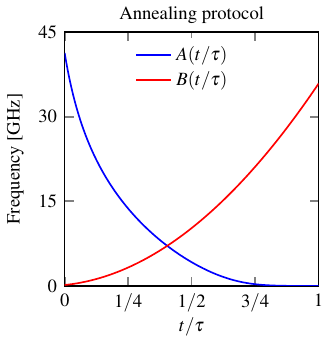}
  \caption{A typical shape of the $A$ and $B$ curves defining the annealing protocol on D-Wave annealers.
    Here, $\tau$ denotes the annealing time. Actual values of the functions vary between the devices.
  }
  \label{fig:annealingfuncs}
\end{figure}

More precisely, the time-dependent Hamiltonian implemented by the D-Wave
devices is of the form:
\begin{equation}
  \label{eq:dwaveham}
  \HH(t) =  -\frac{A(t)}{2}\sum_{i=1}^N \ssigma^{(i)}_x + \frac{B(t)}{2}\left(\sum_{i=1}^N h_i \ssigma^{(i)}_z + \sum_{<i, j>} J_{ij} \ssigma^{(i)}_z \ssigma^{(j)}_z\right).
\end{equation}
where $t \in [0, \tau]$ \cite{dwavedocs}. The tunneling energy curve $A(t)$ is
monotonically decreasing and it vanishes as $t$ approaches $\tau$. Similarly
$B(t)$ is monotonically increasing, and the functions satisfy $A(0) \gg B(0)$
and $B(\tau) \gg A(\tau)$. Illustrative plots of the functions $A$ and $B$ are presented in Fig. \ref{fig:annealingfuncs}.

Since the variables in the spin--glass being solved have to correspond to the
physical qubits, it is clear that the number of qubits of the device limits the
size of the input problem. However, it is not the only factor restricting
problems that can be directly submitted to the annealer. To implement quadratic
terms in the Ising hamiltonian, the qubits have to be physically connected via
a \emph{coupler}. The available connectivity on the device depends on two
factors. The first one is the \emph{topology} of the chip, i.e. graph
describing its qubits layout. The topology is the same for all devices in the
same generation. However, due to manufacturing errors and calibration problems,
some qubits and/or couplers might be unavailable to the end user. The graph
describing qubits' connectivity of an actual device is called its \emph{working
  graph}. Understanding annealer topologies is crucial for understanding how to
program these devices. Hence, in the next section, we will describe topologies
of all currently available D-Wave annealers

\begin{table}
  \footnotesize
  \centering

  \begin{tabular}[pos]{|>{\columncolor{tsubheader}}l|c|c|c|c|}
    \hline
    \rowcolor{theader}
    \textbf{Series}       &
    \textbf{Release year} &
    \textbf{Topology}     &
    \textbf{Num. qubits}  &
    \textbf{Num. couplers}                                      \\
    \hline
    D-Wave One            & 2011      & $C_{4}$  & 128  & 352   \\
    \hline
    D-Wave Two            & 2013      & $C_{8}$  & 512  & 1472  \\
    \hline
    D-Wave 2X             & 2015      & $C_{12}$ & 1152 & 3360  \\
    \hline
    D-Wave 2000Q          & 2017      & $C_{16}$ & 2048 & 6016  \\
    \hline
    Advantage             & 2020      & $P_{16}$ & 5640 & 40484 \\
    \hline
    Advantage 2           & 2023-2024 & $Z_{15}$ & 7440 & 71736 \\
    \hline
  \end{tabular}
  \caption{
    Comparison of different generations of D-Wave annealers. For topologies, $C_n$,
    $P_n$ and $Z_n$ refers to Chimera, Pegasus and Zephyr of size $n$ respectively.
    The numbers of qubits and couplers are given for a perfectly manufactured chip
    with full yield. Actual devices typically have a lower number of qubits or
    couplers.}. \label{tab:dwave}
\end{table}


\subsection{Annealer topologies}

The first topology that we will discuss in this chapter is the \emph{Chimera}
topology, used for all generations of D-Wave devices up to D-Wave 2000Q series.
We decided to describe the Chimera before moving towards newer topologies
because it serves as a building block for its successors.

While discussing the topologies of the D-Wave annealers, we will not discuss
the physical structure of the chip. We decided to do so because, for this
thesis, the \emph{logical} structure of the chip is far more important than the
underlying physical one. However, one consequence of this choice is that the
distinction between two types of couplers (external and internal) will become
less intuitive once we reach beyond the Chimera topology. Nevertheless, we
believe that this will not impair the reader's ability to understand the layout
of qubits in the newer devices. For the description of the physical chip
layouts, we refer the reader to \cite{dwavedocs}.

\subsubsection{The Chimera topology}

\begin{figure}
  \begin{subfigure}[b]{0.5\textwidth}
    \centering
    \includegraphics[width=0.8\textwidth]{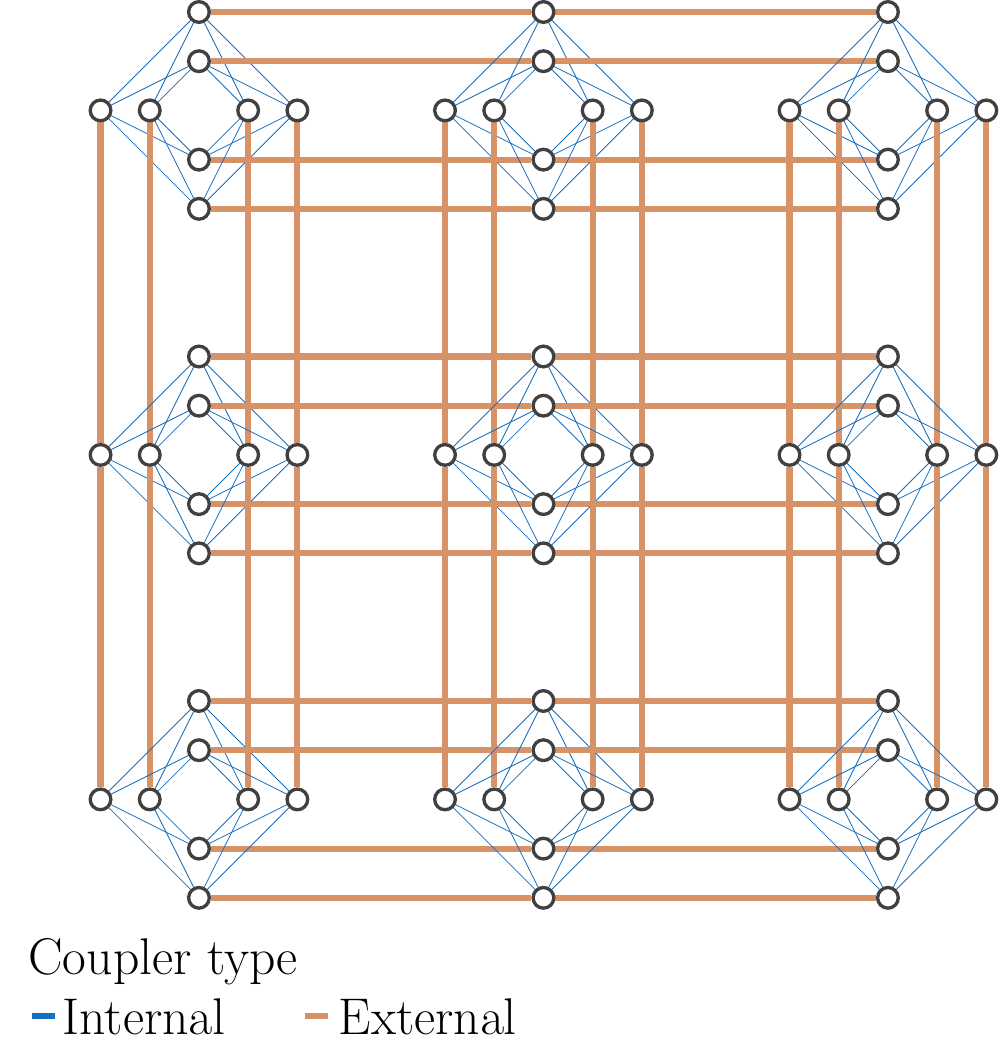}
    \caption{}\label{fig:chimera-cross}
  \end{subfigure}
  \begin{subfigure}[b]{0.45\textwidth}
    \centering
    \includegraphics[width=0.8\textwidth]{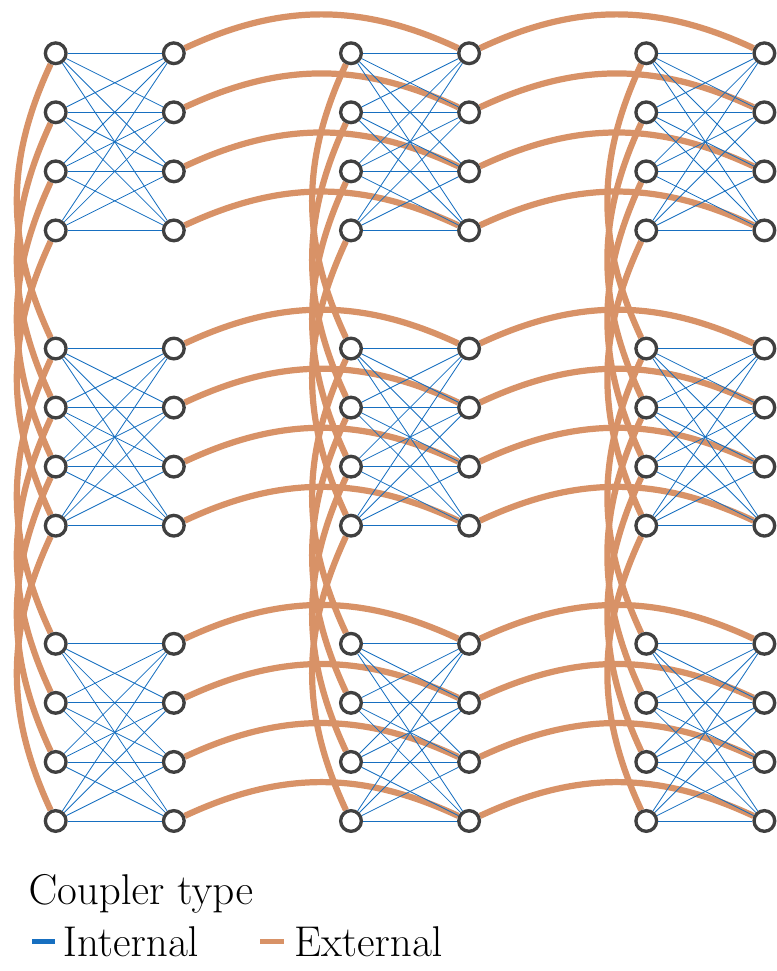}
    \caption{}\label{fig:chimera-shore-column}
  \end{subfigure}
  \caption{
    Chimera $C_3$ topology drawn using different layouts.
    \subref{fig:chimera-cross} The cross layout. \subref{fig:chimera-shore-column}
    The grid layout. The internal couplers are marked with {\color{RoyalBlue}
      blue}, and the external couplers are marked with bold {\color{Tan} orange}. }
  \label{fig:chimera}
\end{figure}

In Chimera topology, depicted in Fig. \ref{fig:chimera}, the qubits are placed
on a rectangular grid of \emph{unit cells}. Every unit cell is a complete
bipartite graph $K_{t,t}$. Each group in the bipartition is called
\emph{shore}, and hence the parameter $t$ is called the \emph{shore size}. Each
qubit in the unit cell (except the ones in the cells on the border) connects to
two qubits on the same shore in the neighboring cells. Hence, the whole Chimera
graph is also bipartite, and the maximum degree of a node is $t+2$. The
couplers connecting qubits in the same unit cell are called \emph{internal} and
the couplers connecting qubits belonging to different cells are called
\emph{external}.

Typically, the devices using Chimera topology utilize a square grid with a
shore size of $4$. Such layouts are denoted by $C_n$, where $n$ is the width
(and the height) of the grid. In such devices, each qubit is connected to a
maximum of 6 qubits, the total number of qubits is $8n^2$ and the total number
of couplers is $16n^2 + 8(n-1)n$.

The Chimera topology is often visualized using two distinct layouts, both of
which are exemplified in Fig. \ref{fig:chimera}. In the cross layout, the
shores of the unit cell form a cross, with one shore being placed vertically
and the second shore being placed horizontally. In the grid layout, each unit
cell is depicted as two columns of qubits, corresponding to both shores of the
cell. While the grid layout might be more intuitive in some applications, cross
layouts are in turn convenient when presenting the Pegasus topology, which we
will discuss next.

\subsubsection{The Pegasus topology}
\begin{figure}
  \includegraphics[width=\textwidth]{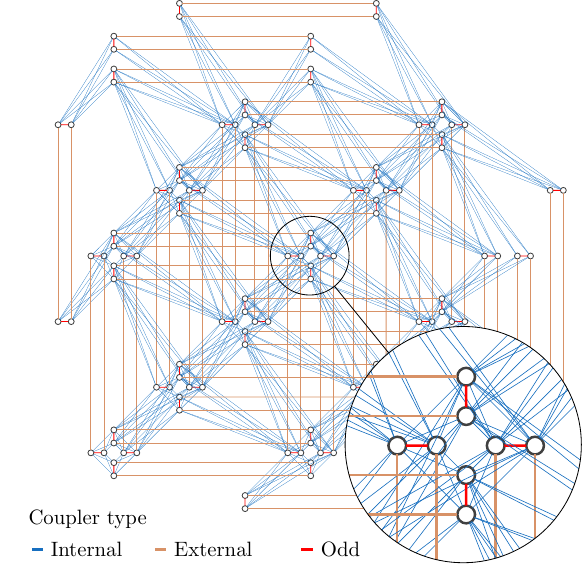}
  \caption{
    The $P_3$ graph, an example of the Pegasus topology. The magnified portion of
    the image shows a part of the graph containing a Chimera unit cell. Observe the
    odd couplers, marked in red, connecting qubits that are not connected in a unit
    cell of Chimera topology. } \label{fig:pegasus}
\end{figure}

The current generation of D-Wave devices, dubbed the Advantage System, uses a
topology called Pegasus \cite{boothby}. An example of this topology is
presented in Fig. \ref{fig:pegasus}. The unit cell of Pegasus comprises 24
qubits grouped into the 3 Chimera unit cells. The topology features several
improvements regarding the qubit connectivity. Firstly, the internal couplers
connect not only the qubits in the same Chimera unit cell but also connect some
neighboring Chimera unit cells. Secondly, inside the Chimera unit cells new
type of connection, called \emph{the odd} couplers, is introduced.
Interestingly, those modifications mean that the Pegasus graph is no longer
bipartite. The Pegasus topology having $n$ rows and $n$ columns of unit cells
is denoted by $P_n$ and contains $24n(n-1)$ qubits.

Observe that a graph in Pegasus topology features subgraphs isomorphic to
Chimera graphs. This fact is important for the annealer users, as all problems
instances compatible with a device using the Chimera topology are automatically
compatible with annealers using a sufficiently large Pegasus topology.

\subsection{The Zephyr topology}
The upcoming generation of D-Wave annealers, called Advantage 2 System, will
use the Zephyr topology \cite{zephyr}. This topology utilizes Chimera unit
cells with a shore size of 8 and, compared to the Pegasus topology, contains
more odd couplers. Overall, the maximum degree of a qubit in Zephyr topology is
20. The Zephyr topology containing $n$ unit cells is denoted $Z_n$ and contains
$16n(2n+1)$ qubits.

\begin{figure}
  \includegraphics[width=\textwidth]{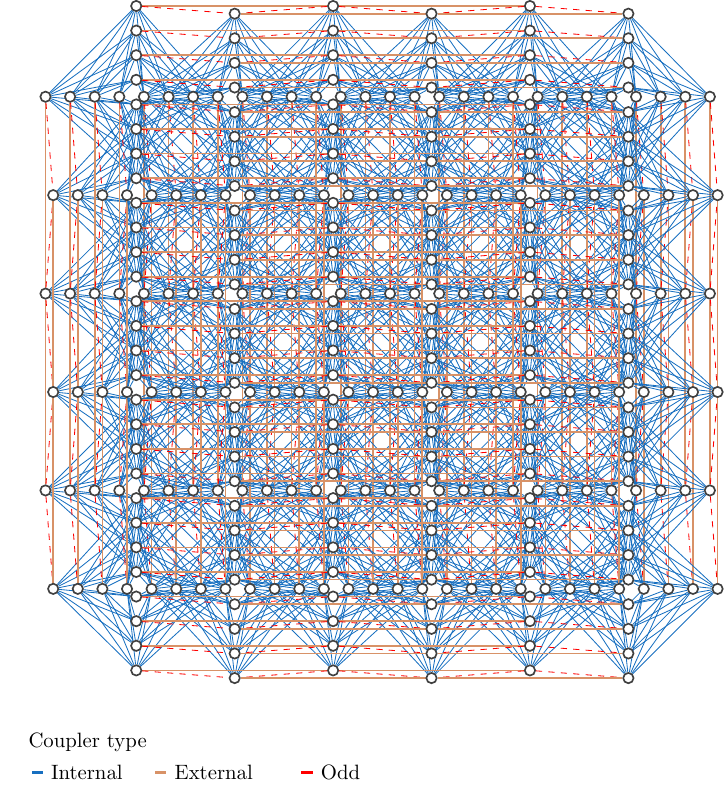}
  \caption{
    The $Z_3$ graph, an example of the Zephyr topology. Different types of couplers
    are color--coded. Observe that, similarly to Pegasus, the Zephyr topology
    contains Chimera subgraphs. However, the shore size of the Chimera unit cells
    in Zephyr is 8 instead of 4. } \label{fig:zephyr}
\end{figure}

\subsection{Minor embeddings}

Oftentimes, even small (relatively to the available number of qubits) instances
are not compatible with the annealer because of its restricted connectivity.
This issue can sometimes be mitigated using a procedure called the \emph{minor
  embedding}, in which the number of qubits is sacrificed for an improvement in
connectivity. Informally, the minor embedding relies on constructing a new
\emph{logical} graph with which the Ising instance to be solved is compatible.
This, in turn, is achieved by introducing \emph{logical} qubits built from
several physical qubits (a process called \emph{contraction}). For the reasons
explained later in this section, we will require all qubits forming the logical
qubit to be connected in a chain. The logical qubit constructed this way
inherits all the neighbors of its physical qubits, and thus one ends up with a
more densely connected graph, albeit with a lower number of qubits. Before
formalizing this idea, let us first present an example of minor embedding.

\begin{example}[Minor embedding]
  \label{ex:minor-embedding}
  \begin{figure}[b]
    \includegraphics[width=\textwidth]{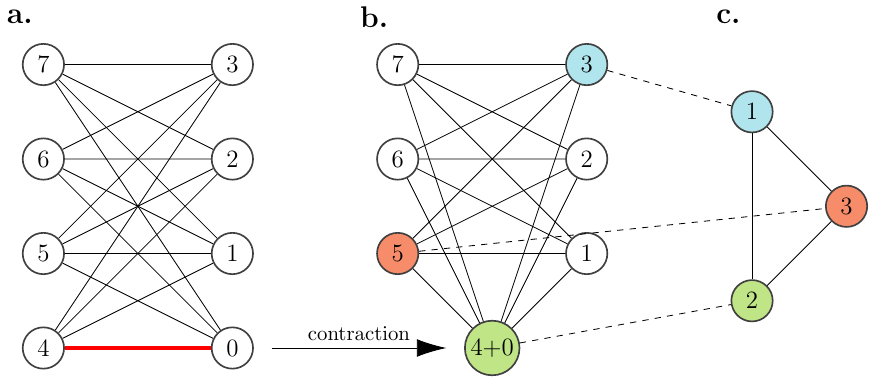}
    \caption{
      Example of minor embedding. Spin--glasses defined on the graph $G$
      (\textbf{c.}) cannot be directly solved on an annealer with $C_1$ topology
      (\textbf{a.}). By contracting neighboring vertices 4 and 0 one obtains a new
      logical graph $C_1'$ (\textbf{b.}), which contains problem graph $G$ as a
      subgraph. } \label{fig:minor-embedding}
  \end{figure}

  Consider an annealer with $C_1$ topology and an (arbitrary) Ising spin--glass
  instance defined on a triangular graph $G$ as depicted in Fig.
  \ref{fig:minor-embedding}. No such instance is compatible with $C_1$, because
  graph $G$ is not bipartite. Combining qubits $0$ and $4$ into a single logical
  qubit yields a graph depicted in Fig. \ref{fig:minor-embedding}\textbf{b.},
  with which $G$ is compatible. If one could use an annealer with this logical
  graph, then instances defined on $G$ could be solved directly. Note that
  vertices 0 and 4 are not the only choice in this case. Indeed, every
  contraction of two vertices in $C_1$ would be sufficient to embed $G$.
\end{example}

As demonstrated in the example, contracting vertices in the annealer's working
graph can make it compatible with an Ising instance otherwise unsolvable by the
annealer. It only remains to explain how this new logical graph can be used
with the actual device.

The idea is to make all the physical qubits in the chain behave like a single
qubit. Since the qubits are connected, one can couple them, including a penalty
large enough that violating qubits' alignment would prohibitively increase the
energy of the solution. The next example presents this idea.

\begin{example}[Minor embedding, continued]
  Consider an Ising spin--glass instance with Hamiltonian:

  \begin{equation}
    H(s_1, s_2, s_3) = s_1 + s_2 + s_3 - s_1s_2 - s_2s_3
  \end{equation}
  with an unique optimal solution $\mathbf{s} = (-1, -1, -1)$. Suppose we want to
  solve it on annealer with $C_1$ topology, using the minor embedding presented
  in Example \ref{ex:minor-embedding}. The problem submitted to the annealer will
  have the following Hamiltonian:
  \begin{equation}
    H'(z_0, z_3, z_4, z_5) = \underbrace{z_3 + z_5}_{s_1 + s_3} + \underbrace{0.5(z_0 + z_4)}_{s_2} - \underbrace{z_3z_4}_{s_2s_1} - \underbrace{z_0z_5}_{s_2s_3} + \underbrace{P(z_0, z_4)}_{\text{penalty}},
    \label{eq:embeddedexample}
  \end{equation}
  where $z_i$ is a spin variable associated to $i$-th qubit and $P(z_0, z_4)$ is
  a \emph{penalty term} for chain comprising qubits 0 and 4. The penalty is of
  the form:
  \begin{equation}
    P(z_0, z_4) = -\alpha z_0z_4
  \end{equation}
  where $alpha$ is a positive constant. Let us examine all possible
  configurations of $(z_0, z_3, z_4, z_5)$ and observe how the penalty term
  influences their energy:
  \begin{table}[h]
    \centering
    \begin{tabular}{|c|c|c|c|}
      \hline
      \rowcolor{theader} \multicolumn{2}{|c|}{feasible solutions} & \multicolumn{2}{c|}{infeasible solutions}                                          \\
      \hline
      \rowcolor{tsubheader} $z_0, z_3, z_4, z_5$                  & energy                                    & $z_0, z_3, z_4, z_5$ & energy          \\
      \hline
      -1, -1,-1, -1                                               & $-5.0 -\alpha$                            & 1, 1,1, -1           & $1.0 + \alpha$  \\
      -1, 1,1, -1                                                 & $-2.0 -\alpha$                            & 1, -1,1, -1          & $1.0 + \alpha$  \\
      -1, -1,1, -1                                                & $-2.0 -\alpha$                            & -1, -1,-1, 1         & $-1.0 + \alpha$ \\
      1, 1,-1, 1                                                  & $2.0 -\alpha$                             & 1, 1,-1, -1          & $2.0 + \alpha$  \\
      1, -1,-1, 1                                                 & $-2.0 -\alpha$                            & 1, -1,-1, -1         & $-2.0 + \alpha$ \\
      -1, 1,-1, -1                                                & $-1.0 -\alpha$                            & -1, -1,1, 1          & $2.0 + \alpha$  \\
      1, -1,1, 1                                                  & $1.0 -\alpha$                             & -1, 1,1, 1           & $2.0 + \alpha$  \\
      1, 1,1, 1                                                   & $1.0  -\alpha$                            & -1, 1,-1, 1          & $3.0 + \alpha$  \\
      \hline
    \end{tabular}
    \caption{All possible configurations for the instance from the equation
      \eqref{eq:embeddedexample}.} \label{tab:embedded}
  \end{table}

  If $\alpha$ is large enough, e.g. $\alpha=2$, all feasible solutions (left part
  of the table) have energy lower than any solution in which qubits $z_0$ and
  $z_4$ are misaligned (right part of the table), which increases a chance of
  sampling them on a physical device. On the other hand, if $\alpha=1$, the
  feasible solution $(1,1,1,1)$ has higher energy than the infeasible solution
  $(1, -1, -1, -1)$.
\end{example}

As demonstrated in the example above, when performing minor embedding it is
important to correctly choose the chain strengths. Typically, it is not
possible to choose a correct $\alpha$ with certainty. In practice, one
typically tries different chain strengths and tests how well they perform for
the given problem.

Since the annealers are inherently heuristic devices, even with carefully
chosen chain strength one might obtain solutions that cannot be decoded into
feasible solutions to the original problem because the qubits forming chains
are misaligned. This situation is known as a \emph{chain break}, and there are
two most commonly used strategies for dealing with it:

\begin{itemize}
  \item discarding the incorrect samples. This is the simplest method, but it reduces
    the total number of samples. Hence, the experiments needing some fixed number
    of samples have to adapt and e.g. sample from the annealer multiple times until
    the desired number of feasible samples is collected.
  \item \emph{majority voting}: whenever the chain of qubits is misaligned, choose the most common value
    among the chain and use it as a value of the logical qubit. In case of a tie, choose
    -1 or 1 with equal probability.
\end{itemize}

Having presented all the necessary information about quantum annealers, we can
conclude this section with a discussion on how quantum annealing differs from
the classical model of computation.

\subsection{Comparison to the classical model of computation}
It is clear that quantum annealing is different from classical computations.
One of the most obvious differences is the computational model. On classical
computers, one essentially writes programs as a series of instructions to be
executed by the CPU. On typical machines, the CPU is capable of performing
arithmetic operations, computing values of some special functions, managing
execution flow, controlling I/O and much more. In comparison, chips in quantum
annealers are capable of executing a single operation: annealing a given
optimization problem. Therefore, programming these devices boils down to
defining an optimization problem and tuning the annealing parameters.

Another difference between classical computers and quantum annealers is the
lack of working memory in the former. Classical computers use working memory
(typically in the form of RAM) to store machine code and data. However, quantum
annealers do not need to store neither code nor data, and hence they do not
feature an analogous component. Similarly, quantum annealers, being purely
computationally oriented devices, do not have mass storage.

A slightly less obvious difference between classical computers and quantum
annealers is their model of parallelism. Classical computers are capable of
running several threads of execution at the same time. However, every
non-trivial classical algorithm involving parallelism must necessarily also
include a serial part which limits speedup gained for introducing more
execution units (CPU cores or CPUs). In contrast, quantum annealers are capable
of annealing multiple qubits at the same time, making their operation
inherently parallel.

\section{Nvidia CUDA}

Quantum annealing, introduced in the previous section, is a heuristic process.
Like many heuristic algorithms, it cannot certify that the solution it found is
optimal. One way to assess the performance of such algorithms is to compare
their results with known low-energy spectra of some test instances. Another
viable approach is to compute the exact low energy spectra of some test
instances, which in turn requires an exact solver. In particular, one might
perform an exhaustive search over all possible states and extract only the
selected number of the ones with the lowest energy, the approach also known as
the brute-force approach. In Chapter \ref{chapter:bruteforce} we demonstrate a
massively parallel implementation of the brute-force algorithm using Nvidia
CUDA, but before we do this, in this section we will introduce the basic
principles of using CUDA-enabled graphic processing units.

\subsection{Brief history of Graphics Processing Units}
The history of specialized hardware for manipulating graphics ranges as far as
the 1970s \cite{framebuffer}. Initially, these devices, which later became
known as Graphics Processing Units (GPUs), offered limited functionalities.
Increasing demand for performance in the gaming industry and professional
graphics processing drove the evolution of GPUs, which eventually became highly
sophisticated devices supporting advanced 2D and 3D image manipulation.
Performing such arithmetically intensive operations requires enormous
computational power, and it was only a matter of time until it was realized
that the power of these devices could be harnessed for the general purpose
computations (so-called GPGPU - General Purpose computing on GPU).

The early forms of GPGPU required framing of computational problems in terms of
operations performed on graphical primitives, as this was the only way for
using specialized API of GPUs \cite{earlygpgpu,fastmatrixmultiplies}. This
changed with the development of devices, toolkits and frameworks that supported
operations needed for the general-purpose computations out of the box. Notable
examples of such computational frameworks include Nvidia CUDA \cite{CUDAguide,
  CUDAvsOpenCL, CUDAExperience} (released in 2007), ATI/AMD FireStream
\cite{firestream1,firestream2} (2006) and ROCm \cite{ROCm1,ROCm2,ROCm3} (2016)
or OpenCL \cite{CUDAvsOpenCL} (2009). The research presented in this thesis was
conducted using Nvidia CUDA-capable devices, which is why in the rest of this
chapter we focus solely on CUDA framework.

\subsection{Differences between CPU and GPU}
The principles behind the operation of CUDA-enabled GPUs are fundamentally
different from the ones governing the execution of a program on traditional
CPU-only architecture. In current x86--based computers, the CPU runs a given
sequence of instructions (so-called thread of execution) using one of its
cores. Such a processor is the ''brain'' of a computer, and it can perform a
wide variety of tasks ranging from arithmetic operations, through accessing the
system's RAM, to performing IO operations and controlling other components of
the system. Typical CPUs are optimized for sequential execution, and as such
are usually equipped with moderate (as compared to the GPUs) number of
high-performance cores.

On the other hand, GPUs are more specialized. They are well suited for
performing numerous arithmetic operations and accessing memory in parallel.
They typically have more cores than a traditional CPU (with even modern
commodity GPUs boasting thousands of them). Although those cores are less
performant than their CPU counterparts and support a much narrower set of
operations, their large number combined with fast memory access gives modern
GPUs an advantage over CPUs in multiple areas.

\subsection{Processing flow on CUDA}
\begin{figure}[ht]
  \centering
  \includegraphics[width=\textwidth]{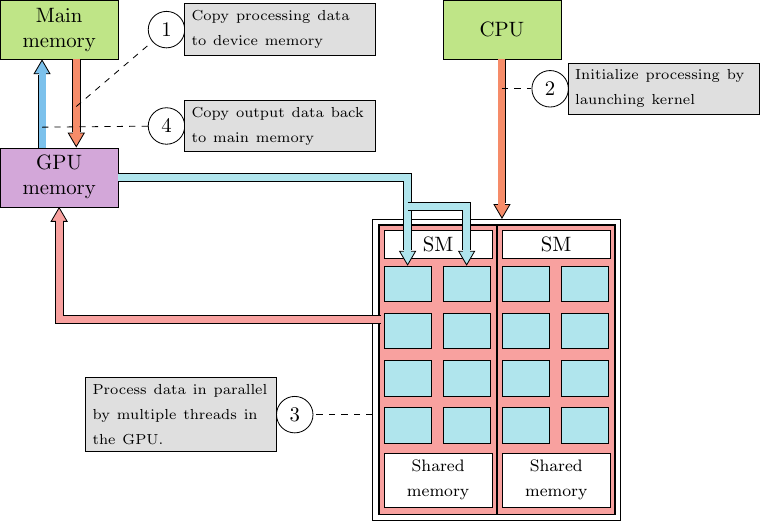}
  \caption{Processing flow on CUDA. The CPU sends input data to the GPU memory and
    launches the computational kernel. The kernel's code is executed, in parallel,
    using multiple threads on the GPU. Once the execution is done, results are
    copied from the GPU memory to the system's RAM.} \label{fig:cuda_flow}
\end{figure}

Considering the architectural differences between CPUs and GPUs, it is hardly
surprising that both of these types of devices are programmed quite
differently. The first major difference is that GPUs cannot operate on their
own and are themselves controlled by the CPU. This is why CUDA is a type of
\emph{heterogenous} architecture as opposed to CPU-only \emph{homogenous}
architecture. The processing flow on CUDA is summarized in Figure
\ref{fig:cuda_flow}.

\begin{figure}[ht]
  \centering
  \includegraphics[width=0.9\textwidth]{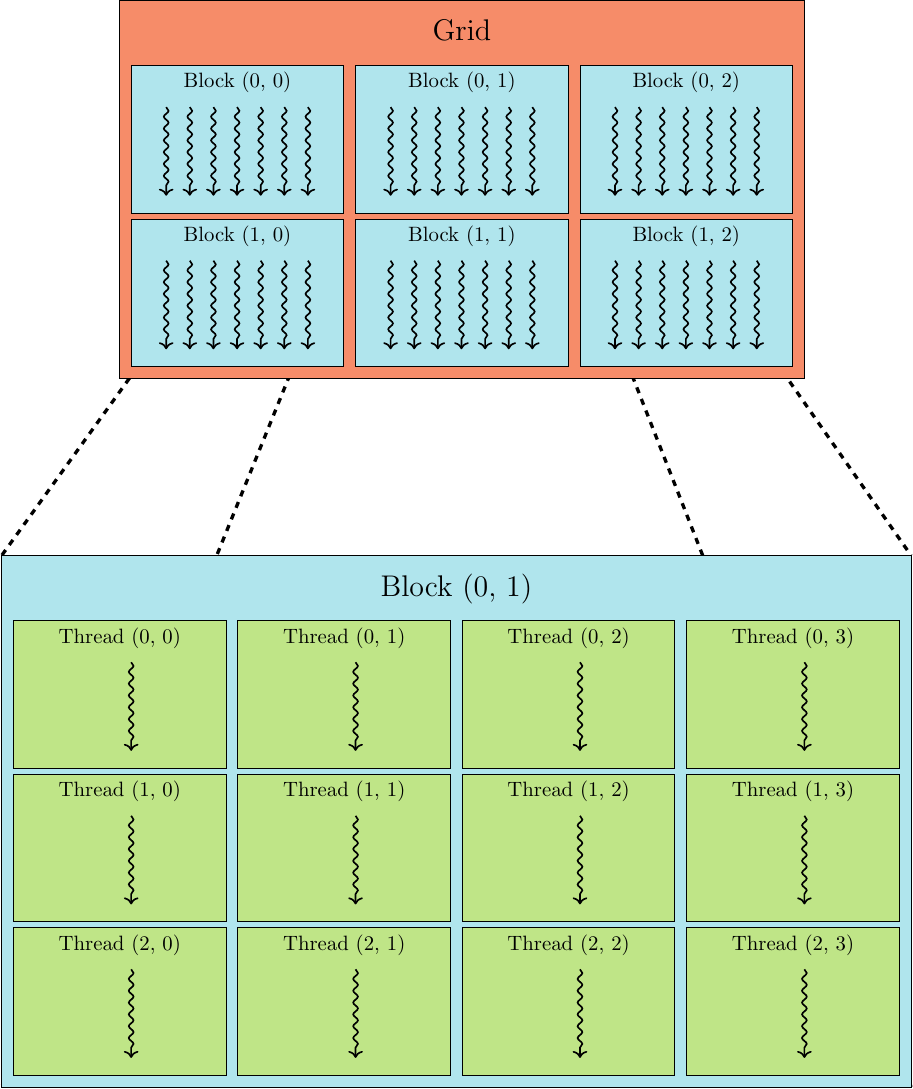}
  \caption{A schematic view of an example two-dimensional CUDA grid. Presented here is a 2
    $\times$ 3 grid of 3 $\times$ 4 blocks.} \label{fig:cuda_grid}
\end{figure}

Programs run on GPU are organized in \emph{kernels}. For the most part, kernels
might be viewed as functions or subroutines (which is indeed how they are
implemented) that don't have a return value. On a CPU, such a function would be
executed by some core as a part of a thread. In CUDA however, the very same
kernel is executed by multiple threads. Executing a kernel requires specifying
a \emph{grid} that will be used for running it. A grid can be 1, 2-- or
3--dimensional and is itself divided into blocks. Each block is in turn also
organized in 1, 2--, or 3-dimensional structure of threads, which has to be the
same for every block in the grid. A schematic view of a two-dimensional grid is
presented in Fig. \ref{fig:cuda_grid}.

As already mentioned, each thread in the grid executes \emph{precisely the
  same} kernel with \emph{precisely the same} parameters. It might therefore seem
surprising that, nevertheless, they can access different parts of memory or
otherwise handle a different part of the computational task. This is possible
because each thread is identified by its indices in both the grid and the
block. Those indices can be used for computing offsets in arrays that are being
processed or (as we will demonstrate in Chapter \ref{chapter:bruteforce})
otherwise used for performing computations.

\subsection{SIMT architecture}
CUDA-enabled GPUs employ an architecture called SIMT (Single Instruction,
Multiple Threads)\footnote{One can contrast SIMT architecture used by CUDA with
  SIMD instructions (Single Instruction, Multiple Data) available on modern
  CPUs.}. As implied by the name, in SIMT architecture, multiple threads execute
the same instruction. Threads are executed in blocks by computational units
called Streaming Multiprocessors (SMs), and blocks are distributed to
multiprocessors on kernel launch. When a block is distributed to SM, it is
further partitioned into \emph{warps}, groups of 32 threads each. All threads
in a warp are scheduled for execution together. Nevertheless, each of them has
a separate program counter and thus their execution flow can diverge. At any
given time, a thread in a warp can be either active (executing \emph{the same}
instruction as the rest of the active threads in a warp) or inactive (not
executing any instruction at all). A thread may be inactive because its
execution diverged from the rest of the warp or because it terminated earlier.
It is interesting to note that starting from the Volta architecture, threads
can be scheduled on a finer level of granularity, allowing them to diverge and
reconverge on the sub-warp level. Each multiprocessor manages a set of 32-bit
registers and a parallel data cache, called \emph{shared memory}, distributed
among the thread blocks. Since those resources are limited, the number of warps
that can run in parallel on any SM is heavily dependent on the resource usage
of the kernel being executed.

\subsection{Memory hierarchy}

On CUDA-enabled devices, threads can access several memory types during kernel
execution, including global memory, local memory, constant and texture memory
and shared memory \cite{CUDAguide}. Physically, those different memory types
can be divided into device memory (global memory, local memory, constant
memory) and on-chip memory (shared memory). SM's on-chip memory also serves as
the L1 cache.

Global memory is a device memory available to all threads. All accesses to
global memory are serviced in 32-, 64-, or 128- bytes memory transactions.
Accesses made from a single warp are coalesced into as many such transactions
as necessary, depending on the device's architecture and access pattern. Reads
and writes targeting global memory are always cached in L2 and (depending on
configuration, compute capability and access pattern) may also be cached in L1
cache.

Local memory in CUDA is only a logical concept. Physically, it resides in the
off-chip memory just like global memory and thus offers the same bandwidth and
latency. Just like global memory, it is always cached in L2 cache. This type of
memory is never used directly by the programmer. Instead, the compiler might
decide to use it for local variables of a thread in case there is not enough
register (so-called \emph{register spilling}) or for dynamically indexed local
arrays. Local memory is arranged in such a way, that accesses are always fully
coalesced as long as all threads access the same relative address (e.g. the
same local variable, the same position of a local array etc.).

Constant and texture memory are two types of read-only
memory\footnote{Read-only here means ``Not writeable from inside the kernel''.}
residing in global memory. Accesses to constant memory are cached in constant
cache and serialized. Therefore, each request is split into as many
transactions as there are different memory addresses in the original request.
Texture memory is cached in the texture cache, which is optimized for accessing
spatial data. Hence, the best performance is achieved if threads in a warp read
or write to the addresses that are placed closely on 2D tiles.

Threads can cooperate and share data through the use of the on-chip
\emph{shared memory}. The amount of allocated shared memory is directly
controlled by the programmer either on the kernel definition level or during
its launch. Shared memory is organized in banks that can be accessed
simultaneously, and the best performance is achieved if each thread in a warp
accesses memory in a different bank. Otherwise, a \emph{bank conflict occurs},
and the request is split into as few conflict-free requests as possible.

\subsection{Programming environment}
CUDA devices can be programmed directly using either C/C++ or Fortran. For both
languages a Nvidia compiler is required to compile the CUDA program, as CUDA
extends C/C++ and Fortran languages with a syntax for defining and launching
kernels. The C/C++ CUDA code can be compiled using Nvidia's \texttt{nvcc}
compiler, shipped out of the box with the CUDA toolkit. For CUDA Fortran code,
the Nvidia High-Performance Computing (HPC) suite contains \texttt{nvfortran}
compiler\footnote{The \texttt{nvfortran} compiler was previously a third--party
  program called \texttt{pgfortran}, developed by PGI \cite{PGI}}. Giving a
comprehensive walkthrough of using either C/C++ or Fortran with CUDA is well
beyond the scope of this thesis, but for the sake of completeness, below we
present a short example of the CUDA C/C++ and CUDA Fortran code.

\begin{example}[Implementing parallel vector addition with CUDA]
  Listings \ref{lst:cuda_fortran} and \ref{lst:cuda_c} below present an example
  implementation of a parallel vector addition using CUDA. The \texttt{addVec}
  defined with \texttt{global} attribute is a kernel and accepts two input
  vectors \texttt{x}, \texttt{y} (in the form of arrays) and their length
  \texttt{n}. Since CUDA kernels cannot return a value, both versions of the code
  accept an additional argument \texttt{res} designating where the result will be
  stored. The \texttt{addVec} kernel can be launched on any one--dimensional grid
  of one--dimensional blocks. Hence, some threads may need to handle more than
  one position in the input arrays. The pattern presented here, where $i$--th
  thread handles positions $i$, $i+N$, $i+2N$, $\ldots$ with $N$ equal total
  number of blocks is called a \emph{grid-stride loop} and is illustrated in Fig.
  \ref{fig:strided-loop}.

  There are several differences between the two code examples stemming from the
  languages used. In CUDA Fortran we can copy data from the host to the GPU using
  array assignment. On the other hand, the equivalent code in C++ requires
  manually calling \texttt{cudaMemcpy} function. Similarly, the GPU arrays in C++
  are declared as pointers, for which the memory has to be manually allocated and
  later deallocated using the combination of \texttt{cudaMalloc} and
  \texttt{cudaFree}. Lastly, C/C++ uses zero-based indexing, whereas Fortran uses
  one-based indexing. This affects how the global thread index, stored in
  \texttt{tid} variable, is computed.

  \begin{listing}
    \begin{Verbatim}[commandchars=\\\{\}]
\PYG{k}{def} \PYG{n+nf}{find\PYGZus{}bit\PYGZus{}to\PYGZus{}flip}\PYG{p}{(}\PYG{n}{i}\PYG{p}{):} \PYG{c+c1}{\PYGZsh{} i starts from 0}
    \PYG{k}{return} \PYG{n}{ffs}\PYG{p}{(}\PYG{n}{gray}\PYG{p}{(}\PYG{n}{i}\PYG{p}{)} \PYG{o}{\PYGZca{}} \PYG{n}{gray}\PYG{p}{(}\PYG{n}{i}\PYG{o}{+}\PYG{l+m+mi}{1}\PYG{p}{))}
\end{Verbatim}
    \caption{Example code in CUDA Fortran implementing parallel addition of vectors on GPU.}
    \label{lst:cuda_fortran}
  \end{listing}

  \begin{listing}
    \begin{Verbatim}[commandchars=\\\{\}]
\PYG{k}{module }\PYG{n}{addVec}
\PYG{k}{contains}
\PYG{k}{   }\PYG{n}{attributes}\PYG{p}{(}\PYG{n}{global}\PYG{p}{)}\PYG{+w}{ }\PYG{k}{subroutine }\PYG{n}{addVec}\PYG{p}{(}\PYG{n}{x}\PYG{p}{,}\PYG{+w}{ }\PYG{n}{y}\PYG{p}{,}\PYG{+w}{ }\PYG{n}{res}\PYG{p}{,}\PYG{+w}{ }\PYG{n}{n}\PYG{p}{)}
\PYG{+w}{      }\PYG{k+kt}{real}\PYG{p}{,}\PYG{+w}{ }\PYG{k}{dimension}\PYG{p}{(}\PYG{o}{*}\PYG{p}{)}\PYG{+w}{ }\PYG{k+kd}{::}\PYG{+w}{ }\PYG{n}{x}\PYG{p}{,}\PYG{+w}{ }\PYG{n}{y}\PYG{p}{,}\PYG{+w}{ }\PYG{n}{res}
\PYG{+w}{      }\PYG{k+kt}{integer}\PYG{p}{,}\PYG{+w}{ }\PYG{k}{value}\PYG{+w}{ }\PYG{k+kd}{::}\PYG{+w}{ }\PYG{n}{n}\PYG{p}{,}\PYG{+w}{ }\PYG{n}{i}\PYG{p}{,}\PYG{+w}{ }\PYG{n}{tid}\PYG{p}{,}\PYG{+w}{ }\PYG{n}{gridsize}

\PYG{+w}{      }\PYG{n}{tid}\PYG{+w}{ }\PYG{o}{=}\PYG{+w}{ }\PYG{p}{(}\PYG{n}{blockidx}\PYG{p}{\PYGZpc{}}\PYG{n}{x}\PYG{+w}{ }\PYG{o}{\PYGZhy{}}\PYG{+w}{ }\PYG{l+m+mi}{1}\PYG{p}{)}\PYG{+w}{ }\PYG{o}{*}\PYG{+w}{ }\PYG{n}{blockdim}\PYG{p}{\PYGZpc{}}\PYG{n}{x}\PYG{+w}{ }\PYG{o}{+}\PYG{+w}{ }\PYG{n}{threadidx}\PYG{p}{\PYGZpc{}}\PYG{n}{x}
\PYG{+w}{      }\PYG{n}{gridsize}\PYG{+w}{ }\PYG{o}{=}\PYG{+w}{ }\PYG{n}{blockdim}\PYG{p}{\PYGZpc{}}\PYG{n}{x}\PYG{+w}{ }\PYG{o}{*}\PYG{+w}{ }\PYG{n}{griddim}\PYG{p}{\PYGZpc{}}\PYG{n}{x}

\PYG{+w}{      }\PYG{k}{do }\PYG{n}{i}\PYG{+w}{ }\PYG{o}{=}\PYG{+w}{ }\PYG{n}{tid}\PYG{p}{,}\PYG{+w}{ }\PYG{n}{n}\PYG{p}{,}\PYG{+w}{ }\PYG{n}{gridsize}
\PYG{+w}{         }\PYG{n}{res}\PYG{p}{(}\PYG{n}{i}\PYG{p}{)}\PYG{+w}{ }\PYG{o}{=}\PYG{+w}{ }\PYG{n}{x}\PYG{p}{(}\PYG{n}{i}\PYG{p}{)}\PYG{+w}{ }\PYG{o}{+}\PYG{+w}{ }\PYG{n}{y}\PYG{p}{(}\PYG{n}{i}\PYG{p}{)}
\PYG{+w}{      }\PYG{k}{end do}
\PYG{k}{   end subroutine}
\PYG{k}{end module}

\PYG{k}{program }\PYG{n}{testAddVec}
\PYG{+w}{   }\PYG{k}{use }\PYG{n}{addVec}
\PYG{+w}{   }\PYG{k}{use }\PYG{n}{cudafor}
\PYG{+w}{   }\PYG{k}{implicit none}
\PYG{k}{   }\PYG{k+kt}{integer}\PYG{p}{,}\PYG{+w}{ }\PYG{k}{parameter}\PYG{+w}{ }\PYG{k+kd}{::}\PYG{+w}{ }\PYG{n}{N}\PYG{+w}{ }\PYG{o}{=}\PYG{+w}{ }\PYG{l+m+mi}{100000}
\PYG{+w}{   }\PYG{k+kt}{real}\PYG{+w}{ }\PYG{k+kd}{::}\PYG{+w}{ }\PYG{n}{x}\PYG{p}{(}\PYG{n}{N}\PYG{p}{),}\PYG{+w}{ }\PYG{n}{y}\PYG{p}{(}\PYG{n}{N}\PYG{p}{),}\PYG{+w}{ }\PYG{n}{res}\PYG{p}{(}\PYG{n}{N}\PYG{p}{)}
\PYG{+w}{   }\PYG{k+kt}{integer}\PYG{+w}{ }\PYG{k+kd}{::}\PYG{+w}{ }\PYG{n}{i}\PYG{p}{,}\PYG{+w}{ }\PYG{n}{nBlocks}\PYG{o}{=}\PYG{l+m+mi}{256}\PYG{p}{,}\PYG{+w}{ }\PYG{n}{nThreads}\PYG{o}{=}\PYG{l+m+mi}{128}
\PYG{+w}{   }\PYG{k+kt}{real}\PYG{p}{,}\PYG{+w}{ }\PYG{n}{device}\PYG{+w}{ }\PYG{k+kd}{::}\PYG{+w}{ }\PYG{n}{x\PYGZus{}d}\PYG{p}{(}\PYG{n}{N}\PYG{p}{),}\PYG{+w}{ }\PYG{n}{y\PYGZus{}d}\PYG{p}{(}\PYG{n}{N}\PYG{p}{),}\PYG{+w}{ }\PYG{n}{res\PYGZus{}d}\PYG{p}{(}\PYG{n}{N}\PYG{p}{)}

\PYG{+w}{   }\PYG{k}{do }\PYG{n}{i}\PYG{+w}{ }\PYG{o}{=}\PYG{+w}{ }\PYG{l+m+mi}{1}\PYG{p}{,}\PYG{n}{N}
\PYG{+w}{      }\PYG{k}{call }\PYG{n+nb}{random\PYGZus{}number}\PYG{p}{(}\PYG{n}{x}\PYG{p}{(}\PYG{n}{i}\PYG{p}{))}
\PYG{+w}{      }\PYG{k}{call }\PYG{n+nb}{random\PYGZus{}number}\PYG{p}{(}\PYG{n}{y}\PYG{p}{(}\PYG{n}{i}\PYG{p}{))}
\PYG{+w}{   }\PYG{k}{end do}

\PYG{k}{   }\PYG{n}{x\PYGZus{}d}\PYG{+w}{ }\PYG{o}{=}\PYG{+w}{ }\PYG{n}{x}
\PYG{+w}{   }\PYG{n}{y\PYGZus{}d}\PYG{+w}{ }\PYG{o}{=}\PYG{+w}{ }\PYG{n}{y}

\PYG{+w}{   }\PYG{k}{call }\PYG{n}{addVec}\PYG{o}{\PYGZlt{}\PYGZlt{}\PYGZlt{}}\PYG{n}{nBlocks}\PYG{p}{,}\PYG{+w}{ }\PYG{n}{nThreads}\PYG{o}{\PYGZgt{}\PYGZgt{}\PYGZgt{}}\PYG{p}{(}\PYG{n}{x\PYGZus{}d}\PYG{p}{,}\PYG{+w}{ }\PYG{n}{y\PYGZus{}d}\PYG{p}{,}\PYG{+w}{ }\PYG{n}{res\PYGZus{}d}\PYG{p}{,}\PYG{+w}{ }\PYG{n}{N}\PYG{p}{)}
\PYG{+w}{   }\PYG{n}{res}\PYG{+w}{ }\PYG{o}{=}\PYG{+w}{ }\PYG{n}{res\PYGZus{}d}

\PYG{+w}{   }\PYG{k}{write}\PYG{p}{(}\PYG{o}{*}\PYG{p}{,}\PYG{o}{*}\PYG{p}{)}\PYG{+w}{ }\PYG{l+s+s1}{\PYGZsq{}Max error: \PYGZsq{}}\PYG{p}{,}\PYG{+w}{ }\PYG{n+nb}{maxval}\PYG{p}{(}\PYG{n+nb}{abs}\PYG{p}{(}\PYG{n}{res}\PYG{+w}{ }\PYG{o}{\PYGZhy{}}\PYG{+w}{ }\PYG{p}{(}\PYG{n}{x}\PYG{+w}{ }\PYG{o}{+}\PYG{+w}{ }\PYG{n}{y}\PYG{p}{)))}
\PYG{k}{end program }\PYG{n}{testAddVec}
\end{Verbatim}

    \caption{Example code in CUDA C/C++ implementing parallel addition of vectors on GPU.}
    \label{lst:cuda_c}
  \end{listing}

  \begin{figure}
    \includegraphics[width=\textwidth]{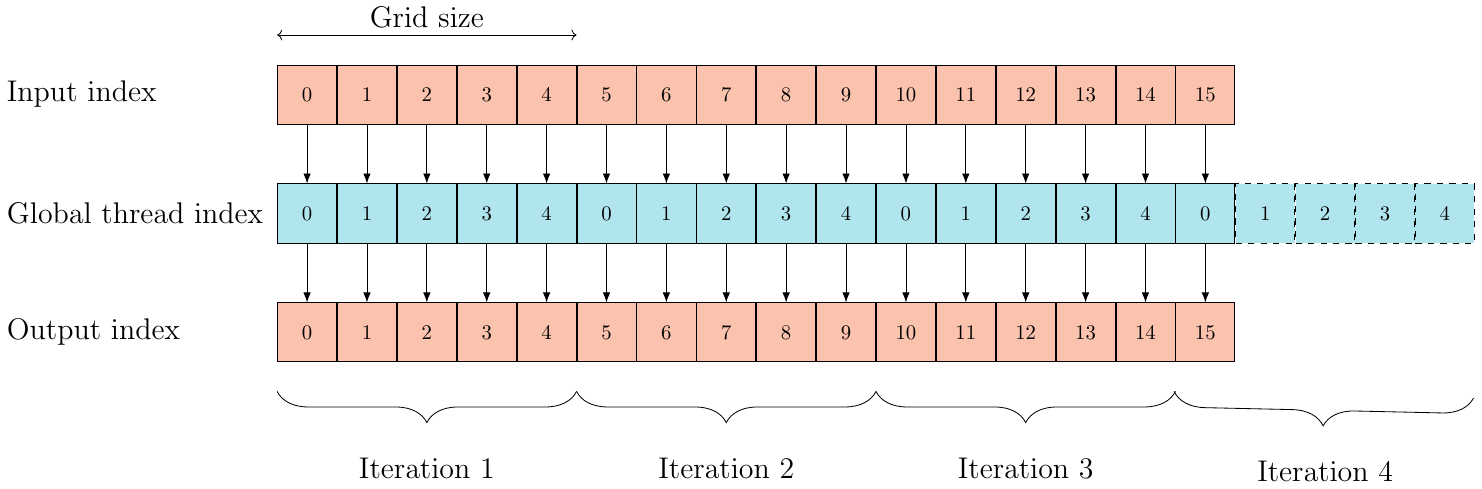}
    \caption{A schematic representation of a GPU kernel transforming an input array into an
      output array of the same size using a grid-stride loop pattern. Here, both
      arrays are of size $k=16$ and the hypothetical grid comprises $l=5$ threads
      (the exact grid configuration is irrelevant). During the first iteration, the
      $i$-th thread accesses $i$-th input element, transforms it and stores the new
      value in $i$-th element of the output array. In subsequent iterations, each
      thread advances the index it processes by the stride equal to the total grid
      size. Observe that for the last iteration only the first thread needs to do
      processing and remaining 4 threads, marked with dashed line, remain inactive. }
    \label{fig:strided-loop}
  \end{figure}

\end{example}

\subsection{Software ecosystem}
Along with the \texttt{nvcc} compiler, the CUDA toolkit contains several, more
specialized libraries. Among others, those include:
\begin{itemize}
  \item cuBLAS \cite{cublas} -- CUDA Basic Linear Algebra Subroutines library,
  \item cuFFT \cite{cufft} -- CUDA Fast Fourier Transform library,
  \item cuRAND \cite{curand} -- CUDA Random Number Generation library,
  \item cuSPARSE \cite{cusparse} -- CUDA library for manipulating sparse matrices,
  \item thrust \cite{thrust} -- parallel algorithms library. Thrust also provides
    parallel implementations of its algorithms that can be run on traditional CPU,
    making it usable even without CUDA.
\end{itemize}
For some high-level languages, there exist third-party libraries enabling the
usage of CUDA. For Python, one could mention e.g. \texttt{PyCUDA}
\cite{pycuda}, \texttt{CuPy} \cite{cupy} or recently introduced
\texttt{setuptools\_cuda} \cite{setuptoolscuda} created by the author of this
thesis. In Julia, integration with CUDA can be achieved with \texttt{CUDA.jl}
\cite{CUDAjl} package.


%% file: simulating-dynamics-with-dwave.tex
\chapter{Simulating dynamics of quantum systems using quantum annealing}
\chaptermark{Simulating dynamics}
\label{chapter:simulating}

One of the leading motivations behind the development of quantum computing
devices is simulating quantum systems intractable by classical computers. But
how far are we from this goal? To answer this question, one might design an
algorithm for conducting such a simulation of a physical system and then test
how it performs on the current generation of quantum computers. In this
chapter, we follow this idea and present a possible approach for simulating
quantum systems (or any time-dependent dynamical system) that can be used with
annealing devices such as D-Wave quantum annealers and similar devices. To
illustrate the working of our algorithm, we simulate the simplest single-qubit
system and demonstrate that already near-term annealing devices are capable of
capturing its dynamics in a narrow regime of parameters. Furthermore, the class
of physics-inspired problem instances proposed in this chapter
can be valuable in benchmarking other (not necessarily quantum) solvers.

\section{Parallel in time simulation of dynamical systems}
Optimization problems that can be solved using quantum annealers exhibit no
time--dependence. Therefore, simulating any time-dependent phenomena using
those devices requires reformulating the problem as one that is static in
nature. In our case, it is possible by enlarging the Hilbert space of the
system under consideration, so that the states of this larger space encode also
temporal information \cite{feynmanclock}.

Let us start by precisely defining the problem we want to address.
Consider an $L$--dimensional real or complex system, whose state at time $t$ is
described by the vector $\ket{\psi(t)}$ evolving according to a differential
equation of the form:
\begin{equation}
  \label{eq:dynamical-system}
  \frac{\partial \ket{\psi(t)}}{\partial t} = K(t) \ket{\psi(t)}.
\end{equation}

Here, $K$ is the so-called Kamiltonian \cite{goldstein2002classical} and can be
any linear operator acting on $\mathbb{R}^L$ or resp. $\mathbb{C}^L$. Observe
that any isolated quantum system can be described by equation
\eqref{eq:dynamical-system}, as putting $K=-\frac{i}{\hbar}H$, where $H$ is its
Hamiltonian, transforms the equation \eqref{eq:dynamical-system} into
Schr\"{o}dinger equation.

Given an initial state, $\ket{\psi(t_0)}$ the equation
\eqref{eq:dynamical-system} admits a unique solution:
\begin{equation}
  \ket{\psi(t)} \coloneqq U(t, t_0) \ket{\psi(t_0)},
\end{equation}
where operator $U(t, t_0)$ is a propagator transforming the initial state of
the system into its state at time $t$ and is given by:
\begin{equation}
  \label{eq:propagator}
  U(t, t_0) = \mathcal{T} \exp \left( \int_{t_0}^t K(\tau)d\tau \right).
\end{equation}
Here, $\mathcal{T}$ denotes the time-ordering operation \cite{chronological}.
Note that in the case when $K(t)$ commutes with $K(t')$ for every $t' \ne t$,
the time-ordering can be omitted. In particular, this is the case if $K$ is
time-independent.

Given the initial state, we are interested in finding the state of the system
at some time $t > t_0$. Numerical methods for solving this problem usually
start by partitioning the interval $[t_0, t]$ into $N$ distinct time points
$t_0 < t_1 < \ldots < t_{N-1} = t$. Then, the desired state $\ket{\psi(t)}$ can
be computed as:
\begin{equation}
  \ket{\psi(t)} = U_{N-1} \cdots U_1 \ket{\psi(t_0)},
\end{equation}
where $U_i$ is a shorthand notation for $U(t_i, t_{i-1})$. This is purely a
rearrangement of computations, which by itself gives no benefit over applying
$U(t, t_0)$ directly. However, shortening the interval allows for a more
efficient approximation of propagators, which can be done using a variety of
methods, including Suzuki-Trotter approximation \cite{suzuki}, commutator-free
expansion \cite{commutatorfree} or tensor-networks based approaches
\cite{dmrg}.

This procedure, common to many sequential methods, gives a starting point for a
class of the so-called parallel in--time methods based on the Feynman clock
operator. In these approaches, one starts by suitably enlarging the state space
so that it can encode the temporal data \cite{feynmanclock}. This can be done
by considering a tensor product of a state space with the new Hilbert space
spanned by the orthonormal basis $\{\ket{0}, \ket{1}, \ldots, \ket{N-1}\}$.
Then, the following superposition encodes states of the system in all $N$
moments of time:
\begin{equation}
  \ket{\Psi} = \sum_{n=0}^{N-1} \ket{n} \otimes \ket{\psi(t_n)}.
\end{equation}
Consider now the following \emph{clock operator} $\clockop$:
\begin{equation}
  \label{eq:clock2}
  \begin{split}
    \clockop
    =
    \sum_{n=0}^{N-2}
    &\ketbra{n+1}{n+1} \otimes I - \ketbra{n+1}{n} \otimes U_n + \\
    &\ketbra{n}{n} \otimes I - \ketbra{n}{n+1}\otimes U_{n}^{\dagger}.
  \end{split}
\end{equation}
One can see that $\ket{\Psi}$ is a solution (although not unique) to the
eigenequation:
\begin{equation}
  \label{eq:clock-eigenequation}
  \clockop \ket{\mathbf{x}} = 0.
\end{equation}
The non--uniqueness of the solution of \eqref{eq:clock-eigenequation} follows
from the fact that the definition of the clock operator $\clockop$ does not
depend on the initial state. We can fix this problem by adding a \emph{penalty}
term $\clockop_0 = \ketbra{0}{0}\otimes(I-\ketbra{\psi_0}{\psi_0})$ to the
left-hand side. The equation to solve becomes then:
\begin{equation}
  \label{eq:clock-eigenequation2}
  (\clockop + \clockop_0) \ket{\mathbf{x}} = 0.
\end{equation}
If one puts $\coefmatrix=\clockop + \ketbra{0}{0} \otimes I$ and $\ket{\Phi} =
  \ket{0}\otimes \ket{\psi_0}$, the equation \eqref{eq:clock-eigenequation2}
becomes:
\begin{equation}
  \label{eq:gsys}
  \coefmatrix \mathbf{x} = \ket{\Phi}.
\end{equation}
Thus, using an approximation of evolution operators, we constructed a system of
linear equations encoding the solution to the equation
\eqref{eq:dynamical-system} under the given initial condition. At this point,
however, it is not possible to solve it using a quantum annealer yet. To do so,
one first needs to convert this system into an optimization problem with
dichotomous variables, which will be the topic of the next section.
\section{Solving systems of linear equations as an optimization problem}
There is a straightforward way of converting equation \eqref{eq:gsys} into an
optimization problem. One can observe that the solution minimizes the norm
$\left\Vert \coefmatrix\ket{\mathbf{x}} - \ket{\Phi}\right\Vert$. Since the
norm is non-negative, it follows that solving equation \eqref{eq:gsys} is
equivalent to the following optimization problem:
\begin{equation}
  \label{eq:optimize_1}
  \ket{\Psi} = \argmin_{\mathbf{x}} f(\mathbf{x}), \quad f({\mathbf x}) = \left\Vert \coefmatrix\ket{\mathbf{x}} - \ket{\Phi}\right\Vert^2.
\end{equation}
However, $f$ in the equation \eqref{eq:optimize_1} is not the only choice of a
target function. If $\coefmatrix$ is positive-definite, one can consider the
following function $h$ instead:
\begin{equation}
  \label{eq:optimize_2}
  h(\mathbf{x})=\frac{1}{2}\bra{\mathbf{x}}\coefmatrix\ket{\mathbf{x}} -
  \braket{\mathbf{x}}{\Phi}.
\end{equation}
Indeed, one can verify that solution to \eqref{eq:gsys} also minimizes $h$ by
computing its gradient and Hessian:
\begin{equation}
  \nabla h(\mathbf{x}) = \coefmatrix\ket{\mathbf{x}}-\ket{\Phi}, \quad \nabla^2 h({\mathbf
    x})=\coefmatrix>0
\end{equation}
Since Hessian is positive and $\ket{\Psi}$ is the only vector at which $\nabla
  h$ vanishes, it follows that $\ket{\Psi}$ is indeed a global minimum of $h$.

\section{Discretizing variables}
Thus far, we have been working with continuous variables. The next necessary
step before solving optimization problems \eqref{eq:optimize_1} and
\eqref{eq:optimize_2} using annealer is converting them in such a way that all
unknowns are dichotomous. To this end, we will follow a strategy presented in
\cite{fixedpoint,chang}. The idea is to express each of the unknown
coefficients of $\ket{\mathbf{x}} = $[$x_1, \ldots, x_{LN}]^T$ in fixed-point
approximation. While this strategy was originally described for real matrices,
it works for complex matrices as well, since one can employ the natural embedding of $\CC$
into $\RR^{2 \times 2}$, $a+bi \mapsto a\hat{I} + ib\ssigma_{y}$. Henceforth, we assume
that the considered systems are real.

If one assumes (binary) order of magnitude of coefficients of
$\mathbf x$ to be $D$ (i.e. $x_i \in [-2^D, 2^D]$ for each $i$), then it can be
approximated up to $R$ bits of precision using the formula:
\begin{equation}
  \label{eq:fixed}
  x_i \approx 2^D \left(2 \sum_{\alpha=0}^{R-1}2^{-\alpha}q_i^{\alpha} -1\right).
\end{equation}
Here variables $q_i^\alpha$ are consecutive bits of the fixed-points expansion
of $x_i$. Note that approximation of $x_i$ in \eqref{eq:fixed} is a linear
combination of its bits, therefore plugging it into optimization problems
\eqref{eq:optimize_1} and \eqref{eq:optimize_2} yields quadratic unconstrained
optimization problems of the form:
\begin{align}
  \label{eq:qubo_f}
  \argmin_{\mathbf{q}} f(\mathbf{q}) = \argmin_{\mathbf{q}} \sum_{i,\alpha} c_i^{\alpha} q_i^r + \sum_{i,j,\alpha,\beta} d_{ij}^{\alpha\beta} q_i^{\alpha} q_j^{\beta} + f_0, \\
  \label{eq:qubo_h}
  \argmin_{\mathbf{q}} h(\mathbf{q}) = \argmin_{\mathbf{q}} \sum_{i,\alpha} a_i^{\alpha} q_i^r + \sum_{i,j,\alpha,\beta} b_{ij}^{\alpha\beta} q_i^{\alpha} q_j^{\beta} + h_0.
\end{align}
Coefficients in equations \eqref{eq:qubo_f} and \eqref{eq:qubo_h} can be
straightforwardly computed by appropriate substitutions into equations
\eqref{eq:optimize_1} and \eqref{eq:optimize_2}. For brevity, here we present
only the formulas for the equation \eqref{eq:qubo_h}, which reads:
\begin{eqnarray}
  \begin{split}
    b_{ij}^{\alpha\beta} &= \coefmatrix_{ij} 2^{1-\alpha-\beta+2D} \\
    a_i^\alpha &= \left( 2^{-\alpha+D}\coefmatrix_{ii} - 2^D\sum_{j}\coefmatrix_{ij}- \Phi_i\right)2^{1-\alpha+D},
    \\
    h_0 &= 2^D\left( 2^{D-1}\sum_{ij}\coefmatrix_{ij}+\sum_i \Phi_i\right).
  \end{split}
  \label{eq:coeff}
\end{eqnarray}

Our approach requires the order of magnitude $D$ and precision $R$ in equation
\eqref{eq:fixed} to be chosen beforehand. Choosing the right $D$ requires
knowledge of the range in which coefficients lie. If its value is too small,
the approximations will fail to capture the most significant bits of the real
solution. On the other hand, choosing $D$ that is too large will result in
wasting variables for encoding insignificant zeros. Fortunately, for many
systems, a suitable $D$ can be determined. For instance, for qubit and
multi-qubit systems, each $x_i$ is bounded by $\pm 1$ which makes $D=0$ the
optimal choice for this case.

QUBOs in the equations \eqref{eq:qubo_f} and \eqref{eq:qubo_h} are defined on
the graph of size $N \cdot R \cdot L$. The number of edges (i.e. non-zero
quadratic terms) depends on the number of non-zero off-diagonal elements of the
matrix $\coefmatrix$. It is interesting to note that the overall density of the
graph is an increasing function of $R$ (bigger precision requires a denser
graph) while, on the other hand, it tends to decrease with increasing $L$.

We converted the original problem of finding the dynamics of the system into a
binary optimization problem suitable for input to the quantum annealer. In the
next section, we will discuss experiments that we performed using D-Wave
2000Q$_{2.1}$ and D-Wave 2000Q$_5$ machines to test the approach we described.
The results we discuss here were originally reported in \cite{parallelintime}.

\section{Parallel-in-time simulations with quantum annealer}
\sectionmark{Parallel-in-time simulations}
\label{sec:parallel-in-time}
Before discussing the results of our experiments, let us focus first on its
design. To exemplify our approach, we chose to simulate the dynamics of a
two-level system with an initial state $\ket{0}$ and a Hamiltonian $\mathcal{H}$:
\begin{equation}
  \mathcal{H} = \frac{\pi}{2}\ssigma_y,
\end{equation}
where $\ssigma_y$ is a Pauli spin operator in the $y$-direction. This particular
choice of Hamiltonian and initial state makes the system suitable for
implementation on present-day quantum annealers for several reasons. One can
easily see that the evolution of the system is real (as opposed to complex),
which halves the number of needed variables. Secondly, for integral time points
$t_0=0, t_1=1, \ldots$ coefficients of the wave function can be expressed
\emph{exactly} using only two bits of precision per coefficient, which further
reduces the number of variables.

\begin{figure}[!h]
  \centering
  \includegraphics{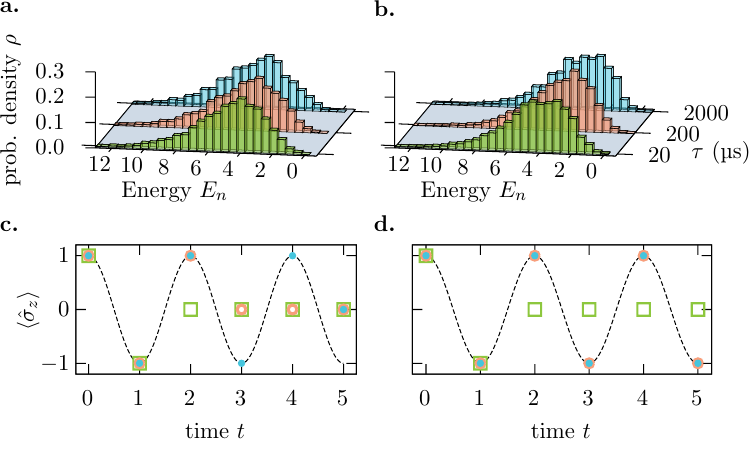}
  \caption{Results of simulating dynamics of two-level system on D-Wave 2000Q$_{2.1}$
    (left) and low-noise D-Wave 2000Q$_{5}$ (right). \textbf{a.}--\textbf{b.}
    Energy distribution of samples obtained from D-Wave annealers for different
    annealing times $\tau$. Notice a slight shift of distributions towards the
    ground state for the 2000Q$_{5}$ device. \textbf{c.}--\textbf{d.} Rabi
    oscillations of the simulated system. The obtained samples were normed before
    plotting. As can be seen in panel \textbf{d.}, the low-noise device was able to
    faithfully capture oscillations for $\tau=200, 2000$. The annealing time is
    color-coded: $\tau=$ \tikzquad\,\,\, -- 20\textmu{}s, \,\tikzcircle\,\,\,--
    200\textmu{}s, \,\tikzdot\,\,-- 2000\textmu{}s. } \label{fig:energy-hist}
\end{figure}

We simulated the above system using values of $R=2, 3$ and for several numbers
of time points $N$. We used annealing time $\tau$ spanning several orders of
magnitude, namely $\tau=20\mu s$, $200 \mu s$ and $2000 \mu s$. Since the
resulting graphs were dense, we decided to use standard embedding of the
complete graph $K_n$ on Chimera \cite{chimeraclique}. To assess the quality of
solutions obtained from the annealer, we sampled each problem $10^4$ times on
DW-2000Q$_{2.1}$ device as well as its low-noise version, DW-2000Q$_{5}$.
Energy distributions of samples obtained for $N=6$ are depicted in Fig.
\ref{fig:energy-hist}. The same figure also illustrates the dynamics of the
expected value of $\ssigma_z$ for the lowest energy sample obtained for a given
annealing time. Note that to preserve the physical meaning of the decoded
solution, the state vector was normed before plotting.

\begin{figure}
  \centering
  \includegraphics[width=\textwidth]{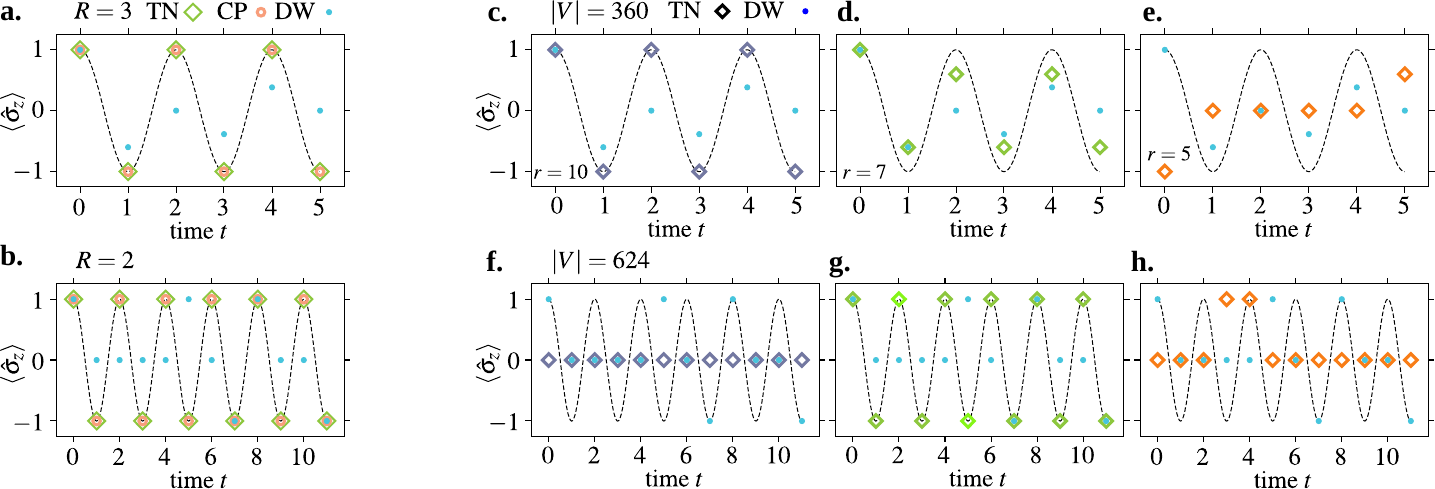}
  \caption{ \textbf{a.}--\textbf{b.} Performance of the two state-of-the-art heuristic
    algorithms: the CPLEX optimizer (CP) and a tensor networks-based (TN) solver
    (see Chapter \ref{chapter:tn}) in comparison to the D-Wave $2000$Q quantum annealer (DW),
    cf. Fig~\ref{fig:energy-hist}. The graphs on which the problems were defined
    had respectively $|V|=360$ (\textbf{a.}) and $|V|=624$ (\textbf{b.}) vertices. The annealing time was set
    to $\tau=200$\textmu{}s. The numerical precision of the solution vector is
    denoted as $R$. 
    \textbf{c.}--\textbf{h.} Degradation of the solution quality resulting from perturbing the
    problem by truncating its coefficients to a given
    numerical precision denoted as $r$. The reference ground state obtained
    with tensor networks (TN) is compared to the  experimental data from the
    D-Wave $2000$Q quantum annealer (DW). This effect, expected to be predominant
    in the current quantum annealing technology, is already visible on
    Fig.~\ref{fig:energy-hist}\textbf{a.}--\textbf{d.} and Fig.~\ref{fig:cplex_tn_dwave}\textbf{a.}--\textbf{b.}.
  }
  \label{fig:cplex_tn_dwave}
\end{figure}

To put these results into context, we also compare them to the ones obtained
using two purely classical methods: CPLEX optimizer and recently developed
tensor network-based algorithm (which we describe later in Chapter
\ref{chapter:tn}. The results of this comparison are depicted in Fig.
\ref{fig:cplex_tn_dwave}.

Results depicted in figures \ref{fig:energy-hist} and \ref{fig:cplex_tn_dwave}
show that the DW-2000Q$_{5}$ was able to faithfully capture dynamics of qubit
if the state of the system was encoded using $R=2$ bits of precision per
coefficient when the annealing time $\tau=200$ was used. For larger values of
$N$ and $R$ one can observe that the quality of solution degrades. Both CPLEX
and tensor networks-based solvers outperformed D-Wave annealers in terms of the
quality of solutions. The differences were especially noticeable for problem
instances with larger graph sizes, i.e. ones with higher precision ($R \ge 3,
  N=6)$), or with extra time points ($N > 6, R = 2$). The observed degradation of
the solution quality is consistent with the results obtained in other works,
especially for the problems requiring complete graphs, see e.g.
\cite{Hamerly2019}.

\subsection{Discussion of error sources}

The poor performance of D-Wave annealer is something certainly to be expected
from such early-stage devices. Annealers are prone to errors stemming from
multiple sources \cite{dwavedocs}, and it is hard to judge which of those
sources contributed most to the lackluster performance of a particular problem
instance. One of the possible sources of errors is DAC quantization, which
essentially limits the precision of both the quadratic and linear coefficients
passed to the annealer. As a result, the problem that the annealer physically
solves is slightly different than the problem the programmer intended to solve.

One can see that such quantization errors would mostly affect problems with
coefficients lying in close proximity to one another. Indeed, suppose that DAC
quantization errors limit the precision of the linear coefficients to $d$
decimal digits. Then any two coefficients, say $h_{i}, h_{{j}}$ lying closer to
each other than $d$ digits, i.e. $|h_{i} - h_{j}| < 10^{-d}$, become physically
undistinguishable to the annealer. The issue also affects coefficients that are
further apart, by possibly diminishing their relative differences.

While it is hard to pinpoint which source contributed the most to the errors in
the case of the optimization problems discussed in this chapter, we argue that
in our case the poor performance of the annealer can be largely explained by
DAC quantization. Indeed, looking at the \eqref{eq:coeff} one can immediately
see that the optimization problem can contain coefficients arbitrarily close to
each other as long as a large enough $R$ is chosen. To justify this reasoning,
we studied how the tensor network solver performs when the coefficients of the
problem are perturbed by truncating their coefficients to a predefined
number of digits $r$. The results of this experiment are presented in Fig.
\ref{fig:cplex_tn_dwave}\textbf{c.--h.}. One can immediately observe that the
error patterns resemble the ones obtained from D-Wave, which might suggest that
DAC quantization might indeed be a significant source of errors in our case.
However, we would like to point out, that our analysis is by no means
conclusive, and further analysis of error patterns is still needed.


%% file: tensor_networks.tex
\chapter{Solving spin-glass problems using tensor networks}
\chaptermark{Solving spin-glass problems}

\label{chapter:tn}

Benchmarking quantum annealers requires adequate algorithms for providing
baselines for the obtainable solutions. While there exists a plethora of
general-purpose optimization algorithms, one might hope to achieve better
results by exploiting the topology of the problem's underlying graph and thus
locality therein. In this chapter, we describe a recent, tensor network-based
algorithm \cite{tn} for finding the low-energy spectrum of Ising spin-glasses,
designed for problems defined on Chimera-like quasi-two-dimensional graphs. The
algorithm exploits the sparsity and locality of the Chimera graph by
representing the Boltzmann distribution of spin-glass as a tensor network,
whose approximate contraction can be used for computing marginal probability
distributions. This procedure can then be combined with the well-known branch
and bound algorithm to iteratively select the most promising partial solutions,
finally producing an approximation of the low-energy spectrum.

\section{Exploring the probability space}

In the algorithm we are going to present in this chapter, we perform the search
in the probability space rather than in the energy space. This physics-inspired
approach is closely tied to the quantum computing paradigm. To explain why, let
us begin by replacing classical Ising Hamiltonian $H(s)$ with its quantum
counterpart $\mathcal{H} = H(\boldsymbol{\sigma}^{z})$ (i.e. replacing each
variable $s_{i}$ with a Pauli operator $\ssigma_{z}$ acting on the $i$-th spin.
Naturally, there exists a one-to-one correspondence between the eigenstates of
$\mathcal{H}$ and the possible classical states. If one now wishes to
find the low-energy spectrum of size $k \ll 2^{N}$, the task is equivalent to
finding the $k$ most probable states according to the Gibbs distribution $\rho
  \sim \exp(-\beta \mathcal{H})$. One way to achieve this is to prepare the
system in a Gibbs state:
\begin{equation}
  \ket{\rho} \sim \sum_{\mathbf{s}} \exp(-\beta \mathcal{H}/2)\ket{\mathbf{s}}
\end{equation}
and then perform a measurement. If repeated multiple times, this procedure
would yield the desired low-energy spectrum with high probability.

While the above procedure is useful conceptually, it clearly cannot be directly
used on a classical computer, as it would require preparing a dense vector of
$2^{N}$ elements. Instead, in our algorithm we represent the Gibbs distribution
approximately via a suitable tensor--network. Then, instead of performing a
quantum measurement, we extract the needed information by traversing the
probability tree using the branch-and-bound method. In what follows, we
describe the procedure in detail, starting with the branch-and-bound
part.

\section{Branch and bound}
Let us first consider an Ising spin-glass problem defined on a square lattice.
The state space of such a
system can be viewed as a tree, in which $k$-th level contains all partial
configurations $(s_1, \ldots, s_k)$. This representation allows one to explore
the state space incrementally in search for low energy states, and possibly
prune the less promising branches. In the approach described here, we use
marginal probability $p(s_1, s_2, \ldots, s_k)$ as a criterion for deciding
which partial configurations are most promising. More precisely, we explore the
solution tree in a top-down manner, keeping at most $M$ states at $k$-th level
and branching them into $2M$ new partial configurations at level $k+1$. The
new marginal probability distributions can be computed using the formula:
\begin{equation}
  \label{eq:conditional-prob}
  p(s_1, s_2, \ldots, s_k, s_{k+1}) = p(s_1, s_2, \ldots, s_k)p(s_{k+1}|, s_1, \ldots, s_k).
\end{equation}
Importantly, in the Appendix \ref{sec:probability} we prove that the conditional
probability in equation \eqref{eq:conditional-prob} can be effectively computed
by exploiting the locality of the problem. The parameter $M$ can be made iteration-dependent by keeping only the states
whose marginal probability divided by the maximal probability is larger than
sum \emph{probability cutoff} $\delta_{p}$.

\begin{figure}
  \centering
  \includegraphics[width=\textwidth]{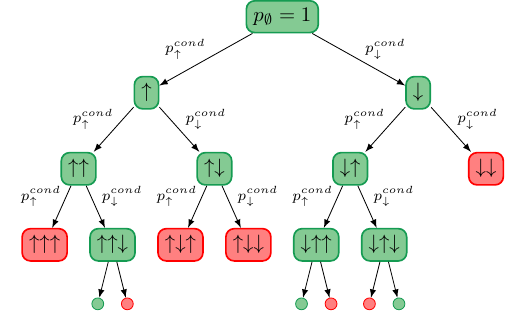}
  \caption{An illustration of the branch and bound method. The state space is explored one spin at a time.
    At each tree level, we branch each of at most $M$ configurations into $2M$ new configurations. Then,
    the tree is pruned, and only $M$ most promising branches are kept. In the depicted example $M=3$.
    As a criterion for pruning the tree, we use the marginal probability of the partial configurations
    corresponding to each node. The marginal probabilities are computed using the equation \eqref{eq:conditional-prob}.
  }
  \label{fig:tree}
\end{figure}

\begin{figure}
  \includegraphics[width=\textwidth]{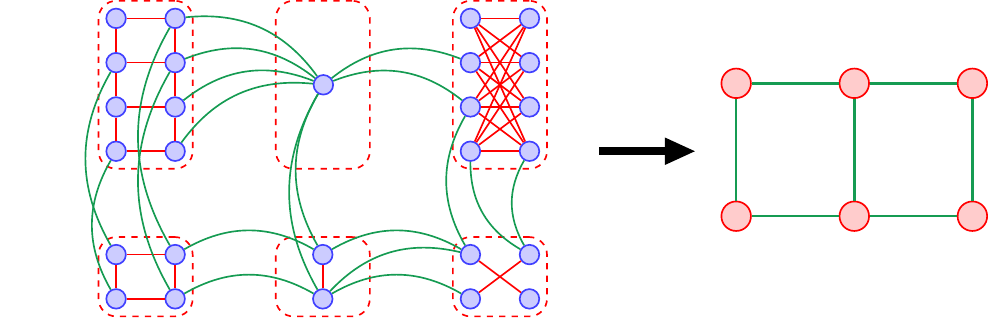}
  \caption{Grouping spins into clusters in a quasi-two-dimensional graph. Here, spins in
    the original graphs are grouped together to form a square lattice. Each site in
    the new lattice then effectively serves as a higher-dimensional system.}
  \label{fig:clustering}
\end{figure}

Before discussing how probabilities in \eqref{eq:conditional-prob} can be
computed, let us first extend the above approach to the more general case of a
quasi-two-dimensional graph, i.e. one in which nodes can be grouped into
\emph{clusters} forming a two-dimensional square lattice (see Fig.
\ref{fig:clustering}). One can easily see, that again we can construct a
tree-like structure representing state space, this time considering joint
configurations of spins in a single cluster. Therefore, for most of the time,
we might "forget" the underlying spin-glass structure and consider square
lattices in which spin clusters act like higher-dimensional systems.
Furthermore, by the same argument, it is clearly visible that our approach is
not limited to the Ising systems, but could be also used for systems of higher
dimensions.

\section{PEPS network construction}
\begin{figure}
  \includegraphics[width=\textwidth]{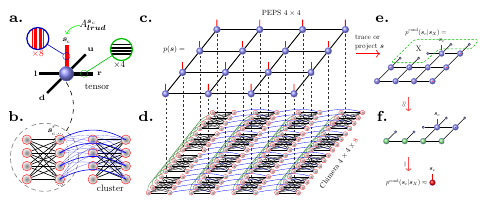}
  \caption{
    Tensor network formalism for solving Ising spin-glasses on Chimera-like
    graphs. \textbf{a.}, \textbf{b.} Assignment of PEPS tensors to groups of $l$ spins (clusters).
    Each PEPS tensor has four virtual legs of dimension $D = 2^{\min(m,n)}$ and
    one physical leg of dimension $2^{l}$. Here, $m$ is the number of spins in one
    cluster interacting with $n$ of those in the neighboring cluster. For the Chimera
    graph, depicted in panel \textbf{d.}, $n=m=4$. Note that adding more complicated interactions not present
    in the Chimera topology as in panel \textbf{b.} would not increase the bond dimension $D$. \textbf{c.} The resulting tensor network
    used to represent probability distribution $p(\mathbf{s})\sim \exp(-\beta H(\mathbf{s}))$.
    \textbf{e.} The conditional probabilities $p(\mathbf{s}_{c}|\mathbf{s}_{X})$ are obtained
    by projecting the physical degrees of freedom in the region $X$ to given
    configuration $\mathbf{s}_{X}$ and tracing out the remaining ones. Next, the
    approximate MPO-MPS scheme is used to collapse the network in a bottom-top fashion
    until only two rows remain. Finally, as in panel \textbf{f.}, the remaining tensors can be exactly contracted to
    obtain the desired conditional probability.
  } \label{fig:tensors}
\end{figure}

We begin the construction of a PEPS network for a quasi-two-dimensional graph
by considering two spins at sites $i$ and $j$ connected by an edge $J_{ij}$.
This edge can be decomposed as:

\begin{equation}
  e^{-\beta J_{ij}s_i s_j} = \sum_{\gamma = \pm 1} B^{s_{i\phantom{j}}}_\gamma C^{s_j}_\gamma
\end{equation}
where
\begin{equation}
  \label{eq:decomposition}
  B^{S_i}_\gamma = \delta_{\gamma s_i} \quad C^{s_j}_\gamma = e^{-\beta \gamma J_{ij} s_j}
\end{equation}
Note that decomposition \eqref{eq:decomposition}, although not unique, has the
advantage of comprising only non-negative coefficients, which positively
affects numerical stability. Next, with each cluster we associate a PEPS tensor:
\begin{equation}
  \label{eq:peps}
  A^{\mathbf{s_c}}_{\mathbf{lrud}} = e^{-\beta H(\mathbf{s_c})} B^{\mathbf{s_c}^l}_\mathbf{l}C^{\mathbf{s_c}^r}_\mathbf{r}B^{\mathbf{s_c}^u}_\mathbf{u}C^{\mathbf{s_c}^d}_\mathbf{d}
\end{equation}
Here, $\mathbf{s_c}$ collects all spins in a given cluster, and
$\mathbf{s_c}^l$, $\mathbf{s_c}^r$, $\mathbf{s_c}^u$, $\mathbf{s_c}^d$ collect
spins interacting with it from the left, right, up and down respectively. Each
such tensor has five legs: the physical one $\mathbf{s_c}$ of dimension $2^m$,
where $m$ denotes the number of spins in the cluster, and the virtual ones $l,
  r, u, d$ with dimensions depending on the number of inter-cluster edges. Note
that $H$ in \eqref{eq:peps} is restricted to the graph induced by spins
belonging to the considered cluster. The construction is depicted in Fig.
\ref{fig:tensors}. Combining the tensors gives an exact representation of the
Gibbs distribution as:
\begin{equation}
  \exp(-\beta H(\mathbf{s})) \sim \sum_{\mathbf{k}}\prod_{c^{i}}A_{\mathbf{k}^{i}}^{\mathbf{s}_{c^{i}}}
\end{equation}
Despite our tensor network representation of the Gibbs distribution being
exact, contracting the network to obtain the information is still a difficult
task. In principle, one could use some approximation schemes \cite{lewenstein}.
However, in our approach, we decided to use another procedure exploiting the
locality of the problem graph. Namely, we employ a matrix product state (MPS)
-- matrix product operator (MPO) based approach \cite{murg} approach. One
starts by considering the first row of the lattice as a vector in high
dimensional space having a natural decomposition in the form of MPS. Then, we
add another row, viewed as MPO, which enlarges the MPS representation. Adding
subsequent rows would require an exponential growth of the bond dimension
$\chi$. To prevent this, a sequence of truncation is performed, which results
in a series of boundary MPS. The new MPS are found by minimizing their distance
from the enlarged previous ones. The MPS-MPO construction is depicted in Fig.
\ref{fig:tensors}(e)--(f). In the end, the network can be contracted exactly
resulting in the desired conditional probability.

\section{Benchmarks}

\begin{figure}
  \includegraphics[width=\textwidth]{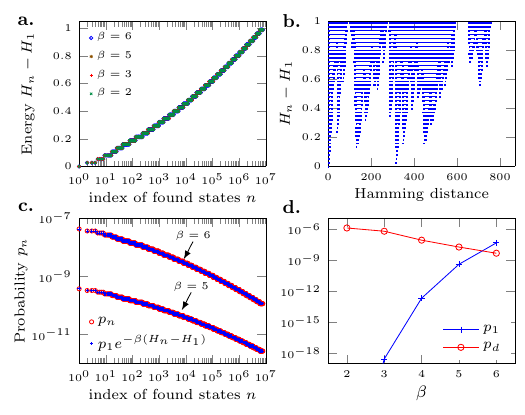}
  \caption{Example result of running our algorithm on a droplet instance with $N=2048$.
    \textbf{a.} Low energy spectrum found by a single run of our algorithm. Observe
    consistency between different values of $\beta$. \textbf{b.} Hamming distance
    of solutions presented in \textbf{a.} from the ground state. \textbf{c.}
    Probabilities of each configuration found for least numerically stable values
    of $\beta$. In the depicted example, we can see full consistency between the
    probabilities obtained from contracting PEPS network $p_{n}$ and the Boltzmann
    weights calculated from the configuration's energy. \textbf{d.} Comparison of
    largest discarded probability $p_{d}$ and the ground state probability $p_{1}$.
    With increasing $\beta$ we were able to achieve $p_{d} < p_{1}$. This indicates
    that the algorithm was indeed able to reach the ground state. }
  \label{fig:tn-single-state}
\end{figure}

To fully investigate the performance of our algorithm, we performed several
benchmarks, testing various metrics quantifying both execution time, as well as
the quality of the found solutions. We tested our algorithm for sets of
\emph{droplet} instances specifically designed to be hard for classical
heuristic solvers, especially ones relying on local updates. We benchmarked our
algorithm against classical solvers based on Parallel-Tempering, and D-Wave
Quantum annealer DW-2000Q$_6$. As it is hard to directly compare samples
obtained from the D-Wave annealer with the output of our deterministic
algorithm, we decided to use time-to-solution as a metric. The time to solution
$\mbox{TTS}$ is defined as:
\begin{equation}
  \label{eq:tts}
  \mbox{TTS} = T \frac{\log(1 - \ptarget)}{\log(1 - \psucc)},
\end{equation}
where $\ptarget$ is the desired probability of obtaining solution, $\psucc$
is the empirical probability of obtaining the solution and $T$ is the running
time of the solver. In addition, for D-Wave annealers, we multiply $\mbox{TTS}$
by the ratio $N/\mbox{num\_qubits}$, to account for the possibility of fitting
multiple instances of the problem on the device at the same time. Naturally,
one might consider $\mbox{TTS}$ metric not only for finding a ground state, but
also for finding a solution approximating a ground state with a given
approximation ratio (i.e. solution lying in the desired lowest fraction of the
full energy spectrum). The results of these benchmarks are presented in Table
\ref{tab:tnvspt}. For all instances, our algorithm was able to find the ground
state, which was not the case for other solvers. However, if one is not
necessarily interested in finding the ground state, both D-Wave annealers and
classical parallel tempering solver might be a better choice, as they were able
to find a satisfying solution in a shorter time. In Fig.
\ref{fig:tn-single-state}, we show an example solution for a single instance
with discreet values of $J_{ij}$ with $dJ=\frac{1}{75}$. One can observe that
increasing the $\beta$ allows for obtaining tighter bounds on the possible
error. It is also visible that the algorithm demonstrates consistency between
the probabilities obtained from the tensor network contractions and the ones
obtained from the Boltzmann weights calculated from the corresponding
configuration energies.

\begin{table}[b]
  \centering
  \begin{tabular}{|l|c|ccc|}
    \hline
    \rowcolor{theader}  Method & approx. ratio & $N=512$   & $N=1152$  & $N=2048$  \\
    \hline
    TN                         & g.s.          & 30s       & 150s      & 450s      \\
    \hline
    \hline
    PT (adaptive)              & g.s.          & 800s      & ---       & ---       \\
    \hline
    PT (geometric)             & $0.01$        & 0.53s     & 4.16s     & ---       \\
    PT (geometric)             & $0.005$       & 2.51s     & 56.4s     & ---       \\
    PT (geometric)             & $0.001$       & 158.4s    & timed-out & ---       \\
    PT (geometric)             & $0.0001$      & 897.6s    & timed-out & ---       \\
    \hline
    \hline
    DWave 2000Q$_6$            & $0.01$        & $0.003$s  & $0.006$s  & $0.02$s   \\
    DWave 2000Q$_6$            & $0.005$       & $0.2$s    & timed-out & timed-out \\
    DWave 2000Q$_6$            & $0.001$       & timed-out & timed-out & timed-out \\
    \hline
    \hline
  \end{tabular}
  \caption{Comparison of time-to-solution metric for our tensor network-based algorithm,
    in-house Parallel Tempering implementation and D-Wave 2000Q$_{6}$. The
    \emph{adaptive} and \emph{geometric} terms refer to the distribution of inverse
    temperature $\beta$ in Parallel Tempering replicas. We bounded the running time
    of our solver to 30 minutes with bond dimension $\chi = 16$, $\beta=3$ and
    probability cutoff $\delta_{p} = 10^{-3}$. For PT, the $T$ in the equation
    \ref{eq:tts} is inferred from the running time and number of performed MC
    sweeps: a single MC sweep took 0.00005s for N=512 and 0.00011s for $N=1024$.
    For the adaptive PT, we used 12 replicas. For geometric PT, we used 25 replicas
    with geometrically distributed $\beta$, with $\beta_{\min}=0.0001$ and
    $\beta_{\max}=10$. For all probabilistic samplers, we used target probability
    $p_{\mbox{target}}=0.99$. In the case of D-Wave annealers, we modified
    instances by dropping inactive qubits. To obtain the reference ground state, we
    once again used our algorithm. We optimized time to solution over annealing
    times of $5\mu s$, $20\mu s$ and $200\mu s$. For each instance and each
    annealing time, we gathered 1000 samples for $N=512$ and 2500 for other values
    of $N$. Also, we used $T=\tau$ for the D-Wave annealers, i.e. we considered
    only annealing time and disregarded other factors contributing to overall
    solution time. This choice is justified by the fact that the other
    contributions are minuscule. The ``timed-out'' string indicates that the given
    algorithm could not find a solution within the given approximation ratio (i.e.
    $\psucc=0$). } \label{tab:tnvspt}
\end{table}

\begin{figure}
  \centering
  \includegraphics[width=0.8\textwidth]{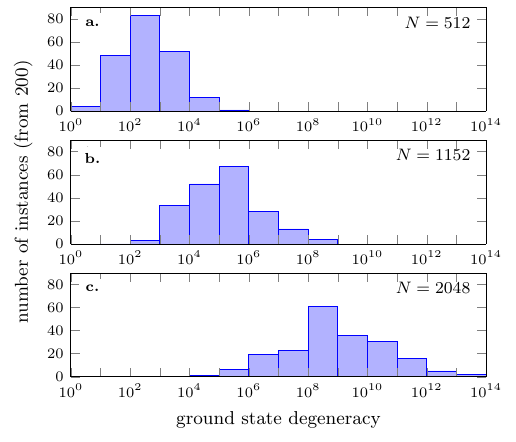}
  \caption{Histogram of ground state degeneracy found by our algorithm for test instances
    constructed by drawing couplings $J_{ij}$ uniformly from a set $\{\pm 1, \pm 2,
        \pm 4\}$ and setting all local fields $h_{i} = 0$.}
  \label{fig:ground-degeneracy}
\end{figure}

We also benchmarked our algorithm on the set of \emph{deceptive cluster loops}
instances \cite{helmut_deceptive_2018}, also expected to be hard for the
classical heuristic solvers. One particular reason for the hardness of these
instances is their enormous ground-state degeneracy. In 97\% of the cases, we
were able to recover the lowest reference energy from
\cite{helmut_deceptive_2018}. In the other 3\% of instances, we were able to
find a better solution.

In our final benchmark, we tested our algorithm with regard to fair sampling.
In order to do so, we solved instances of the Ising model with integer
coefficients and counted the identified ground-state degeneracy. The test
instances had $J_{ij}$ drawn uniformly at random from the set $\{\pm 1, \pm 2,
    \pm 4\}$, following similar tests performed for parallel tempering and parallel
tempering with isoenergetic cluster moves \cite{Zhu_PT+ICM_2015} in Ref.
\cite{zhu_fair_2019}. We present the results in Fig.
\ref{fig:ground-degeneracy}. For smaller system sizes, we observe consistency
with the results reported in \cite{zhu_fair_2019}. For $N=1152$, we observe
some degeneracies approaching the order of $10^{8}$, while the previously
reported numbers were reaching only the magnitude of $10^{6}$. Moreover, we
were able to reach beyond $N=1152$ studied in \cite{zhu_fair_2019}.


%% file: bruteforce.tex
\chapter{Brute--forcing spin--glass problems with CUDA}
\label{chapter:bruteforce}
\chaptermark{Brute--forcing spin--glass problems}
In Chapter \ref{chapter:tn} we presented a tensor network--based heuristic algorithm
tailored for Ising spin--glass problems defined on Chimera graphs. In stark contrast, in this
chapter we will shift our attention to a deterministic algorithm capable of solving problems defined
on arbitrary (but relatively small) graphs.

Conceptually, the simplest approach for solving any optimization problem is a
brute force approach, i.e. an exhaustive search through the set of all possible
solutions. For the QUBO or Ising spin--glass with $N$ variables, this would
require iterating over $2^{N}$ possible states and computing energy for each of
them, resulting in a superexponential algorithm. Although the approach is
clearly infeasible for large problems, it presents several advantages. The
algorithm is deterministic and can certify\footnote{i.e., prove that the found
  solution is in fact optimal} the solution. Moreover, it can be used to compute
a low energy spectrum of arbitrary size $k$ (provided that it can fit into
memory). Lastly, it is trivially parallelizable and hence can be efficiently
accelerated using virtually any parallel computing paradigm, thus significantly
increasing attainable problem sizes.

In this chapter, we discuss such a brute--force algorithm using massively
parallel CUDA architecture. We start by outlining the basic version of the
algorithm and then discuss its recent optimizations for cases when the goal is
to find only the ground state (as opposed to finding a low energy spectrum).
Our implementation is capable of finding the ground state of instances of size
$N=54$ in an hour using a commodity GPU and achieving the same task in less
than 5 minutes on a server-grade NVIDIA DGX H100. Lastly, we present a possible
application of our algorithm, which is validating a recent MPS--based algorithm
for solving Ising spin--glasses.



\section{Finding low--energy spectrum with CUDA}
\subsection{Outline of the algorithm}
An idealized brute force algorithm for solving QUBO problems running on a
hardware with infinite storage and an infinite number of execution units can be
summarized as follows:
\begin{enumerate}
  \item Launch number of threads equal to the total number of possible states.
  \item Let each thread compute the energy of one of the states.
  \item Extract (e.g. by sorting) the desired number of low--energy states.
\end{enumerate}
Naturally, an attempt to implement such an algorithm on real hardware is doomed
to fail. To exemplify this, consider a problem with $N=40$ variables. Assuming
we use 32-bit floating point numbers, one would need an enormous amount of
$2^{40}\cdot 4B = 4398046511104\mbox{B}$, or $4\mbox{TB}$ of working memory to
    store the computed energies. For $N=50$, this number grows to $4096\mbox{TB}$.
    Clearly, such an amount of needed memory is prohibitively large, and that is
    even before we consider some form of storage for system states. Moreover, no
    current hardware can execute $2^{40}$ threads in parallel. Fortunately, we can
    adapt our algorithm to take into account limited memory and parallelism. To do
    so, we introduce the following assumptions:
    \begin{enumerate}
      \item We will process the space of possible solutions in \emph{chunks} that can fit into the GPU
        memory.
      \item Number of states in a chunk can be larger than the total number of threads.
        Should this be the case, the threads will process the chunk using a
        grid--stride loop pattern.
    \end{enumerate}
    As an added benefit of our assumptions, we decouple the grid size from the
    problem size. The number of thread blocks and the block size become parameters of
    our algorithm, which facilitates further fine-tuning of the kernel execution
    parameters.

    The algorithm will keep track of $k$ lowest--energy states computed so far.
    This information will be updated after each new chunk is processed. The
    downside of this approach is that the size of the low-energy spectrum we can
    compute is limited by the chunk size. However, this limitation is not as severe
    as it seems, because in a typical scenario, we have $k \ll 2^{N}$.

In the next section, we discuss another important aspect of our algorithm,
which is efficient storage and representation of system states.

\subsection{Storage and representation of system states}
Implementing efficient algorithms involves choosing the right storage strategy
for the data the algorithm operates on. This is especially the case for
present-day GPUs, which are equipped with fairly limited memory, as compared to
the operating memory available to the traditional CPU. Moreover, memory
transfers between host and GPU induce additional overhead that should be
avoided whenever possible. For this reasons one often aims for designing the
storage strategy such that it reuses information already available on the GPU
as much as possible, thus optimizing resource usage and minimizing the number
of memory transfers.

In principle, each configuration of a $N$--variable QUBO can be represented by
$N$ integers. However, since each variable can be assigned only one of two
possible values, this wastes a lot of available memory, as out of each machine
word only a single bit is used. Instead, one can pack the whole state of the
system into a single integer by identifying each bit of the underlying machine
word with a single spin. In our implementation, we decided to use 64-bit
integers. This particular implementation choice limits attainable problem sizes
to $N=64$. However, considering that solving larger problems using the brute
force approach is not likely to be possible in the near future (as demonstrated
by our benchmarks presented further in this chapter), this is not a significant
limitation. Furthermore, should the need arise, one could extend the
implementation to use multiple 64-bit integers for storing a single
configuration.

Identifying states with integers greatly simplifies their enumeration, as it
boils down to iterating over an appropriate range of natural numbers. More
importantly, it allows GPU threads to identify the system state they have to
process using their index and additional offset designating the chunk. In our
implementation, we restrict ourselves to chunk sizes being power of two, i.e.
chunk size $=2^{M}$ for some $M < N$. We conceptually split each configuration
    into two parts:
    \begin{enumerate}
      \item A \emph{local} part comprising least significant $M$ bits. This is part is
        \emph{different} for each state in the chunk.
      \item A \emph{suffix} comprising the most significant $N-M$ bits. This part is
        \emph{the same} for each state in the chunk.
    \end{enumerate}
    Now, since there are $2^{N-M}$ chunks, we can identify each chunk with a $N-M$
    bit number. Finally we arrive for a formula for an integer representation
  $\mathbf{q}_{i}^{j}$ of an $i$-th configuration in $j$-th chunk:
\begin{equation}
  \mathbf{q}_{i}^{j} = i + 2^{M}\cdot j,\qquad i=0,\ldots,2^{M}-1, \; j=0,\ldots,2^{N-M}-1
\end{equation}
The following example demonstrates the representation described above.
\begin{example}[Processing solution space in chunks]
  Consider QUBO with $N=8$ variables. We decide to use $M=5$. Hence, there are
  $2^{M}=32$ states in each chunk and a total of $2^{N-M}=8$ chunks. The
  \emph{local} part of the first state in each chunk is $0$, or $(00000000)_{2}$
      in binary. The local part of the last state in each chunk is $31$, or
    $(00011111)_{2}$. The table \ref{tab:chunks} below enumerates ranges of
  combined integer representation of states in each chunk.
  \begin{table}[ht!]
    \centering
    \begin{tabular}{|c|c|c|c|c|c|}
      \hline
      \rowcolor{theader}
      \multicolumn{2}{|c|}{Chunk}      &
      \multicolumn{2}{c|}{First state} &
      \multicolumn{2}{c|}{Last state}                                                                        \\
      \hline
      \rowcolor{tsubheader} Index      & Binary    & Decimal & Binary           & Decimal & Binary           \\
      \hline
      0                                & $(000)_2$ & 0       & $(00000000)_{2}$ & 31      & $(00011111)_{2}$ \\
      1                                & $(001)_2$ & 32      & $(00100000)_{2}$ & 63      & $(00111111)_{2}$ \\
      2                                & $(010)_2$ & 64      & $(01000000)_{2}$ & 95      & $(01011111)_{2}$ \\
      3                                & $(011)_2$ & 96      & $(01100000)_{2}$ & 127     & $(01111111)_{2}$ \\
      4                                & $(100)_2$ & 128     & $(10000000)_{2}$ & 159     & $(10011111)_{2}$ \\
      5                                & $(101)_2$ & 160     & $(10100000)_{2}$ & 191     & $(10111111)_{2}$ \\
      6                                & $(110)_2$ & 192     & $(11000000)_{2}$ & 223     & $(11011111)_{2}$ \\
      7                                & $(111)_2$ & 224     & $(11100000)_{2}$ & 255     & $(11111111)_{2}$ \\
      \hline
    \end{tabular}
    \caption{An example enumeration of chunks iterated over by brute force algorithm.}
    \label{tab:chunks}
  \end{table}

  \begin{table}{}

  \end{table}
\end{example}

\subsection{Implementation details}

In our approach we decided to store states and their corresponding energies in
arrays of size $k+2^{M}$, where $k$ is the desired size of the low energy
    spectrum and $2^{M}$ is the chunk size. The arrays are always synchronized,
    i.e. at all times $i$-th state corresponds to $i$-th energy. The first $k$
    elements store the lowest energies and corresponding configurations found so
    far. When a new chunk is being processed, the second part of the arrays is
    populated with new states and energies by the energy--computing kernel. Next,
    the best $k$ states from the current chunk are selected and moved into indices
  $k, k+1,\ldots,2k-1$. In this way, the global best solutions from previous
    chunks and the lowest energy states from the current chunk in a continuous
    space in memory, which facilitates updating the best configurations.

\begin{figure}
  \includegraphics[width=\textwidth]{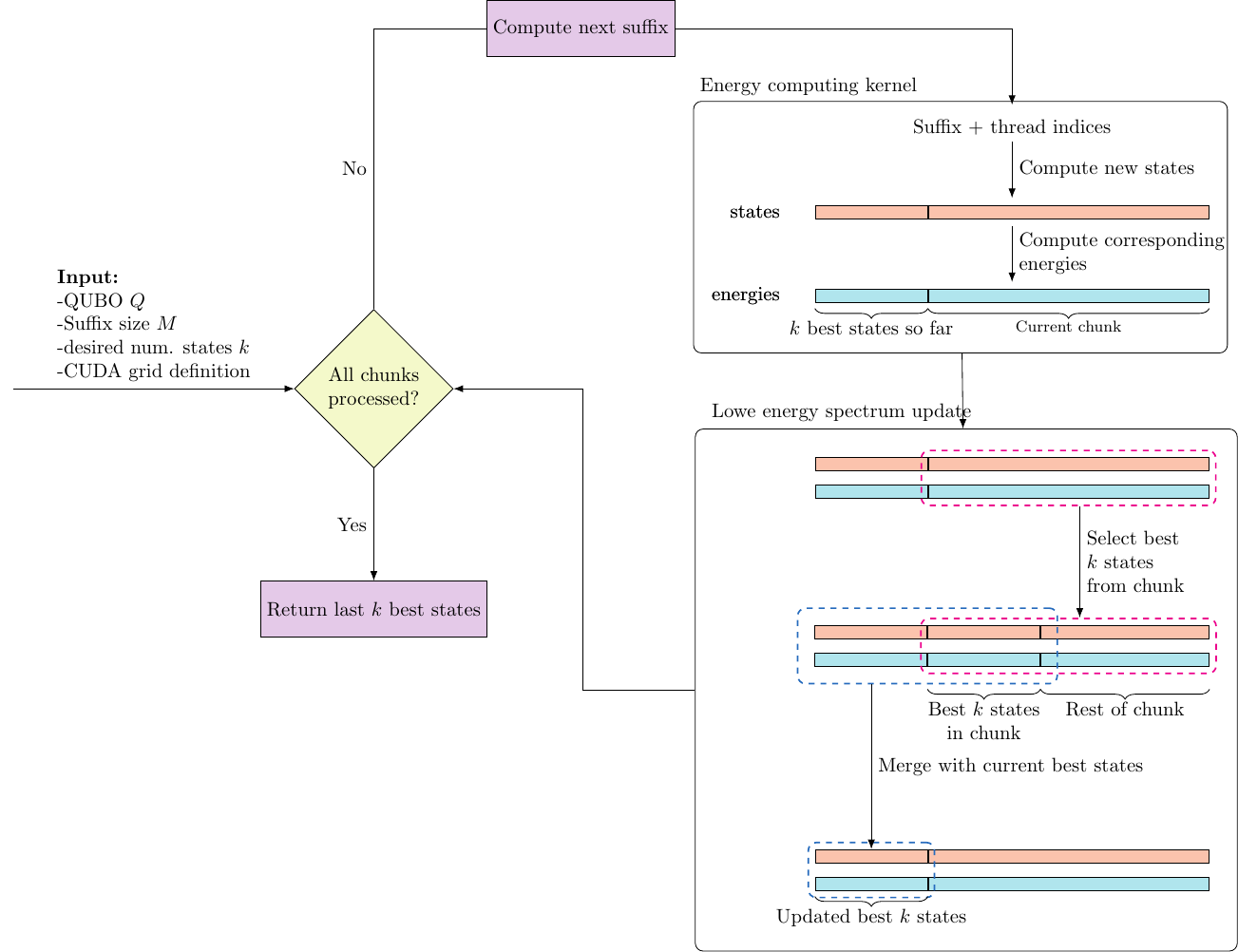}
  \caption{
    Detailed representation of brute force algorithm for finding $k$-lowest energy
    states of a QUBO. The algorithm iterates over the set of all possible states in
    chunks of size $2^{M}$, where $M$ is a user-defined parameter. Throughout the
        algorithm execution, we maintain arrays of states and corresponding energies.
        The first part of those arrays stores the $k$ best configurations encountered
        so far, and the second part stores configurations belonging to the currently
        processed chunk. In the first phase of the iteration, an energy-computing
        kernel is launched. Then, the $k$--lowest energy configurations from the given
        chunk are selected and moved towards the part of the array with the current
        best solutions. Finally, the best $k$ states are selected from the first $2k$
    configurations and the algorithm proceeds to the next chunk or terminates if
    all the chunks have been iterated over. } \label{fig:bruteforce}
\end{figure}

    One could use a parallel sorting procedure for extracting the $k$
lowest--energy states at each step. However, for improved performance, we
decided to use a combination of the \texttt{bucketSelect} \cite{bucketselect}
algorithm in tandem with \texttt{thrust::partition\_by\_key} \cite{thrust}. The
\texttt{bucketSelect} algorithm is used to find the pivot configuration that
would reside at $k$--th position in the sorted array. Then,
\texttt{thrust::partition\_by\_key} is used to reorder both arrays such that
the configurations with energies lower than the one of the pivot are moved to
the beginning. The same procedure is used both for extracting the $k$-lowest
energy states in the given chunk, as well as to update the global solution by
extracting $k$-lowest energy configurations from the first $2k$ configurations.
The whole procedure is depicted in Fig. \ref{fig:bruteforce}.

Lastly, we would like to note that the algorithm we just described can also be
implemented on homogenous, CPU-only architectures using any of the available
parallelization approaches. In our implementation, we used the
OpenMP\cite{openmp} for a CPU--only version.

\subsection{Performance benchmarks}
In order to test the performance of our algorithm, we run extensive benchmarks
using the following hardware:
\begin{itemize}
  \item CPU:
    {$10$
      Cores {\rmfamily Intel\textregistered} Core \textsuperscript{TM}i7-6950X};
  \item GPU(1):
    Nvidia
      GeForce GTX $1080$, $8$GB GDDR$5$ global memory, $2560$ CUDA Cores;
  \item  GPU(2): {Nvidia Titan V,
      $12$GB HBM$2$ global memory, $5120$ CUDA Cores}.
\end{itemize}
The hardware listed above is certainly not the most performant
one available on the market at the time of writing this thesis. However, these initial
benchmarks were performed in 2020 and originally published in \cite{bruteforce}.

\begin{figure}
  \centering
  \includegraphics[width=\textwidth]{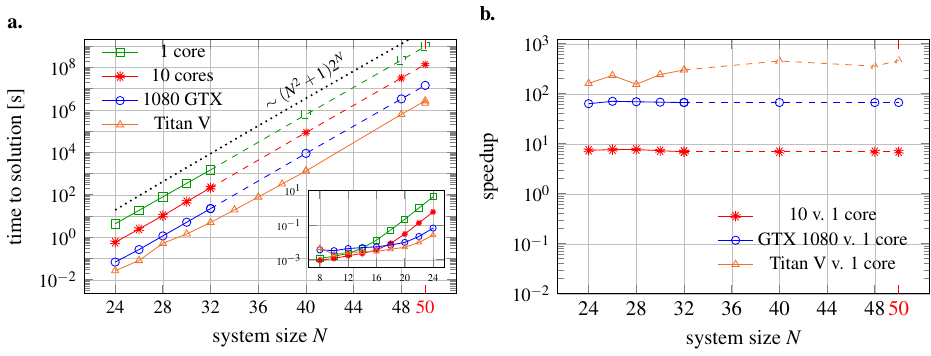}
  \caption{Results of benchmarks of our algorithm. {\textbf{a.}} Time to solution vs.
    system size $N$. {\textbf{b.}} Speedup of multi-core/GPU implementation with
    respect to a single core one vs. system size $N$. The solid lines represent the
    numerical results and the dashed lines present estimates based on results
    obtained for smaller system sizes.} \label{fig:benchmark_results}
\end{figure}

For conducting our benchmarks we generated $100$ spin-glass instances for each
$N=24, 26, \ldots, 30, 32$. Additionally, we generated $100$ instances of size
$N=40$ and single instances of sizes $N=48, 50$ that were feasible to solve
with Titan V GPU (which was the most powerful card available to us at the time
of performing the benchmarks). Coefficients of each spin-glass were drawn
randomly from uniform distributions on the intervals $[-2, 2]$ and $[-1, 1]$
for magnetic fields and couplings respectively. For each instance, we computed
the low energy spectrum of $k=100$ states with our algorithm. We used a maximum
chunk size of $2^{29}$ for Titan V and CPU and the chunk size of $2^{27}$ for
    GTX 1080. As already mentioned, larger instances $(N > 32)$ were solved only
    using Titan V GPU. For GTX 1080 and CPU implementation, the expected time to
    solve those instances was estimated based on the timings for smaller $N$. The
results of our benchmarks are presented in Fig. \ref{fig:benchmark_results}.

\section{Example application: verifying MPS-based optimization algorithm}
\sectionmark{Example application}

As an example application, in this section, we use our brute-force-based
approach to verify the performance of a heuristic algorithm based on Matrix
Product States (MPS). The detailed description of this algorithm is outside the
scope of this thesis, and we refer the interested reader to the Supplementary
Material in \cite{tn}. Nevertheless, before we outline how the algorithm works.

\begin{figure}
  \begin{subfigure}[t]{0.45\textwidth}
    \caption{}\label{fig:mps:boltzmann}
    \includegraphics[width=\textwidth]{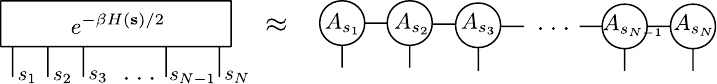}
  \end{subfigure}\hfill
  \begin{subfigure}[t]{0.45\textwidth}
    \caption{}\label{fig:mps:compress}
    \includegraphics[width=\textwidth]{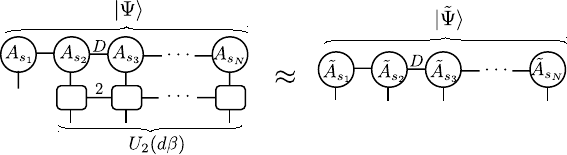}
  \end{subfigure}
  \caption{\subref{fig:mps:boltzmann} Approximation of the Boltzmann distribution using the MPS ansatz. \subref{fig:mps:compress} Compression scheme used in the MPS algorithm.}
  \label{fig:mps}
\end{figure}

Similarly to the algorithm presented in Chapter \ref{chapter:tn}, in the
MPS--based approach one explores the probability distribution (as opposed to
exploring the energy landscape directly). The basic idea behind is to
approximate Boltzmann distribution as
\begin{equation}
  e^{-\beta H(\mathbf{s})/2} \approx A^{s_1} A^{s_2} \ldots A^{s_L} = |\Psi(\beta) \rangle,
  \label{eq:mps}
\end{equation}
for large enough $\beta$. Here, $A^{s_{i}}$ are real matrices of limited
    dimension $\le D$. In this context, the parameter $D$ is referred to as the
\emph{bond dimension}. Fig. \ref{fig:mps:boltzmann} shows a pictorial
representation of such approximation. At each bond, the system is split into
two halves. An exact decomposition would require bond $D$ of exponential
(w.r.t. number of spins) size. Limiting $D$ effectively limits the amount of
entanglement related to given bipartition \cite{Wall18}. Once the approximation
in the equation \eqref{eq:mps} is constructed, it is possible to effectively
compute any marginal or conditional probability, and then systematically search
for the most probable (and thus, ones with lowest energy) classical
configurations using branch-and-bound procedure, constructing tree of most
probable spin configurations one variable at the time.

The search starts with $\beta=0$, for which the MPS decomposition is trivial.
Then, the algorithm subsequently simulates the imaginary time evolution by
applying the sequence of gates:
\begin{equation}
  \label{eq:gate}
  U_i(d\beta) = e^{-d\beta s_i (\sum_{j>i}J_{ij}  s_j+h_i)/2},
\end{equation}
which totals to $\prod_{i=1}^N U_i(d\beta) = e^{-d \beta H(\mathbf{s})/2}$. One
    can observe that applying each gate results in the doubling of the bond
    dimension. Hence, at each step, one has to systematically find an approximation
    maintaining the fixed $D$. The whole procedure is depicted in Fig.
\ref{fig:mps:compress}.

\begin{figure}[th]
  \includegraphics[width=\textwidth]{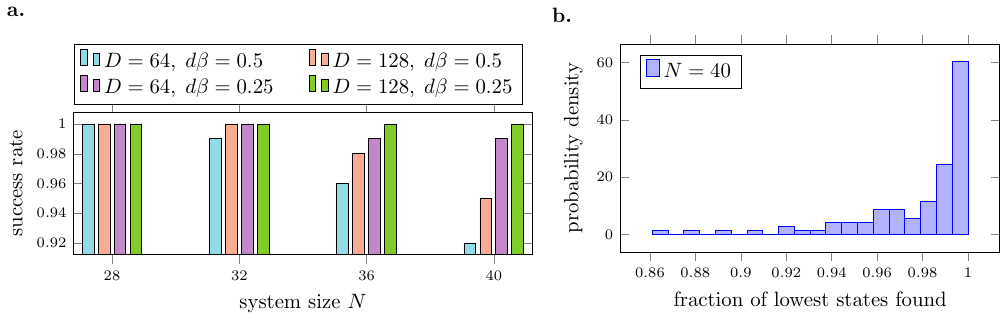}
  \caption{Results of the MPS benchmarks. In both panels $\beta=1$. \textbf{a.} Success
    rate, defined as a fraction of instances (out of 100) for which the MPS
    algorithm found the ground state. \textbf{b.} Normalized histogram showing the
    number of instances for which the MPS algorithm was able to find the given
    fraction of lowest $k=1000$ states.} \label{fig:mpsbench}
\end{figure}

By construction, the MPS-based ansatz outlined above is one-dimensional. Hence,
the question is to what end can it be used to find low-energy solutions of
spin-glasses defined on a complete graph? To answer this question, we ran the
MPS-based algorithm on 100 instances of different sizes and then compared the
results to the output of our brute-force algorithm. Instances were drawn at
random using the same procedure as described in the previous section. The
results of these tests are depicted in Fig. \ref{fig:mpsbench}. One can observe
that bond dimension $D=128$ and inverse temperature $\beta=1$ are already
sufficient to find the ground state of all the test instances, and recover most
of the $k=1000$ lowest energy states. The results also demonstrate the
significance of setting the time-step parameter $d\beta$ to small enough value.
As the last conclusion from our benchmarks, we would like to point out the
magnitude of compression of the relevant information in the MPS representation.
Indeed, an exact MPS decomposition would require the bond dimension $D=2^{N/2}
  \gg 128$.

\section{Improving the algorithm using Gray Code}

The algorithm presented in the previous chapter is already highly performant.
However, we can still improve upon it by altering the order in which we
enumerate the integral representation of states used by our algorithm.

\subsection{Single bit--energy difference}
Suppose we are given a QUBO with $F(q_{1},\ldots,q_{N})$ as in the equation
\eqref{eq:qubo}. Consider two states, say $\bq{1} =
  (\q{1}_{1},\ldots,\q{1}_{N})$ and $\bq{2}=(\q{2}_{1},\ldots,\q{2}_{N})$ such
    that they only differ in the $k$-th bit, i.e. $\q{2}_{k}=1-\q{1}_{k}$ and
  $\q{2}_{i}=\q{1}_{i}$ for $i \ne k$. The energy difference
  $F(\bq{2})-F(\bq{1})$ can be easily computed and the formula reads
\begin{align}
  \begin{split}
    \label{eq:energydiff1}
    F(\bq{2})-F(\bq{1}) &= b_{k}(\q{2}_{k}-\q{1}_{k}) + \sum_{i\ne k}a_{ik}\q{1}_{i}(\q{2}_{k} - \q{1}_{k}) \\
    &= (\q{2}_{k}-\q{1}_{k}) \left(b_{1} + \sum_{i \ne k}a_{ik}\q{1}_{i}\right) \\
    & = (1-2\q{1}_{k}) \left(b_{k} + \sum_{i \ne k}a_{ik}\q{1}_{i}\right).
  \end{split}
\end{align}
Interestingly, computing the difference in equation \eqref{eq:energydiff1}
requires only $O(N)$ multiplications. But how can this be used to improve the
performance of the exhaustive search through QUBO state space?

Moving $F(\bq{1})$ to the right-hand side, we obtain a formula for $F(\bq{2})$,
    which allows for computing it with only $N+1$ instead of maximum of $N(N+1)/2$
    multiplications, provided that $F(\bq{1})$ is known. Remember that this is only
    possible because $\bq{1}$ and $\bq{2}$ differ only by a single bit. If we could
    enumerate states in such a fashion that every consecutive two states differ
    only by a single bit, we could leverage the above formula instead of
    recomputing energy for each state from scratch. Before we describe how the
    procedure works and how to implement this on GPU, let us first introduce the
    necessary notation. Given a state $\mathbf{q} = (q_{1},\ldots,q_{N})$, by
  $\flip{\mathbf{q}}{k}$ we will denote a state resulting from flipping $k$-th
    bit of $\mathbf{q}$, i.e.
    \begin{equation}
      \flip{\mathbf{q}}{k} \coloneq (q_{1}, \ldots, q_{k-1}, 1-q_{k}, q_{k+1}, \ldots, q_{N})
    \end{equation}
    and by $\diff_{F}(\mathbf{q},k)$ we will denote the difference between the
    energies of $\flip{\mathbf{q}}{k}$ and $\mathbf{q}$. Using the equation
\eqref{eq:energydiff1}, we see that the expression for
$\diff_{F}(\mathbf{q},k)$ is
\begin{align}
  \begin{split}
    \diff_{F}(\mathbf{q},k) = F(\flip{\mathbf{q}}{k}) - F(\mathbf{q}).
  \end{split}
\end{align}
The pseudocode for a serial algorithm for solving a QUBO problem using our
observations is outlined in listing \ref{lst:grayserial}. Before we can
implement it on GPU though, we need to answer the following questions:
\begin{listing}
  \begin{Verbatim}[commandchars=\\\{\}]
\PYG{k}{def} \PYG{n+nf}{solve\PYGZus{}qubo}\PYG{p}{(}\PYG{n}{F}\PYG{p}{,} \PYG{n}{q}\PYG{p}{):}
    \PYG{n}{q} \PYG{o}{=} \PYG{p}{[}\PYG{l+m+mi}{0}\PYG{p}{]} \PYG{o}{*} \PYG{n}{N} \PYG{c+c1}{\PYGZsh{} Start with all bits set to 0}
    \PYG{n}{best\PYGZus{}state} \PYG{o}{=} \PYG{n}{current\PYGZus{}state} \PYG{o}{=} \PYG{n}{q}
    \PYG{n}{best\PYGZus{}energy} \PYG{o}{=} \PYG{n}{current\PYGZus{}energy} \PYG{o}{=} \PYG{n}{F}\PYG{p}{(}\PYG{n}{q}\PYG{p}{)}

    \PYG{k}{for} \PYG{n}{i} \PYG{o+ow}{in} \PYG{n+nb}{range}\PYG{p}{(}\PYG{l+m+mi}{2} \PYG{o}{**} \PYG{n}{N} \PYG{o}{\PYGZhy{}} \PYG{l+m+mi}{1}\PYG{p}{):}
        \PYG{n}{k} \PYG{o}{=} \PYG{n}{find\PYGZus{}next\PYGZus{}bit\PYGZus{}to\PYGZus{}flip}\PYG{p}{(}\PYG{n}{i}\PYG{p}{)}
        \PYG{n}{current\PYGZus{}energy} \PYG{o}{=} \PYG{n}{current\PYGZus{}energy} \PYG{o}{+} \PYG{n}{diff}\PYG{p}{(}\PYG{n}{q}\PYG{p}{,} \PYG{n}{k}\PYG{p}{)}
        \PYG{n}{current\PYGZus{}state} \PYG{o}{=} \PYG{n}{flip}\PYG{p}{(}\PYG{n}{q}\PYG{p}{,} \PYG{n}{k}\PYG{p}{)}
        \PYG{k}{if} \PYG{n}{current\PYGZus{}energy} \PYG{o}{\PYGZlt{}} \PYG{n}{best\PYGZus{}energy}\PYG{p}{:}
            \PYG{n}{best\PYGZus{}energy} \PYG{o}{=} \PYG{n}{current\PYGZus{}energy}
            \PYG{n}{best\PYGZus{}state} \PYG{o}{=} \PYG{n}{current\PYGZus{}state}
    \PYG{k}{return} \PYG{n}{best\PYGZus{}state}\PYG{p}{,} \PYG{n}{best\PYGZus{}energy}
\end{Verbatim}

  \caption{Pseudocode for algorithm solving the QUBO problem using energy differences and
    bit flips.} \label{lst:grayserial}
\end{listing}
\begin{enumerate}
  \item How to produce a sequence of $2^{N}-1$ bits such that executing them enumerates
    a rll possible states?
  \item How to divide work among CUDA threads?
\end{enumerate}
The answer to the first question is well-known and involves enumerating
integers using the Gray code, which we will describe now.

\subsection{Gray code}
When one talks about a binary encoding of integers, the first thing that comes
to mind is a usual positional base-2 system. This encoding certainly does not
fit our purpose. Indeed, suppose $N=3$ and we are currently processing state
corresponding to number $3$, whose representation in binary is $(011)_{2}$. The
    next state, corresponding to number $4$, is encoded by the string $(100)_{2}$, which
differs in all three bits.

Instead of using the positional system, we might utilize an encoding called
Gray code, or Reflected Binary Code (RBC) \cite{gray,lucal1959}, which is
primarily used to improve the robustness of electromechanical switches and in
the error correction protocols. In this code, encoding of two successive
integers always differs by at most one bit, which makes it suitable for
application in our algorithm.

\begin{figure}
  \includegraphics[width=\textwidth]{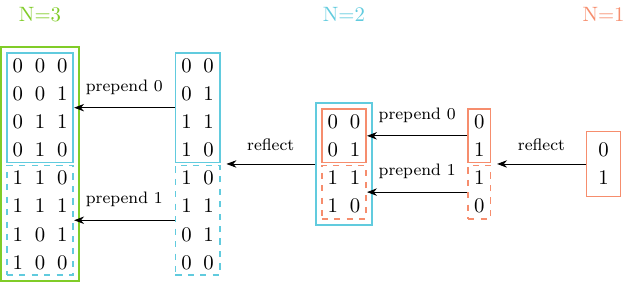}
  \caption{Reflection--based construction of Gray code. The length of the code is denoted
    by $N$. For $N=1$, the code comprises two binary strings, $0$ and $1$. To
    construct the code of length $N>1$, the code of length $N-1$ and its vertical
    reflection are stacked. Then, the first, unreflected half is prepended with $0$
    while the second, reflected half is prepended with $1$.} \label{fig:gray}
\end{figure}

The conceptual construction of the Gray code is straightforward. For Gray code
of length $1$ we have two binary strings: $0$ and $1$. To obtain all Gray codes
for a given length $N > 1$, we first construct an ordered list of codes of
length $N-1$ and call it $L$. Then, we reverse the list of codes and call it
$H$, an operation called \emph{reflection}. Finally, we prepend $0$ to all
elements of $L$ and prepend $1$ to all elements of $H$. The concatenation of
$L$ and $H$ forms the $N$--bit Gray Code. The process is illustrated in Fig.
\ref{fig:gray}. One useful consequence of the construction is that the shorter
Gray code might be viewed as an initial part of the larger one prepended with
enough zeros. Thus, statements like ``$n$-th Gray code'' make sense and are
unambiguous.

An important thing to observe is that in our algorithm we need at most two Gray
code-encoded numbers at the time to determine the bit to be flipped. The
reflection--based construction outlined so far would require precomputing a
large part (if not all) of the encodings at once. Considering the size of the
state space, this is clearly infeasible. However, there exists an explicit
formula for computing $n$-th Gray code, which reads \cite{grayalgo}:
\begin{equation}
  \mbox{gray}(n) = n \oplus (n >> 1),
\end{equation}
where $\oplus$ denotes the bitwise xor operation and $>>$ is right bitshift.

To compute which bit differs between consecutive Gray codes, we can xor them,
and then find the position of the only set bit in the resulting integer. One
can easily implement a function that finds the first set bit in a 64-bit
integer, or use one of the available library or compiler built-in functions.
For instance, POSIX--compatible C standard libraries include \texttt{ffsll}
function \cite{ffs}. In CUDA, there is a \texttt{\_\_ffsll} function available
\cite{CUDAguide}. For both of the above cases, the function counts bits from
$1$. Using this convention, we can write a pseudocode for a function
\texttt{find\_bit\_to\_flip}, occurring in the listing \ref{lst:grayserial}, like in the
listing~\ref{lst:findbit}.

\begin{listing}
\begin{Verbatim}[commandchars=\\\{\}]
\PYG{k}{def} \PYG{n+nf}{find\PYGZus{}bit\PYGZus{}to\PYGZus{}flip}\PYG{p}{(}\PYG{n}{i}\PYG{p}{):} \PYG{c+c1}{\PYGZsh{} i starts from 0}
    \PYG{k}{return} \PYG{n}{ffs}\PYG{p}{(}\PYG{n}{gray}\PYG{p}{(}\PYG{n}{i}\PYG{p}{)} \PYG{o}{\PYGZca{}} \PYG{n}{gray}\PYG{p}{(}\PYG{n}{i}\PYG{o}{+}\PYG{l+m+mi}{1}\PYG{p}{))}
\end{Verbatim}

  \caption{Pseudocode for a function generating bit flips for Gray code construction}
  \label{lst:findbit}
\end{listing}

Now that we know how to construct a correct sequence of bit flips, it is time
we design a parallelization strategy, which is what we will do next.

\subsection{Parallelization using GPU}
\begin{figure}
  \includegraphics[width=\textwidth]{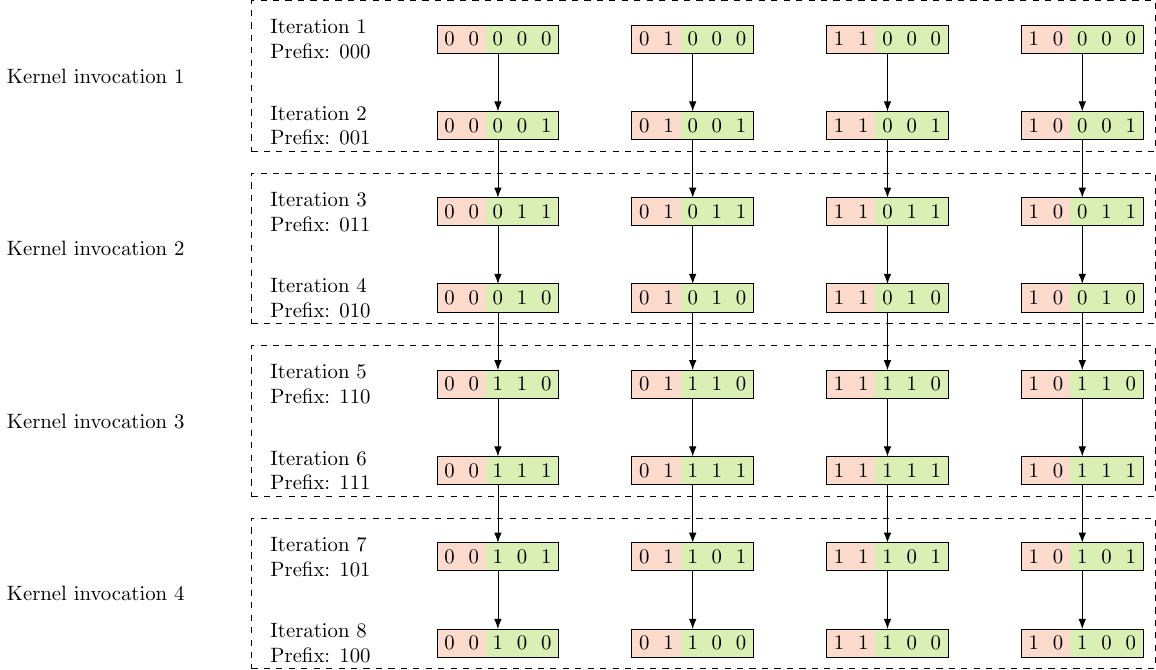}
  \caption{
    Parallel processing of $N=5$--variable QUBO configurations in Gray code order.
    In our example, suffix length $M=2$, and hence $2^{M}=4$ states are processed
        in each iteration. Consequently, there are $2^{N-M}=8$ iterations. For this
        example, we consider a kernel that processes two iterations per kernel
        invocation, resulting in $2^{N-M}/2 = 4$ kernel invocations total. }
  \label{fig:grayparallel}
\end{figure}

The algorithm presented in \ref{lst:grayserial} is fully serial. Our task is
now to parallelize it so that it can be executed on GPU. Unsurprisingly, we
will once again employ the strategy of dividing each state into a suffix and a
prefix part. This time, however, it is the suffix that will stay fixed between
iterations. The prefix part will be updated in each iteration by flipping a
single bit in Gray code order. The process is illustrated in Fig.
\ref{fig:grayparallel}.

Throughout the execution of the algorithm, we maintain four arrays of size
$2^{M}$. In each array, the $i$-th item always corresponds to the $i$-th
suffix. The \texttt{best\_states} and \texttt{best\_energies} arrays store the
best states found so far amongst states with $i$-th suffix. The
\texttt{current\_states} and \texttt{current\_energies} store configuration and
corresponding energy of current state being processed for $i$-th suffix. Each
iteration starts by determining the index of the next bit to be flipped. This
value is the same for all suffixes. Next, the algorithm computes the energy
difference using the equation \ref{eq:energydiff1} and updates the
corresponding energy accordingly. After all $2^{N-M}$ iterations, the
\texttt{best\_states} and \texttt{best\_energies} arrays are used to extract
the ground state.

It is crucial to note that since we are only interested in finding the ground
state, we can group several iterations in one kernel invocation. In fact, it is
entirely possible to implement a kernel that runs all the iterations, which
would avoid kernel launch overhead. Moreover, such kernel could use
thread--local variables to store current state and energy instead of using
global arrays, which would further increase performance. However, as we will
see further in this chapter, we will propose further optimizations that would
require us to split the algorithm into several kernel invocations.

\subsection{Further optimizing parallel execution}

There are two optimizations we can make to further reduce the number of
operations performed in each iteration. Let us first notice that the only bit
flips that can happen, do so in the prefix part. Going back to equation
\eqref{eq:energydiff1}, we can rewrite the expression for $F(\bq{2})-F(\bq{1})$
into a sum of two parts:
\begin{align}
  F(\bq{2})-F(\bq{1}) = & (1-2\q{1}_{k}) \left(b_{k} + \sum_{i=0,i\ne k}^{N-M-1}a_{ik}\q{1}_{i}\right) + \label{eq:energydiff2} \\
                        & (1-2\q{1}_{k}) \left(b_{k} + \sum_{i=N-M}^{N-1}a_{ik}\q{1}_{i}\right)\label{eq:energydiff3}
\end{align}
Since the first summand \eqref{eq:energydiff2} is independent from the suffix,
which means that for each of the considered suffixes in any given iteration, it
has the same value. Since the states in the iteration are processed in parallel
by GPU threads, we have to either redo the same computation multiple times, or
use some synchronization mechanism, e.g. compute the prefix in one thread in
each block and then propagate the result to the whole block through shared
memory. However, there is a third approach. For each iteration, we compute the
prefix part of the energy difference using CPU, and then use it as a kernel
parameter. More precisely, we compute $L$ values of the prefix part of the
energy difference and pass it to the kernel as an additional array. Since the
information about which bit to flip is also relevant, we pass the bit sequence
as another array as well.

As for the \eqref{eq:energydiff3} part, observe that for each given prefix
there are only $N-M$ possible values of $k$ (again, that's because the bit
flips happen only in the prefix part, and there are $N-M$ prefix bits).
However, not all values of $k$ are equally common. Examining the Gray code
construction (c.f. Fig. \ref{fig:gray}) reveals that the least significant bit
flips half of times and the second least significant bit flips a quarter of
times. Generally, for $k=0,\ldots,N-M-1$ the $k$-th bit flips constitutes
approximately \footnote{Approximately, because there is an odd number of
$2^{N-M}$ -1 flips performed, because we do not perform last bit flip which
    would take us back to $(0,\ldots,0)_2$ prefix.} $1/2^{k+1}$ of times.
Therefore, we can cache the value of \eqref{eq:energydiff3} for the $K$ most
commonly--occuring bit flips, where $K$ is a user--controlled parameter.

\begin{figure}
  \includegraphics[width=\textwidth]{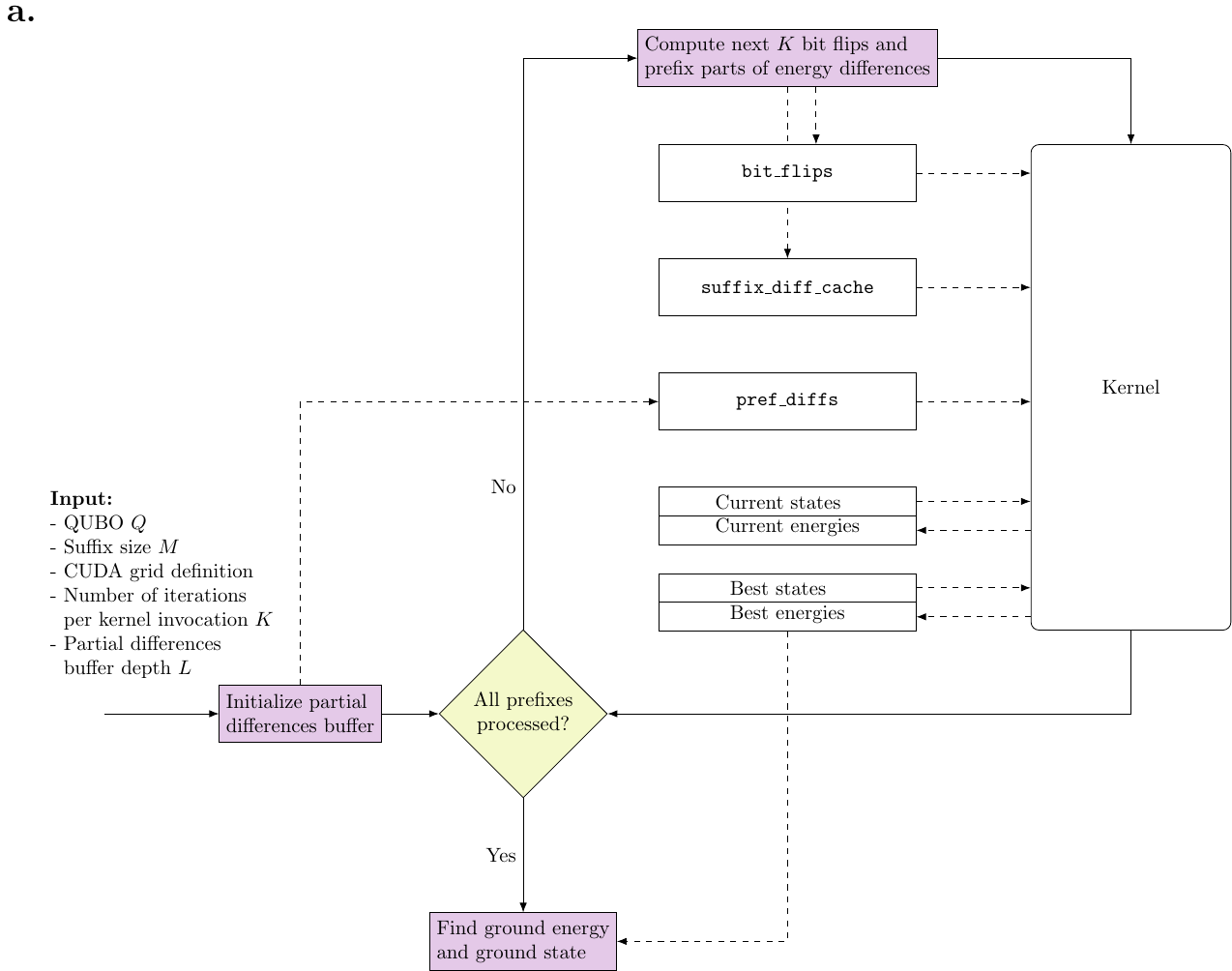}
  \caption{
    Schematic representation of the GPU--enabled brute force algorithm for finding
    ground state of a QUBO problem using Gray codes. } \label{fig:bruteforcegray}
\end{figure}

\begin{figure}
  \includegraphics[width=\textwidth]{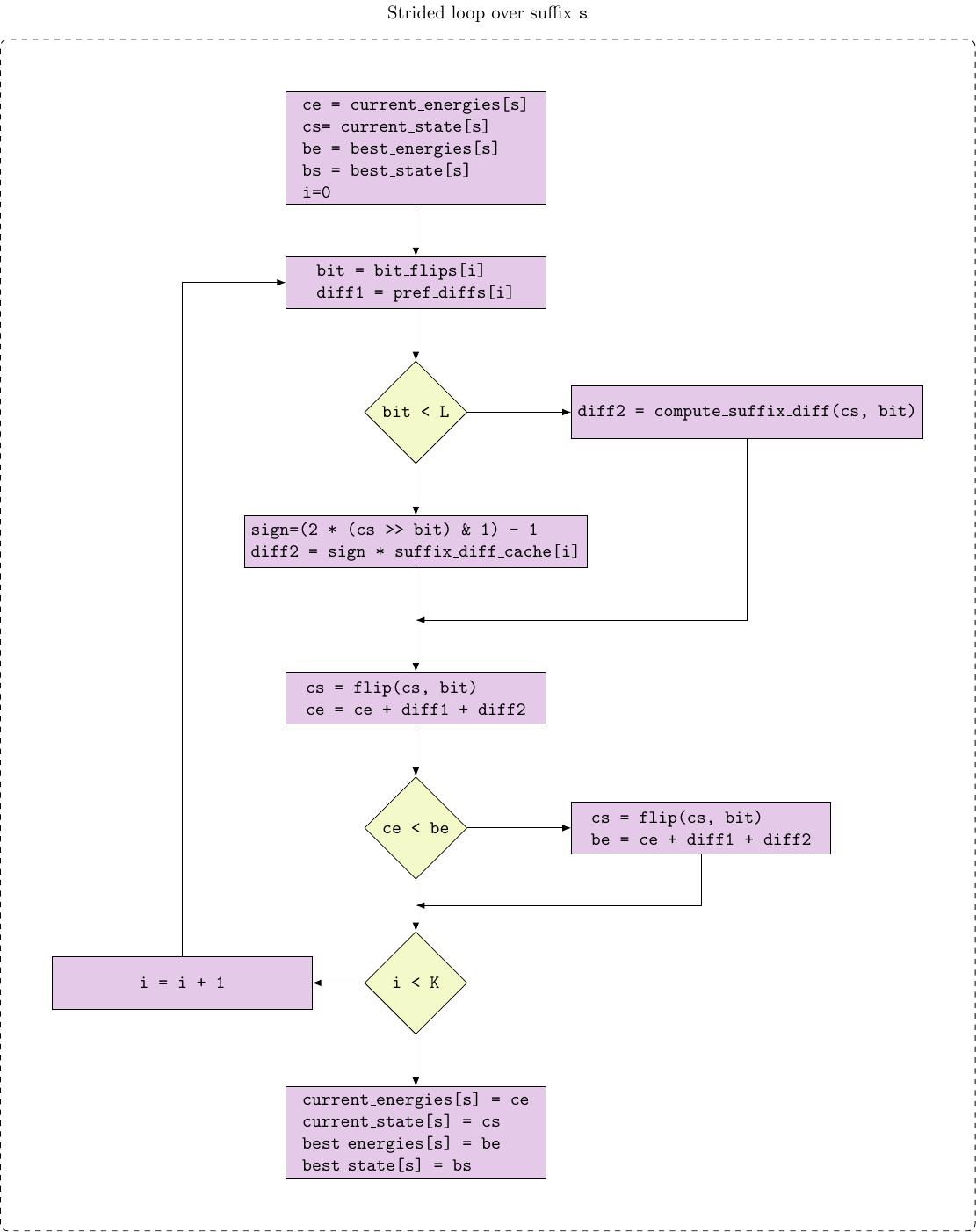}
  \caption{Implementation of the strided loop for kernel in fig \ref{fig:bruteforcegray}.}
\end{figure}

\subsection{Performance evaluation of the optimized algorithm}

We performed a preliminary performance evaluation of the Gray code--based
algorithm for finding the ground state. For our benchmarks, we used several
setups with different Nvidia GPUs. Some setups were equipped with more than one
copy of the same GPU. The summary of test setups, as well as capabilities of
the used GPUs is presented in table \ref{tab:bruteforce-gs-only-table}.

\begin{table}[!ht]
  \footnotesize
  \begin{tabular}{|l|c|c|c|c|c|}
    \hline
    \cellcolor{theader}                                              & \multicolumn{2}{c|}{\cellcolor{theader} Kernel launch parameters} & \multicolumn{2}{c|}{\cellcolor{theader}Algorithm parameters} & \cellcolor{theader}                                                                      \\
    \hhline{~----~}
    \cellcolor{theader} \multirow{-2}{*}{\cellcolor{theader} GPU(s)} & \cellcolor{tsubheader} Block size                                 & \cellcolor{tsubheader} Grid size                             & \cellcolor{tsubheader} Suffix size & \cellcolor{tsubheader} \makecell{\# Steps per       \\ kernel launch} & \multirow{-2}{*}{\cellcolor{theader} \makecell{\# Fixed \\ variables}}\\
    \hline
    A10                                                              & 512                                                               & 4096                                                         & 27                                 & 4096                                          & N/A \\
    \hline
    A100                                                             & 512                                                               & 4096                                                         & 27                                 & 4096                                          & N/A \\
    \hline
    A6000                                                            & 1024                                                              & 4096                                                         & 27                                 & 4096                                          & N/A \\
    \hline
    V100 x8                                                          & 1024                                                              & 4096                                                         & 27                                 & 4096                                          & 3   \\
    \hline
    DGX H100 x8                                                      & 1024                                                              & 8192                                                         & 29                                 & 8192                                          & 3   \\
    \hline
    RTX 4090 x8                                                      & 512                                                               & 4096                                                         & 28                                 & 4096                                          & 1   \\
    \hline
  \end{tabular}
  \caption{Parameters used for benchmarking} \label{tab:bruteforce-gs-only-table}
\end{table}

Since each of the test setups was available to us only for a limited amount of
time, we were only able to measure execution times for a very limited subset of
parameters, and we needed to make some educated guesses. We decided to use a
constant depth of the prefix differences buffer equal to 10. Depending on the
available memory size, we used suffix sizes of 27, 28 and 29. The tested kernel
executions grids included blocks of 256, 512 or 1024 threads, and 4096 or 8192
blocks per grid. We also considered 2048, 4096 and 8192 algorithm steps per
single kernel execution. The best parameters found for each setup are presented
in table \ref{tab:bruteforce-gs-only-table}. We would like to stress out,
however, that such a coarse--grained process of parameter tuning does not
guarantee their global optimality. Further parameter tuning could be achieved
by searching through a finer grid of parameters, possibly combined with
profiling. Lastly, for multi-GPU setups, we solved each instance by
distributing the work equally between GPUs by splitting the problem into chunks
of equal size. The splitting was done by constructing new QUBOs by fixing the
values of $l$ variables, resulting in $2^{l}$ total subproblems.

\begin{figure}
  \includegraphics[width=\textwidth]{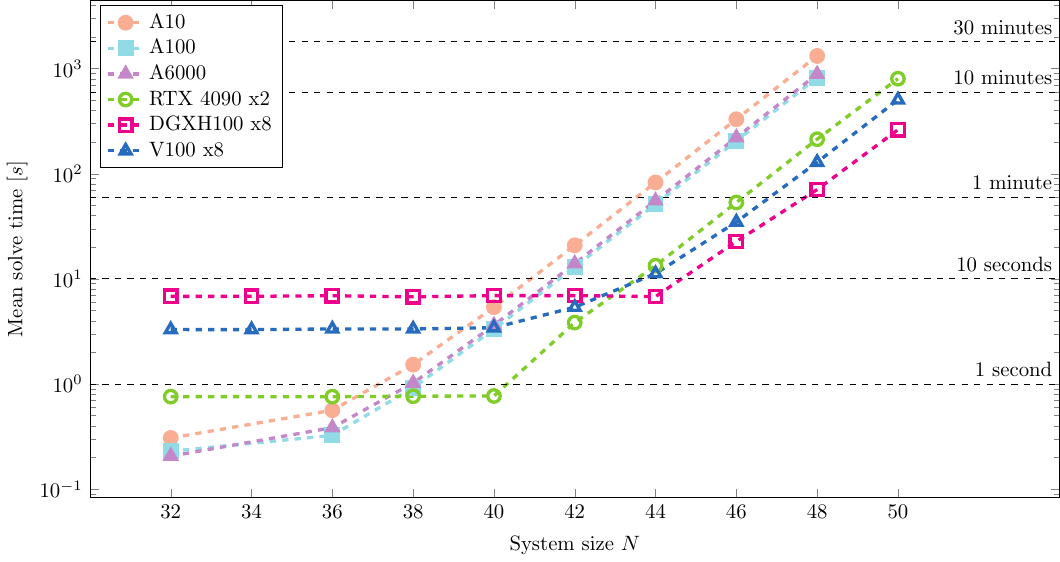}
  \caption{
    Benchmarking results for Gray code--based brute force algorithm for finding a
    ground state of Ising model. The dashed lines between data points are provided
    for visual guidance. For each system size $N$, the solution times were averaged
    over 20 different instances with known ground states. Observe that for setups
    with multiple GPUs and small system sizes, the solution time remains virtually
    constant. This is because, for small system sizes, the execution time is
    dominated by tasks related to distributing work and gathering results. The
    parameters used for benchmarking in each setup are summarized in table
    \ref{tab:bruteforce-gs-only-table}. } \label{fig:bruteforce-gsonly-benchmarks}
\end{figure}

The figure Fig. \ref{fig:bruteforce-gsonly-benchmarks} shows the results of our
benchmarks. Observe that for multi-GPU setups and system sizes $N \le 40$, the
execution time is almost constant. This is because the time presented on the
graph factors in the time of scattering work among GPUs and gathering the
results, which for small system sizes dominates the actual solver execution
times. Aside from these plateaus, as expected, the graphs of measured solution
times resemble exponential curves. On the setup with eight DGX H100 GPUs, the
optimized code was able to find the ground state of instances with system size
$N=50$ in less than 5 minutes.

%% file: trains.tex
\chapter[Railway conflict management]{Application to railway conflict management}
\label{chapter:trains}
As the last point in the thesis, in this chapter we describe how the methods
presented so far can be applied in the field of operational research. Namely,
we propose an approach to solving the railway dispatching problem using quantum
annealing. We benchmark the implementation of our algorithm on the current
generation of D-Wave annealers, using solutions obtained via tensor networks
and exhaustive search as a baseline for comparison.

\section{Overview of the problem}
\begin{figure}[ht]
  \includegraphics[width=\textwidth]{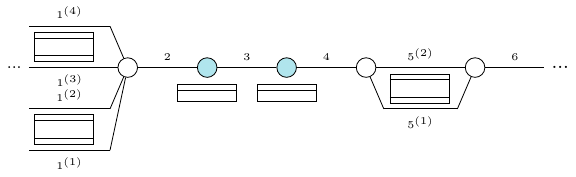}
  \caption{
    An example network. Sections $2, 3, 4$ and $6$ are \emph{line blocks}, while
    sections $1$ and $5$ are \emph{sidings} with respectively $4$ and $2$ tracks.
    Rectangles represent platforms. Circles represent points where a line block and
    a siding join (white) or where two line blocks join (blue). Superscripts denote
    tracks within a siding. } \label{fig:railway-network}
\end{figure}

We will consider a part of a railway network, which we will simply refer to as
a \emph{network}. The network is divided into \emph{block sections} or simply
\emph{blocks}. In our approach, we focused only on the single--track lines,
which means that the network can only comprise the following types of blocks:
\begin{itemize}
  \item \emph{Line blocks}, or \emph{single track} sections, pieces of infrastructure that can be occupied by one train
    at a time.
  \item \emph{Sidings}, or \emph{parallel tracks} (occurring e.g. at stations). At the sidings,
    trains passing in the same direction can meet--and--overtake, and trains passing
    in the opposite directions can meet--and--pass. Each siding comprises two or more
    tracks, each of which can also be occupied by one train at a time. In our examples, the
    sidings will occur at the station, and hence we will also call them \emph{station blocks}.
\end{itemize}
Fig. \ref{fig:railway-network} shows an example network. -

The trains move through the network according to a \emph{timetable}. It is
assumed that this timetable is conflict-free, i.e. at any time no two trains
occupy the same track.

Now, suppose the network is affected by a disturbance, which has prevented some
trains from running according to their original timetable. Examples of possible
disturbances include, but are not limited to, a malfunction of one or more
trains or a malfunction of railway tracks. After the disturbance, some trains
occupy different parts of the networks than they are supposed to and resuming
operations according to the original timetable might not be possible. The problem
might be viewed from
various perspectives, e.g. that of a passenger or transport operation company
\cite{tornquist,lamorgese,Jensen2016}. In this chapter, we look at the problem
from the perspective of the infrastructure manager whose task, in the presence
of a disturbance, is to create a new, conflict-free timetable. Naturally, in
most cases, there will be multiple possible solutions to the arising conflicts,
and hence one has to decide on what criteria make one timetable more appealing
than another. In our approach, we assume that the dispatcher aims to minimize
some function of the trains' delays, which we will describe later. There are
also other possible choices of the objective function \cite{8795577} such as
the total passenger delay or the total cost of operations.

Let us observe that independently from the algorithm for constructing a new
timetable, some delays after the disturbance might be inevitable, e.g. due to
engineering or even physical limits\footnote{For instance, if a train has been broken
for some time, it can only follow the timetable if it is not already late and
it can make up for the time it already lost -- and this can only happen if it
can reach a sufficiently large speed. If this is not the case, the train will
be necessarily delayed.}. Moreover, by taking into account the maximum speed with
which the trains can move through each section, one can calculate the lower
bounds on the delays. These lower
bounds are known as unavoidable, or \emph{primary}, delays \cite{dariano}.

In the ideal case, if all the trains could travel at their maximum speed, all
trains would be delayed only by their primary delays and there would be nothing
to optimize. However, it might not always be possible. Suppose for instance,
that two trains going in the same direction are already delayed at a station
neighboring a line block. From each train's perspective, the optimal solution
is to start its route immediately when possible. However, as only one of them
can do this because a line block can only be occupied by one train at a time.
Hence, at least one of these two trains will have a delay larger than the
primary one. What is important is that this additional delay is not a
consequence of some physical or engineering limitations, but rather a
consequence of the dispatcher's decision made to avoid a potential conflict.
All such delays are called the \emph{secondary} delays and, unlike the primary
ones, they are subject to optimization.

The distinction between primary and secondary delays might seem artificial at
first, but it has profound consequences. Namely, when constructing a function
to be minimized we only need to take into account the secondary delays. For
instance, we might want to minimize their total sum or their weighted sum, with
weights corresponding to the trains' priorities.

The above high-level description of the problem needs now a mathematical
formulation, which we present in the next section.

\section{The mathematical model}
Before we can formulate the optimization problem to be run on D-Wave, we need
first to formally describe the railway model. The first idea that comes to mind
is to define quantities corresponding to the departure and arrival times of each
train and relevant station blocks and express all other quantities in the
model in terms of their difference with respect to the times found in the
original timetable. However, as we will soon see, one can almost completely
forget about arrival and departure times, and instead express all quantities in
the model using delays. Moreover, we will further simplify our model by
assuming all secondary delays are integers falling into some finite range.

We assume that the analyzed network segment is a sequence $N$ blocks, with
first and the last blocks being station blocks. The set of all trains will be
denoted by $\JJ$. This set is naturally partitioned into the set $\JJ_0$ of
trains going into one direction and the set $\JJ_1$ of trains going into the
opposite direction. This is a proper partition, i.e.:
\begin{equation}
  \JJ_0 \cup \JJ_1 = \JJ \quad \JJ_0 \cap \JJ_1 = \emptyset
\end{equation}
We assume that each train travels through the whole analyzed network segment.
The route $\Bj$ of a train $j$ comprises sequence of blocks:
\begin{equation}
  \Bj \coloneq (b_{j,1}, b_{j,2}, \ldots, b_{j,N})
\end{equation}
Our model forbids recirculation, i.e. each train passes every block in its
route exactly once. Therefore, the route of each train is uniquely identified
by a sequence of station blocks $\Sj$:
\begin{equation}
  \Sj \coloneq  \left(s_{j,1}, s_{j, 2}, \ldots, s_{j, \eend}\right).
\end{equation}
We will denote the time at which the train $j$ should leave a block $b \in \Bj$ according to the
original timetable by $\ttout(j, b)$. Similarly, the time at which the train
$j$ is supposed to enter block $b$ will be denoted by $\ttin(j, b)$. In our
model, we assume that the time at which a train leaves one block is precisely
the same as the time it enters the next block, i.e.
\begin{equation}
  \ttout(j, b_{j,k}) = \ttin(j, b_{j,k+1}).
\end{equation}
It is clear that the original timetable determines how long it takes for a
train $j$ to travel through a given block $b \in \Bj$. We call this time the
\emph{passage time}, and denote it by $\pt(j, b)$:
\begin{equation}
  \pt(j, b) \coloneq \ttout(j, b) - \ttin(j, b).
\end{equation}
An important observation is that the passage times defined by the timetable may
not be the minimum physically achievable passing times $\pmin(j, b)$.
Therefore, one can define a time reserve $\alpha(j, b)$ which can be used by
train $j$ to compensate for the delay when traveling through block $b$:
\begin{equation}
  \label{eq:pt}
  0 \le \alpha(j, b) \coloneq \pt(j, b) - \pmin(j, b).
\end{equation}
The time reserve will become important when discussing the propagation of the
primary delays.

\subsection{Delay representation}
Suppose the disturbance happened, resulting in some trains not being able to
meet the schedule. Hence, the actual leaving and arrival times (denoted by
$\tout$ and $\tin$) differ from the scheduled ones. The delay $d(j, s)$ of the
train $j$ at station block $s \in \Sj$ is defined as the difference:
\begin{equation}
  \label{eq:djs}
  d(j, s) \coloneqq \tout(j, s) - \ttout(j, s) 
\end{equation}
As already mentioned, $d(j, s)$ can be expressed as a sum:
\begin{equation}
  d(j, s) = d_U(j, s) + d_S(j, s)
\end{equation}
where $d_U$ denotes the primary (or unavoidable) delay, and $d_S$ denotes the
secondary delay \cite{dariano}. In the absence of time reserve, one would simply have $d_U(j,
  s) = d_U(j, s')$ for any given train $j$ and blocks $s,s' \in S_{j}$. However,
    the time reserve allows to somewhat compensate delays, and hence we have
    \begin{equation}
      d_U(j, s_{j,k+1}) = \max\left\{0, d_U(j, s_{j,k}) - \sum_{b}\alpha(j, b)\right\},
    \end{equation}
    where the sum runs over all blocks starting from the one following $s_{j,k}$
    and ending on $s_{j,k+1}$. The secondary delays can be, in principle,
    arbitrarily large. However, it is convenient to assume that all secondary
    delays for the train $j$ are bound from above by some constant $d_{\max}(j)$.
One can find a reasonable upper bound by running some fast heuristic, or
determine it manually (e.g. there might be an \emph{a priori} established
maximum allowable delay of the train). Henceforth, we will consider
$d_{\max}(j)$ to be parameters of the model. With this assumption, we have the
following bounds on the overall delays:
\begin{equation}
  d_U(j, s) \le d(j, s) \le d_U(j, s) + d_{\max}(j).
\end{equation}

\section{Discretizing delays}
Formulation of the problem presented so far can facilitate the construction of
a linear, constrained model of the dispatching problem. However, since the
secondary delay values are continuous variables, such a model would not be
compatible with the quantum annealer. We circumvent this issue by discretizing
the delays. One way to do it is to require all secondary delays to be natural
numbers, i.e.:
\begin{equation}
  \forall_{j \in \JJ} \forall_{s \in S_{j}}\quad  d_{S}(j, s) \in \{0, 1, \ldots, d_{\max}(j)\}.
\end{equation}
As a consequence, the total delays become discretized as well. We will denote
the set of possible values for $d(j, s)$ by $A_{j, s}$, i.e.:
    \begin{equation}
      \forall_{j}\forall_{s \in \Sj}  A_{j, s} \coloneq \{d_{U}(j, s), d_{U}(j, s) + 1, \ldots, d_{U}(j, s) + d_{\max}(j)\}.
    \end{equation}
    Notice that this discretization is not particularly restrictive, as timetables
    typically have a finite resolution of minutes anyway.

    We can now use one-hot encoding for $d(j, s)$ and introduce the decision binary variables
  $x_{s, j, m}$:

    \begin{equation}
      \forall_{j \in \JJ}\forall_{s \in S_{j}} \forall_{m \in A_{j, s}} \quad x_{s,j,m} = \begin{cases}
        1, & d(j, s) = m      \\
        0, & \mbox{otherwise}
      \end{cases}.
    \end{equation}
    Naturally, possible values for $d(j, s)$ are mutually exclusive, which can be
    expressed as the following constraint:
    \begin{equation}
      \label{eq:onehotconstraint}
      \forall_{j \in \JJ}\forall_{s \in \Sj} \sum_{m \in A_{j, s}} x_{s, j, m} = 1
    \end{equation}
    As for the cost function, we decided to use a simple weighted sum of the
    delays, i.e. the cost function of the form:
    \begin{equation}
      \label{eq:qubo:cost}
      f(\mathbf{x}) = \sum_{j \in \mathcal{J}}\sum_{s \in \Sjs} \sum_{m \in A_{j,s}} w(s,j,m) \cdot x_{j,s,m},
    \end{equation}
    where $\Sjs = \Sj \setminus \{s_{j,\eend}\}$. For instance, choosing $w(s, j,
  m)=m$ would result in an objective of minimizing the sum of all delays. In
general, however, one could take into account the relative importance of the
trains, as we will describe later when introducing the real railway sections
considered in our research.

\section{Dispatching conditions and the penalties}
The cost function \eqref{eq:qubo:cost} together with constraint
\eqref{eq:onehotconstraint} is not enough to construct a meaningful
optimization problem. We also have to take into account other constraints
stemming from dispatching conditions. For instance, we cannot allow a schedule
in which two trains occupy the same track at the same time. We describe the
precise forms of the constraints in detail in the
Appendix~\ref{chapter:dispatching}, and in this section, we will only provide
their brief overview. The dispatching conditions are:
\begin{enumerate}
  \item \textbf{The minimum passing time condition.} Train cannot travel through a block faster than the corresponding minimum passing time.
  \item \textbf{The single block occupation condition.} Two trains cannot occupy the same part of a single railway track.
  \item \textbf{The deadlock condition.} Suppose trains $j$ and $j'$ are
    heading in opposite directions on a route determined by two consecutive
    stations $s_{j,k}$ and $s_{j,k+1}$. In this case, $j$ has to arrive at $s_{j,k+1}$ before $j'$ can leave $s_{j,k+1}$, or vice versa.
  \item \textbf{The rolling stock circulation condition.} Our model assumes that some trains are assigned the same train set. The rolling stock circulation condition ensures there exists some minimum \emph{turnover time}, before a train set can be reused.
\end{enumerate}

The dispatching conditions together with the cost function \eqref{eq:qubo:cost}
and one-hot encoding constraint \eqref{eq:onehotconstraint} define a
constrained $0-1$ problem. However, in order to use a quantum annealer, we must
convert it to QUBO, which means we have to incorporate those constraints into
the objective function.

One might observe that penalties defined by the dispatching conditions (c.f.
Appendix \ref{chapter:dispatching}) are of the form:
\begin{equation}
  \label{eq:quadraticpenalty}
  \sum_{(l, l') \in \mathcal{V}_{p}} x_{l}x_{l'} = 0,
\end{equation}
for some set $\mathcal{V}_{p}$ of pairs of multiindices. For every feasible
solution (i.e. one meeting all the constraints) the sum in the equation
\eqref{eq:quadraticpenalty} is 0, whereas violation of the corresponding
condition gives a strictly positive value. Hence, one can add such a sum to the
cost function, effectively penalizing the infeasible solutions. More generally,
one might multiply the sum by some constant $\ppair > 0$, to further increase
the value of the cost function for the infeasible solutions. Finally, taking
into account all penalties from all dispatching conditions gives the following
term that can be added to the objective function:
\begin{equation}
  \Ppair(\mathbf{x}) = \ppair \sum_{\mathcal{V}_{p}}\sum_{(l, l') \in \mathcal{V}_{p}} x_{l}x_{l'}.
\end{equation}

The same reasoning cannot be applied e.g. to the constraint
\eqref{eq:onehotconstraint}, which comprises equations of the form:
\begin{equation}
  \label{eq:linearpenalty}
  \sum_{l \in \mathcal{V}_{s}}x_{l} = 1.
\end{equation}
If one added sums from the equation \eqref{eq:linearpenalty} to the cost
function, it would favor the infeasible solution comprising all 0s. Instead,
one can consider the following quadratic form of the same penalty:
\begin{equation}
  \label{eq:linearpenalty2}
  \left(\sum_{l \in \mathcal{V}_{s}}x_{l} -1 \right)^{2} = 0
\end{equation}
In contrast to \eqref{eq:linearpenalty}, this time the left-hand side is equal
to 0 for feasible solutions, and takes a positive value for any solution
violating the one-hot encoding constraint. As with previous, quadratic
penalties, we might want to multiply such penalties by some constant $\psum >
  0$. An important thing to mention here is that the expansion of the left-hand
side in \eqref{eq:linearpenalty2} gives a nonzero constant offset, which we
will ignore. Therefore, the final form of the penalty corresponding to the
one-hot encoding constraint is:
\begin{equation}
  \Psum(\mathbf{x}) = \psum \sum_{\mathcal{V}_{s}}\left(\sum_{(l,l') \in \mathcal{V}_{s}^{\times 2}, l\ne l'} x_{l}x_{l'}  - \sum_{l \in \mathcal{V}_{s}}x_{l}\right).
\end{equation}
Lastly, the total objective function for our QUBO reads:
\begin{equation}
  f'(\mathbf{x}) = f(\mathbf{x}) + \Psum(\mathbf{x}) + \Ppair(\mathbf{x}).
\end{equation}
\section{Results}

\subsection{Studied railway segments}
In our work, we considered two single-track railway lines managed by the Polish
state-owned company PKP Polskie Linie Kolejowe:

\begin{itemize}
  \item Railway line No. 216 (Nidzica -- Olsztynek section)
  \item Railway line No. 191 (Goleszów -- Wisła Uzdrowisko section)
\end{itemize}

\begin{figure}
  \begin{subfigure}{\textwidth}
    \caption{}\label{fig:linesmall:line}
    \includegraphics[width=\textwidth]{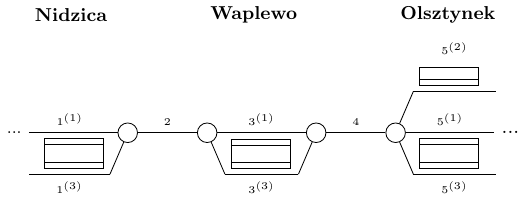}
  \end{subfigure}
  \begin{subfigure}{\textwidth}
    \caption{}\label{fig:linesmall:diagram}
    \includegraphics[width=\textwidth]{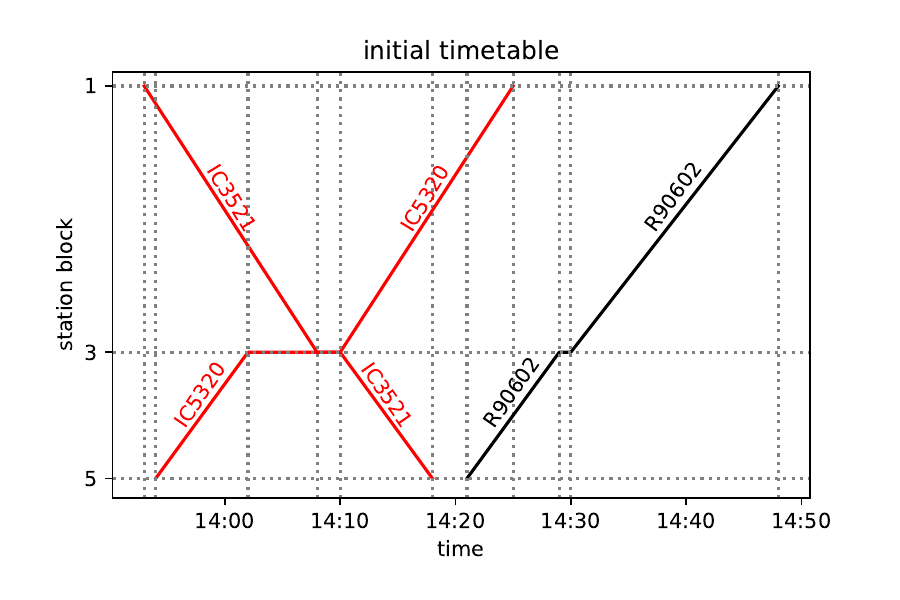}
  \end{subfigure}
  \caption{\subref{fig:linesmall:line} Nidzica -- Olsztynek segment of line No. 216. The segment comprises three
    station blocks (1 -- Nidzica, 3 -- Waplewo, 5 -- Olsztynek), and two line
    blocks (2, 4). We assume that passing through the station block takes the same
    amount of time independently of which track is used. \subref{fig:linesmall:diagram} Train diagram for the undisturbed timetable of the line in \subref{fig:linesmall:line}. The timetable features two
    \emph{Inter--City} trains IC3521 and IC3520 and one \emph{Regio} train R90602. The paths for the
    \emph{Inter-City} trains are marked with red and path of the \emph{Regio} train is marked with black.}
  \label{fig:linesmall}
\end{figure}

\begin{figure}
  \begin{subfigure}{\textwidth}
    \caption{}\label{fig:linelarge:line}
    \includegraphics[width=\textwidth]{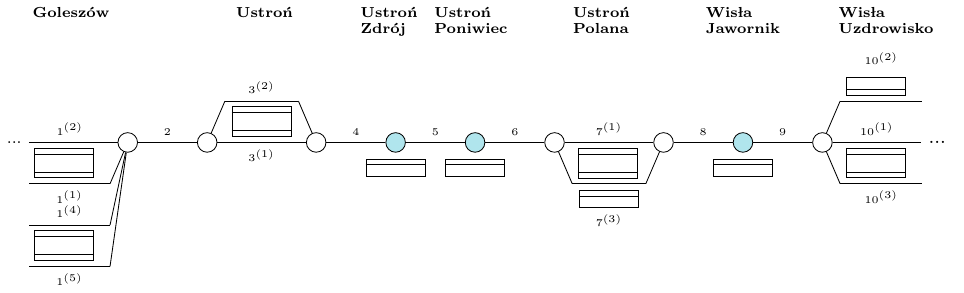}
  \end{subfigure}
  \begin{subfigure}{\textwidth}
    \caption{}\label{fig:linelarge:diagram}
    \includegraphics[width=\textwidth]{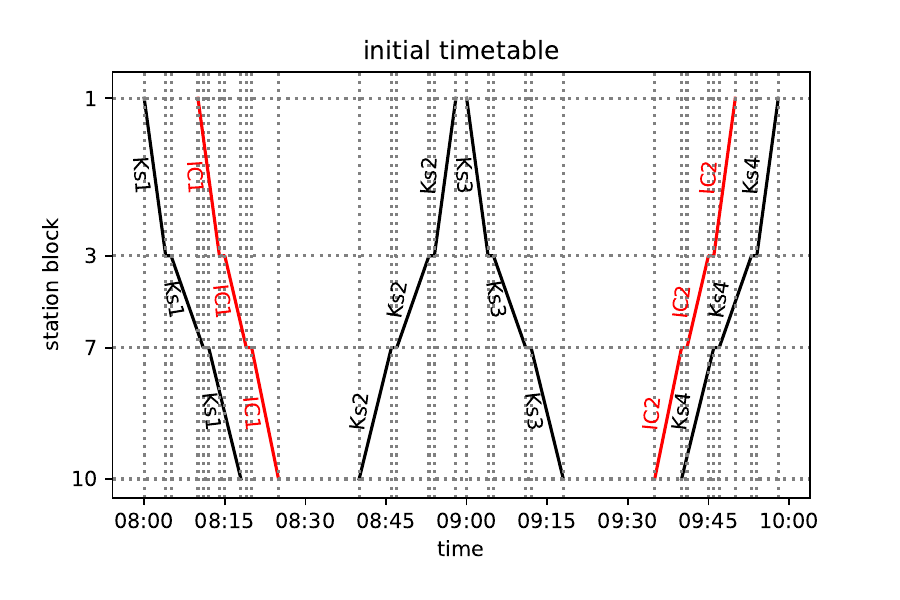}
  \end{subfigure}
  \caption{\subref{fig:linelarge:line} Goleszów -- Wisła Uzdrowisko segment of line No. 191. The segment comprises 4
    station blocks (1 -- Goleszów, 3 -- Ustroń, 7 -- Ustroń Polana, 10 -- Wisła
    Uzdrowisko) and 6 line blocks (2, 4, 5, 6, 8, 9). Between line blocks there are
    additional passenger platforms at Ustroń Zdrój, Ustroń Poniwiec and Wisła
    Jawornik. \subref{fig:linelarge:diagram} Train diagram for the timetable of the line in \subref{fig:linelarge:line}.
    The timetable features two \emph{Inter--City} trains (IC1 and IC2) and four regional trains (Ks1--Ks4).
    The paths of the \emph{Inter--City} trains are marked with red and the paths of the regional trains are
    marked with black.}
  \label{fig:linelarge}
\end{figure}

The segments are depicted in Fig. \ref{fig:linesmall:line} and Fig.
\ref{fig:linelarge:line}. For the railway line No. 216, we considered its
official train schedule (as of April 2020). The line No. 191 was undergoing a
renovation at the time we were conducting our original experiments \cite{railwaydispatching}, and hence it had no available timetable.
Based on the planned parameters of the line, as described in the official
documents \cite{PKPPLK}, we constructed a cyclic timetable. Initial,
undisturbed timetables are depicted in Fig. \ref{fig:linesmall:diagram} and
Fig. \ref{fig:linelarge:diagram}.

Timetable for the network segment of line No. 216 includes two
\emph{Inter-City} trains, IC5320 and IC3521, and a regional \emph{Regio} train
R90602. For the line No. 191, the timetable includes two \emph{Inter-City}
trains IC1, IC2 and four regional trains Ks1--Ks4. We assume both
\emph{Inter-City} trains in line No. 191 are operated with the same train set,
with a minimum turnover time (see Appendix \ref{chapter:dispatching}) of
$R(j,j') = 20$ minutes.

For both network segments, we assume that the minimum waiting times at all
considered stations are 1 minute. Also, we assume that the passing times
through all the line blocks were initially scheduled according to the maximum
permissible speeds. As a result of those assumptions, the only possible nonzero
time reserve occurs at the station blocks.

\subsection{Disturbance scenarios}

\begin{figure}
  \begin{subfigure}[b]{0.5\columnwidth}
    \caption{}\label{conflict}
    \includegraphics[width=\textwidth]{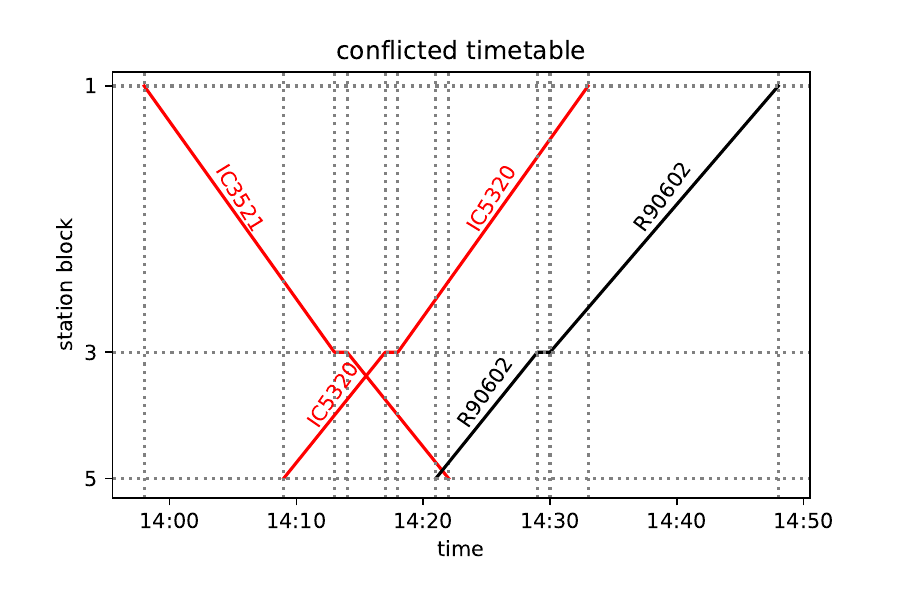}
  \end{subfigure}
  \begin{subfigure}[b]{0.5\columnwidth}
    \caption{}\label{resolution}
    \includegraphics[width=\textwidth]{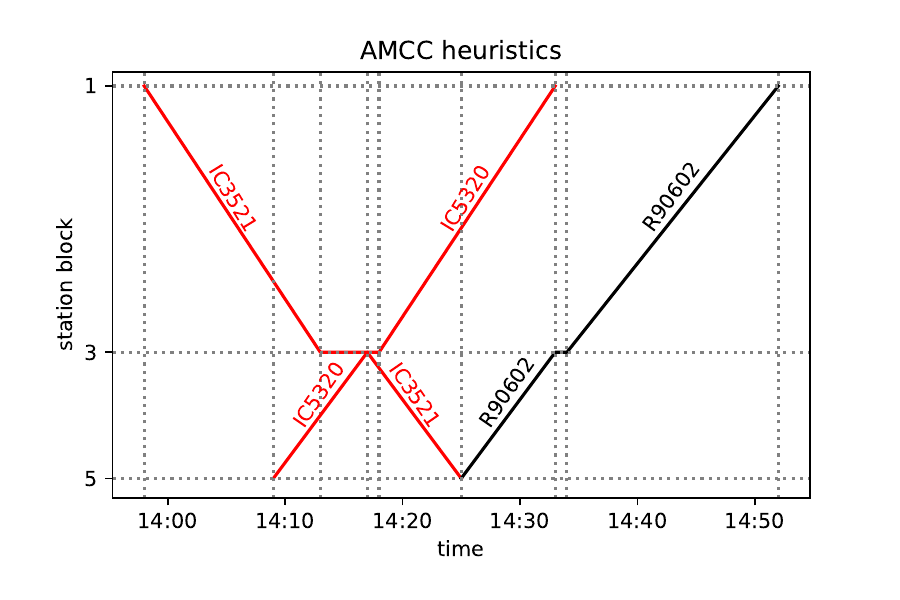}
  \end{subfigure}
  \caption{
    \subref{conflict} Conflicted timetable for railway segment of line No. 216. Compared to the original timetable
    (Fig. \ref{fig:linesmall:diagram}), two trains are delayed, resulting in two conflicts. The conflicts
    can be quickly identified visually as intersections of train paths at line blocks.
    \subref{resolution} Conflict resolution via AMCC heuristics. The same solution was obtained using FCFS and FLFS heuristics.
  }
  \label{fig:smallconflict}
\end{figure}

For the Nidzica--Olsztynek railway segment, we considered a single scenario
with two delays. The purpose of this scenario is to illustrate our approach on
a simple and yet real-world example. The first one is a 15-minute delay of
IC5320 starting from station block 5. The second one is that of the IC3521
leaving first station block 5 minutes late. Considering this and our assumptions,
this creates conflicts where two \emph{Inter--City} trains, as well as an
\emph{Inter--City} train and the \emph{Regio} train, have conflicts at block 4.
The conflicted, infeasible train diagram for this situation is depicted in Fig.
\ref{fig:smallconflict}.

For the Goleszów -- Wisła Uzdrowisko line, we considered several different
scenarios, which were designed to illustrate our approach on a larger example:

\begin{enumerate}
  \item A moderate delay of the \emph{Inter--City} train starting from the station
    block 1. This results in a single conflict between IC1 and Ks2.
  \item A moderate delay of all the trains starting from station block 1, resulting in
    two conflicts.
  \item A significant delay of some trains starting from station block 1. Results in
    two conflicts.
  \item A significant delay of the \emph{Inter--City} train IC1 starting from the
    station block 1. Results in a single conflict, which is straightforward to
    resolve.
\end{enumerate}

\begin{figure}
  \begin{subfigure}[b]{0.5\textwidth}
    \caption{}\label{c1}
    \includegraphics[width=\textwidth]{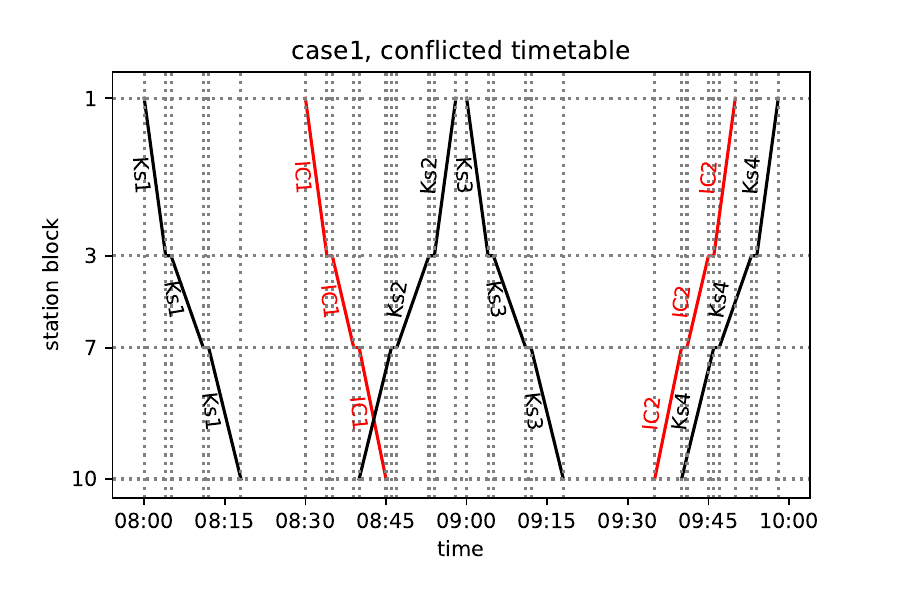}
  \end{subfigure}
  \begin{subfigure}[b]{0.5\textwidth}
    \caption{}\label{c2}
    \includegraphics[width=\textwidth]{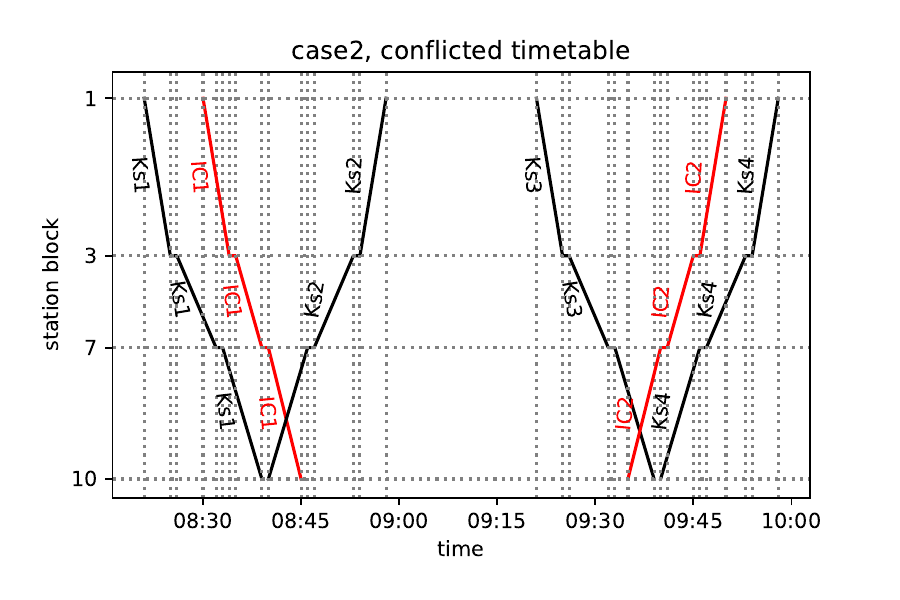}
  \end{subfigure}

  \begin{subfigure}[b]{0.5\textwidth}
    \caption{} \label{c3}
    \includegraphics[width=\textwidth]{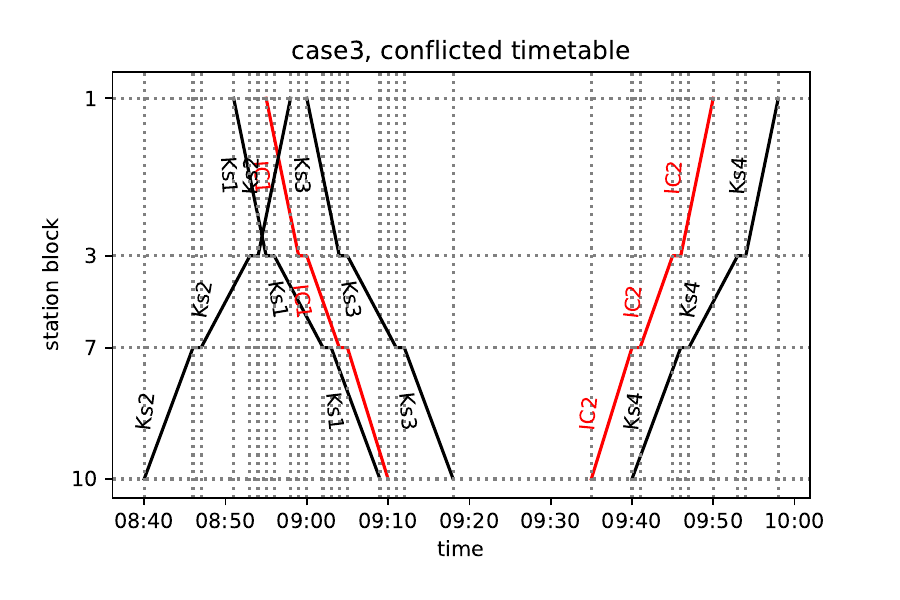}
  \end{subfigure}
  \begin{subfigure}[b]{0.5\textwidth}
    \caption{}\label{c4}
    \includegraphics[width=\textwidth]{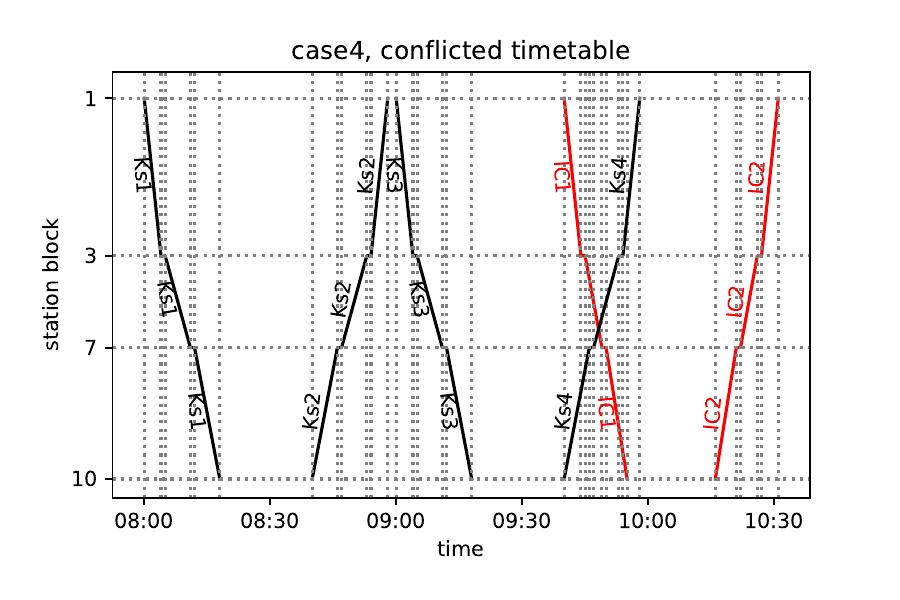}
  \end{subfigure}
  \caption{Conflicted timetables for line No. 216. \subref{c1} Single conflict, observe
    that the additional delay of Ks$2$ will propagate to the delay of Ks$3$.
    \subref{c2} Two conflicts, with no impact of Ks$2$ on Ks$3$. \subref{c3}
    Multiple conflicts. \subref{c4} One conflict, straightforward to resolve.}
  \label{fig:conflictlarge}
\end{figure}

The delays in all the aforementioned scenarios were chosen so that they indeed
result in conflicts. The conflicted timetables are presented in Fig.
\ref{fig:conflictlarge}.

\subsection{Solution using simple heuristics}
To establish a baseline for Quantum Annealing, we solved the problems described
in the previous section using simple heuristics common in the railways
practice. Those heuristics are:

\begin{itemize}
  \item FCFS (First Come First Served),
  \item FLFS (First Leave First Served),
  \item AMCC (Avoid Maximum Current $C_{\max}$).
\end{itemize}

In FCFS (resp. FLFS) the way is given to the train that first arrives (resp.
first leaves) the considered station block at which the conflict occurs. AMCC
\cite{mascis2002job} is slightly more complex. In this heuristic, one tries to
minimize the maximum secondary delays of the trains. We want to stress that
those heuristics facilitate different objective functions, and hence it is not
possible to directly compare them -- nevertheless, it might be useful to
discuss qualitative differences between the solutions they produce. The
solutions provided by the AMCC heuristic also provide a lower bound for the
values of maximum secondary delays $d_{\max}(j)$, which we will use when
constructing QUBO.

For the case of Nidzica--Olsztynek line, all heuristics returned the same
solution, depicted in Fig. \ref{resolution}. The conflict is avoided by
delaying IC3521 by another 3-minutes, and allowing R9062 to enter the block not
earlier than 14:25, i.e. 4 minutes later than in the conflicted timetable. In
this case, the additional 4 minutes constitute the maximum secondary delay of
the solution.

We also applied the aforementioned simple heuristics to all the considered
disturbances in the Goleszów -- Wisła Uzdrowisko segment. For brevity, we
refrain from presenting a detailed discussion of the solutions for all the
cases and limit ourselves to the summary of the maximum secondary delay, which
is presented in Table \ref{tab:simple}:

\begin{table}[bh]
  \centering
  \begin{tabular}{|c|c|c|c|c|}
    \hline
    \rowcolor{theader} Heuristics & case $1$ & case $2$ & case $3$ & case $4$ \\
    \hline
    FLFS                          & 6        & 13       & 4        & 2        \\
    \hline
    FCFS                          & 5        & 5        & 5        & 2        \\
    \hline
    AMCC                          & 5        & 5        & 4        & 2        \\
    \hline
  \end{tabular}
  \caption{The maximum secondary delays, in minutes, resulting from simple heuristics.
    Observe that for each case, there are solutions far below $d_{\text{max}} =
      10$.} \label{tab:simple}
\end{table}

\subsection{Details on QUBO construction}

To formulate our dispatching problems as QUBO and solve them on the D-Wave
annealer (or using any other method), we first need to decide on the values of
several parameters of the model, as well as the precise form of the cost
function. We start with the latter.

We decided on using the cost function proportional to the secondary delays of
all trains entering their last station block. Additionally, we weight the
contributions of each delay with a coefficient depending on the prioritization
of the corresponding train, resulting in the cost function of the form:
\begin{equation}
  f(\mathbf{x}) = \sum_{j \in J}\left(\sum_{m  \in A_{j,s_{\eend-1}}}w_{j} \frac{d(j,s^{*}) - d_{U}(j,s^{*})}{d_{\max}(j)}x_{j,s^{*},m}\right),
\end{equation}
where $s^{*} = s_{\eend-1}$. The priorities $w_{j}$ are chosen specifically
    for both networks. One can immediately observe that larger values of $w_{j}$
    increase contribution stemming from the delay of a given train, and hence the
    objective function favors solutions with smaller delays for the trains with
    larger priorities. For the segment of line No. 216, we assume $w_{j}= 1.5$ for
all \emph{Inter-City} trains, and $w_{j}=1.0$ for the regional train. This
    prioritization coincides with the usual prioritization of trains in Poland (and
    many countries). For the segment of line No. 191, we decided to adopt a
    slightly more complicated prioritization. For the trains heading toward block
    10, we set a lower priority of $w_{j}=0.9$. For the trains heading in the
    opposite direction, we set $w_{j}=1.5$ and $w_{j}=1.0$ for \emph{Inter-City}
and regional trains respectively. This is because the trains heading towards
block 1 (Goleszów) also head towards important junctions in the Polish railway
network (Katowice for regional trains, and the capital city of Warsaw for
\emph{Inter-city} trains). Our strategy therefore tries to avoid larger delays
in this direction to limit further disturbance to the rest of the network.

As for the maximum secondary delay $d_{\max}$, for simplicity, we assume it is
    the same for all trains. On the one hand, its value cannot be smaller than the
    one returned by the AMCC heuristics. On the other hand, setting this value too
    high increases the number of decision variables and complicates the objective
    function, which is especially undesirable because of the limited number of
    qubits on D-Wave annealers. For line No. 216, we set $d_{\max}=7$ and for line
    No. 191 we set $d_{\max}=10$. The total number of decision variables is given
    by
    \begin{equation}
      \label{eq:numvariables}
      \mbox{\#variables} = (\mbox{\#station blocks}-1) \cdot (\mbox{\#number of trains}) \cdot (d_{\max}+1)
    \end{equation}
    Using formula \eqref{eq:numvariables} we get $2\cdot 3 \cdot 8 = 48$ decision variables for line No. 216
    and $3 \cdot 6 \cdot 11 = 198$ variables for line No. 191. Importantly, the
moderately low number of variables for line no. 216 allows us to solve it using
the brute-force algorithm presented in Chapter \ref{chapter:bruteforce}.

\begin{figure}
  \includegraphics[width=\textwidth]{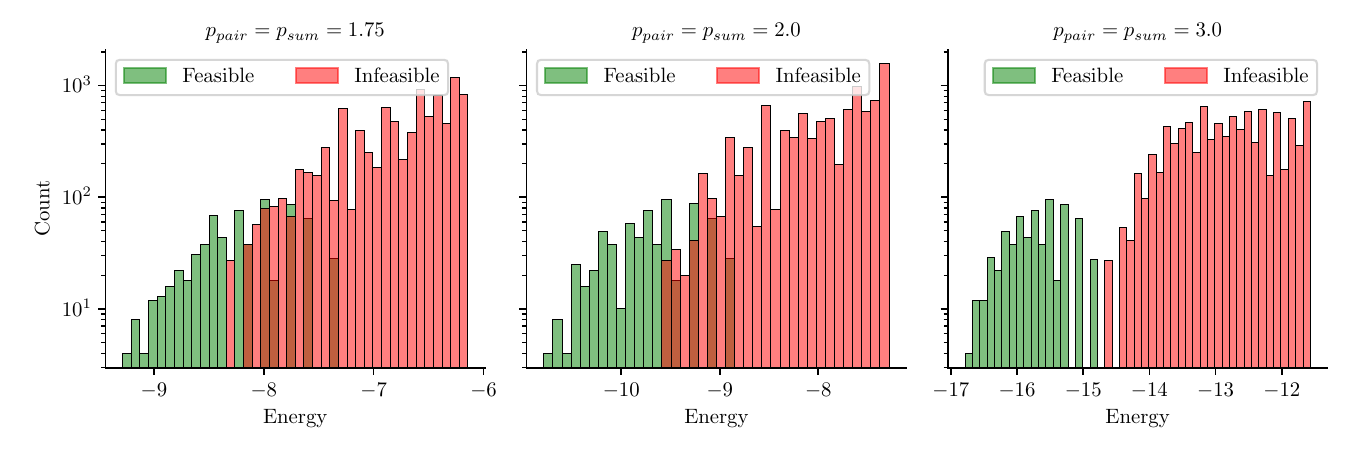}
  \caption{Energy histogram for feasible (green) and infeasible (red) solutions of QUBO
    defined for line No. 216 with varying penalty weights. The figure takes into
    account the first 5000 low-energy states.} \label{fig:penaltyhistogram}
\end{figure}

Lastly, we need to choose values for $\ppair$ and $\psum$ penalty weights.
    This is a very subtle choice. On the one hand, setting it too low may cause
    some of the infeasible solutions to have the value of the objective function
    smaller than that of feasible solutions, which is undesirable. On the other
    hand, if penalty weights are too high, the actual cost function becomes merely
    a perturbation for the penalty terms, which is also undesirable. To illustrate
    the difference those weights make to the energy landscape, we computed the
    low-energy spectrum for the problem defined on line No. 216 for several
    different values of $\ppair$ and $\psum$. The obtained energy histograms
    are presented in Fig. \ref{fig:penaltyhistogram}.

In our experiments, we used several combinations of $p_{pair}$ and $p_{sum}$.
For D-Wave 2000Q series devices, which we used for the experiments reported
in~\cite{railwaydispatching}, we used $p_{\text{pair}}=2.7$,
  $p_{\text{sum}}=2.2$ and $p_{\text{pair}} = p_{\text{sum}} = 1.75$.
    Additionally, in this thesis, we extend these results by running
    experiments with $p_{\text{pair}}=p_{\text{sum}}=n$ for $n=2, 3, 4$ on
Advantage and Advantage2 prototype devices.

\subsubsection{Initial experiments on D-Wave annealers}
In our initial experiments, reported in \cite{railwaydispatching}, we used
mostly the D-Wave 2000Q device. We were able to successfully embed all the
problem instances, except case 3 for Line 191. As for the QUBO parameters, we
used $p_{\text{pair}}=2.2$, $p_{\text{sum}}=2.7$ and
  $p_{\text{pair}}=p_{\text{sum}}=1.75$. We used several values of the chain
    strengths, all of them being a multiplicity of $\max|J_{ij}|$ (computed
separately for each problem before the embedding). Following convention from
\cite{railwaydispatching}, we call the multiplier \emph{chain strength scale}
($css$), and in our experiments, it ranged from $1$ to $9$. The annealing time
varied between $5$--$2000\mu$s.

\begin{figure}
  \begin{subfigure}[b]{0.5\textwidth}
    \caption{}\label{fig:dwtd1}
    \includegraphics[width=\textwidth]{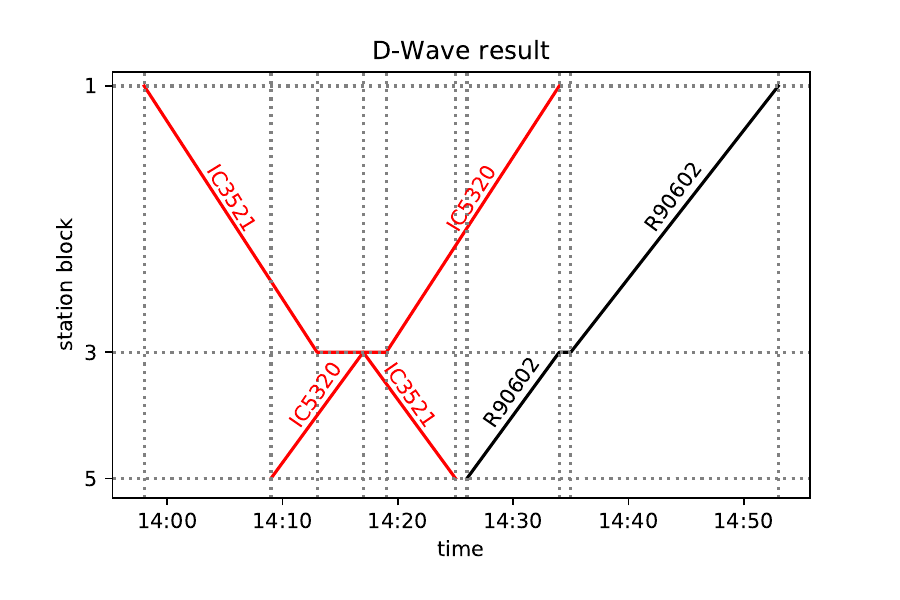}
  \end{subfigure}
  \begin{subfigure}[b]{0.5\textwidth}
    \caption{}\label{fig:dwtd2}
    \includegraphics[width=\textwidth]{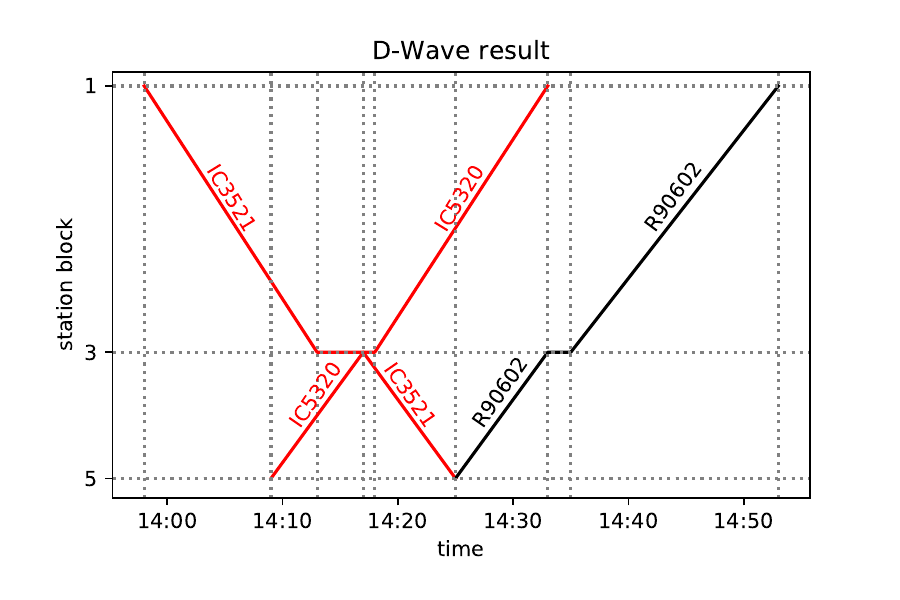}
  \end{subfigure}
  \begin{subfigure}[b]{0.5\textwidth}
    \caption{}\label{fig:dwen1}
    \includegraphics[width=\textwidth]{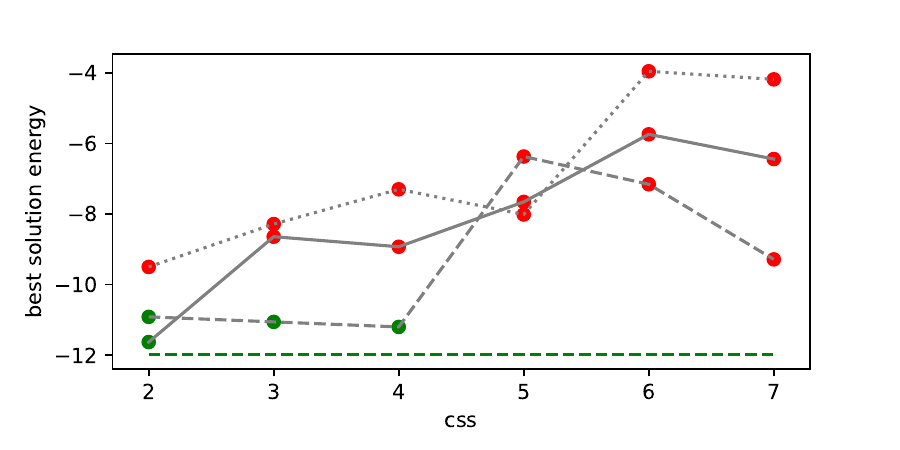}
  \end{subfigure}
  \begin{subfigure}[b]{0.5\textwidth}
    \caption{}\label{fig:dwen2}
    \includegraphics[width=\textwidth]{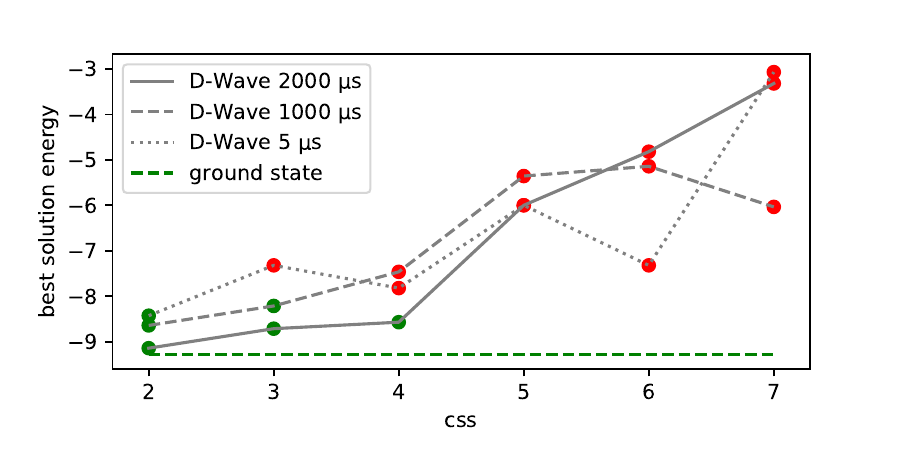}
  \end{subfigure}
  \caption{
    \textbf{a.} -- \textbf{b.} Best solutions obtained with D-Wave 2000Q annealer,
    optimized over all annealing times and chain strength scales.
    \textbf{c.} -- \textbf{d.} Energy of the best D-Wave solution as the function of
    $css$ scale. For panels \textbf{a.} and \textbf{c.} we used $\ppair=2.2$, $\psum=2.7$
    and for panels \textbf{b}, \textbf{d.} we used $\ppair=1.75$, $\psum=1.75$.
  }
  \label{fig:dwtrainsold}
\end{figure}

For QUBO defined for Line 216, the D-Wave annealers failed to reach the ground
state for all the tested parameters. However, the lowest energy--solution found
by the annealer was equivalent to the ground state from the dispatching
perspective\footnote{Here, equivalent from the dispatching perspective means
  that the order of trains leaving any given station is the same.).}. The Fig.
\ref{fig:dwtrainsold} shows the solutions obtained from the D-Wave annealer, as
well as the deviation from the ground state energy. Since the best solution was
obtained for $css=2$, we decided to use the same value for the consecutive
experiments.

\begin{figure}
  \begin{subfigure}[b]{0.5\textwidth}
    \caption{}\label{fig:dwtd1large}
    \includegraphics[width=\textwidth]{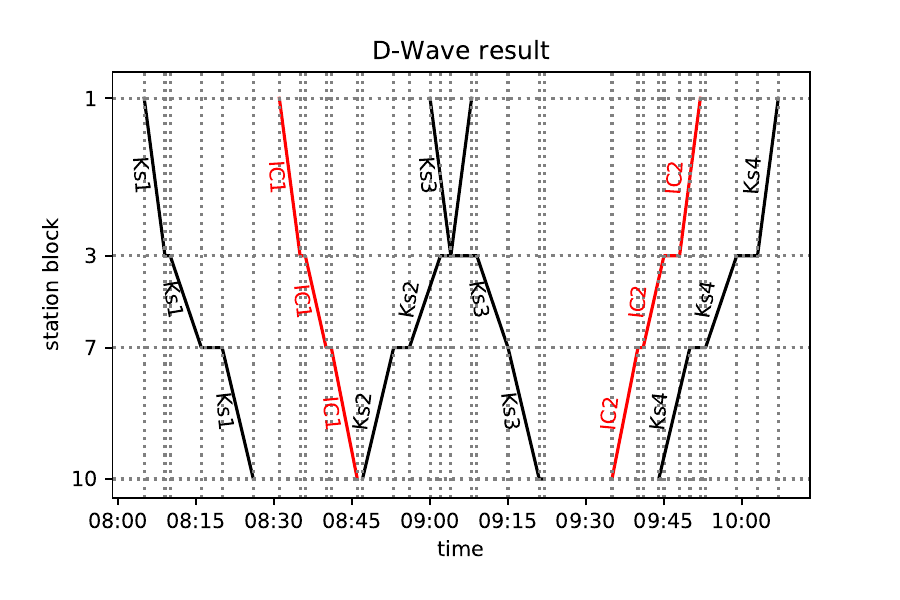}
  \end{subfigure}
  \begin{subfigure}[b]{0.5\textwidth}
    \caption{}\label{fig:dwtd2large}
    \includegraphics[width=\textwidth]{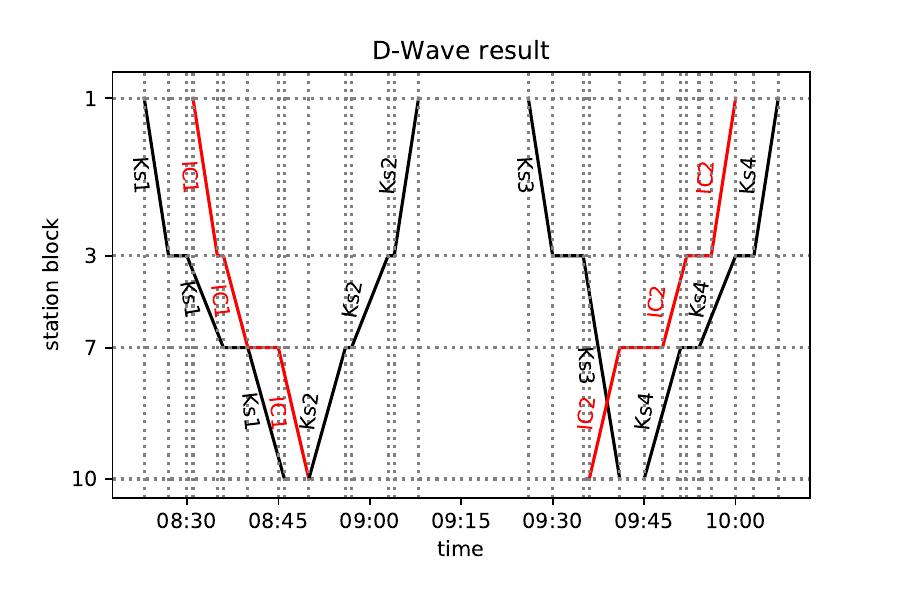}
  \end{subfigure}
  \caption{
    Lowest energy solutions obtained for QUBO problems defined for Line 191. In all
    panels $\ppair = 2.2, \psum = 2.7$ and $css=2.0$. \textbf{a.} Solution obtained
    for case 1 (with annealing time $\tau=1400$). The solution is infeasible
    because the train Ks3 stays at the station block 7 shorter than 1 minute. The
    solution can be turned into a feasible one by prolonging the stay of Ks3 at
    station block 7. \textbf{b.} The best solution obtained for case 2
    ($\tau=1200$). The solution is infeasible as Ks3 does not stop at station 7.
    This solution can be made into an optimal one by shortening the stays of Ks3 at
    station 3 and IC2 at station 7. } \label{fig:dwtrainsoldlarge}
\end{figure}

Finding a feasible solution for QUBOs defined for Line 191 proved to be much
more difficult for the D-Wave 2000Q annealer. Hence, we increased the total
number of obtained samples to $250$k. Still, even with the increased number of
samples we were unable to reach a feasible solution. The best solutions found
by the annealer for case 1 and case 2 violate a single constraint and can be
easily corrected to obtain a feasible (and in case 2, even optimal) solution,
see Fig. \ref{fig:dwtrainsoldlarge}.

In comparison, the QUBOs for cases 1--4 turn out to not be that challenging for
the classical solvers. Both the tensor networks algorithm, described in Chapter
\ref{chapter:tn}, and the IBM CPLEX solver were able to find high-quality
solutions equivalent to the ground state from the dispatching point of view,
with CPLEX slightly outperforming the tensor network algorithm in cases 3 and
4. The values of the cost function obtained from these solvers are presented in
table \ref{tab:line191classical}. In the same table, we also present, for
reference, values of our objective function for solutions obtained with simple
heuristics.

At the time we were conducting experiments presented in
\cite{railwaydispatching}, the new Advantage System 1.1 device was entering the
market, and we were able to run a very limited set of experiments. We decided
to try a slightly larger problem, constructed by extending the timetable for
Line 191 with more trains. Although we were able to embed it on the device, our
attempts to find a feasible solution on this early Pegasus-based system were
futile. For the details of this part of the experiment, we refer the interested
reader to \cite{railwaydispatching}.

\begin{table}
  \centering
  \begin{tabular}{|c|c|c|c|c|c|}
    \hline
    \rowcolor{theader} \multicolumn{2}{|c|}{Method} & Case 1          & Case 2                      & Case 3                      & Case 4                                                    \\
    \hline
    \multirow{2}{*}{QUBO model}                     & CPLEX           & \textcolor{RoyalBlue}{0.54} & \textcolor{RoyalBlue}{1.40} & \textcolor{RoyalBlue}{0.73} & \textcolor{RoyalBlue}{0.20} \\
    \cline{2-6}
                                                    & Tensor Networks & \textcolor{RoyalBlue}{0.54} & \textcolor{RoyalBlue}{1.40} & \textcolor{RoyalBlue}{1.65} & \textcolor{RoyalBlue}{0.29} \\
    \hline
    \multirow{3}{*}{Simple heuristics}              & AMCC            & 0.77                        & \textcolor{RoyalBlue}{1.30} & \textcolor{RoyalBlue}{0.73} & \textcolor{RoyalBlue}{0.20} \\
    \cline{2-6}
                                                    & FLFS            & \textcolor{RoyalBlue}{0.54} & 1.71                        & \textcolor{RoyalBlue}{0.73} & \textcolor{RoyalBlue}{0.20} \\
    \cline{2-6}
                                                    & FCFS            & 0.77                        & \textcolor{RoyalBlue}{1.30} & 0.95                        & \textcolor{RoyalBlue}{0.20} \\
    \hline
  \end{tabular}
  \caption{
    Values of the cost functions obtained by the classical solvers for the QUBO
    problems defined for line 191. Values marked with \textcolor{RoyalBlue}{blue}
    represent solutions equivalent (from the dispatching perspective) to the ground
    state of the corresponding problem. Values for the solutions obtained with
    simple heuristics are provided for reference, but it should be noted that those
    methods use different objective functions and hence cannot be directly compared
    to CPLEX or tensor networks-based solver. } \label{tab:line191classical}
\end{table}

\subsubsection{Extended experiment on the Advantage System annealers}

Since the time of our experiments described in the previous section, new models
of the annealers from the Advantage System generation became available.
Furthermore, the first Advantage2 Prototype devices entered the market. We
decided to extend our experiment and run further tests to investigate the
performance of these devices for a broader range of parameters. To this end, we
decided to test how the newer Pegasus-based devices perform on the QUBO problem
defined on Line 216. We decided that due to the limitation of our resources, we
could not run comprehensive experiments with the problem cases defined on Line
191, and instead opted for a more comprehensive sweep through the parameter
space for the smaller problem.

In this new scenario, we decided to increase $\ppair$ and $\psum$ values, to
investigate if a wider energy separation between feasible and infeasible
solutions will be beneficial for the annealers' performance. As previously, the
annealing times varied from $\tau=5$ to $\tau=2000$. We used chain strengths
varying between $4$ and $12$. We would like to stress, that here we mean
absolute values of the chain strengths and not scales of chain strengths in
relation to the maximum absolute value of quadratic terms of the problem like
in the initial experiment.

\begin{figure}[t]
  \begin{subfigure}[b]{0.5\textwidth}
    \caption{}
    \includegraphics[width=\textwidth]{dwave_line216_ground1}
  \end{subfigure}
  \begin{subfigure}[b]{0.5\textwidth}
    \caption{}
    \includegraphics[width=\textwidth]{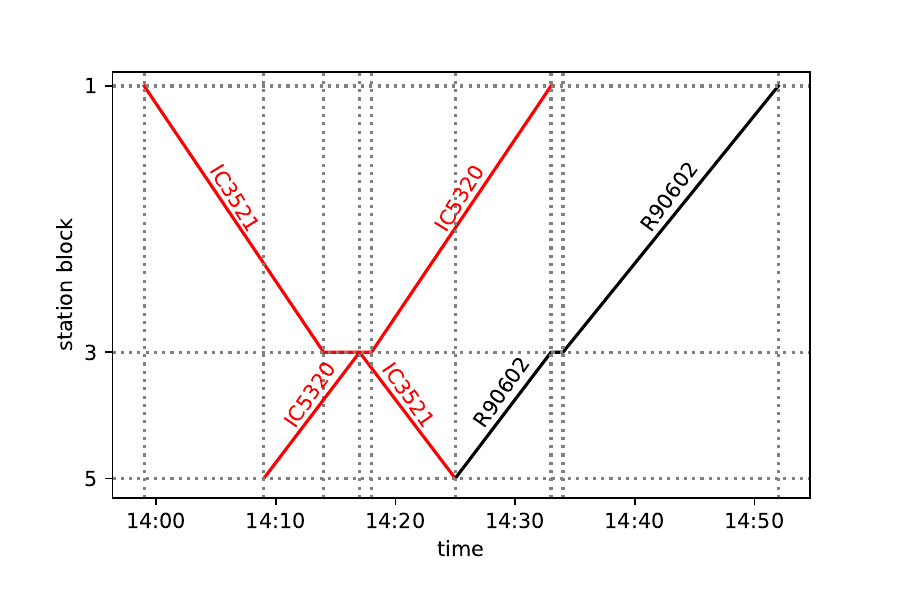}
  \end{subfigure}
  \caption{Example ground state solutions for the conflicted timetable of Line 216. All
    other ground states are equivalent from the dispatching point of view.}
  \label{fig:dwline216grounds}
\end{figure}

\begin{figure}
  \includegraphics[width=\textwidth]{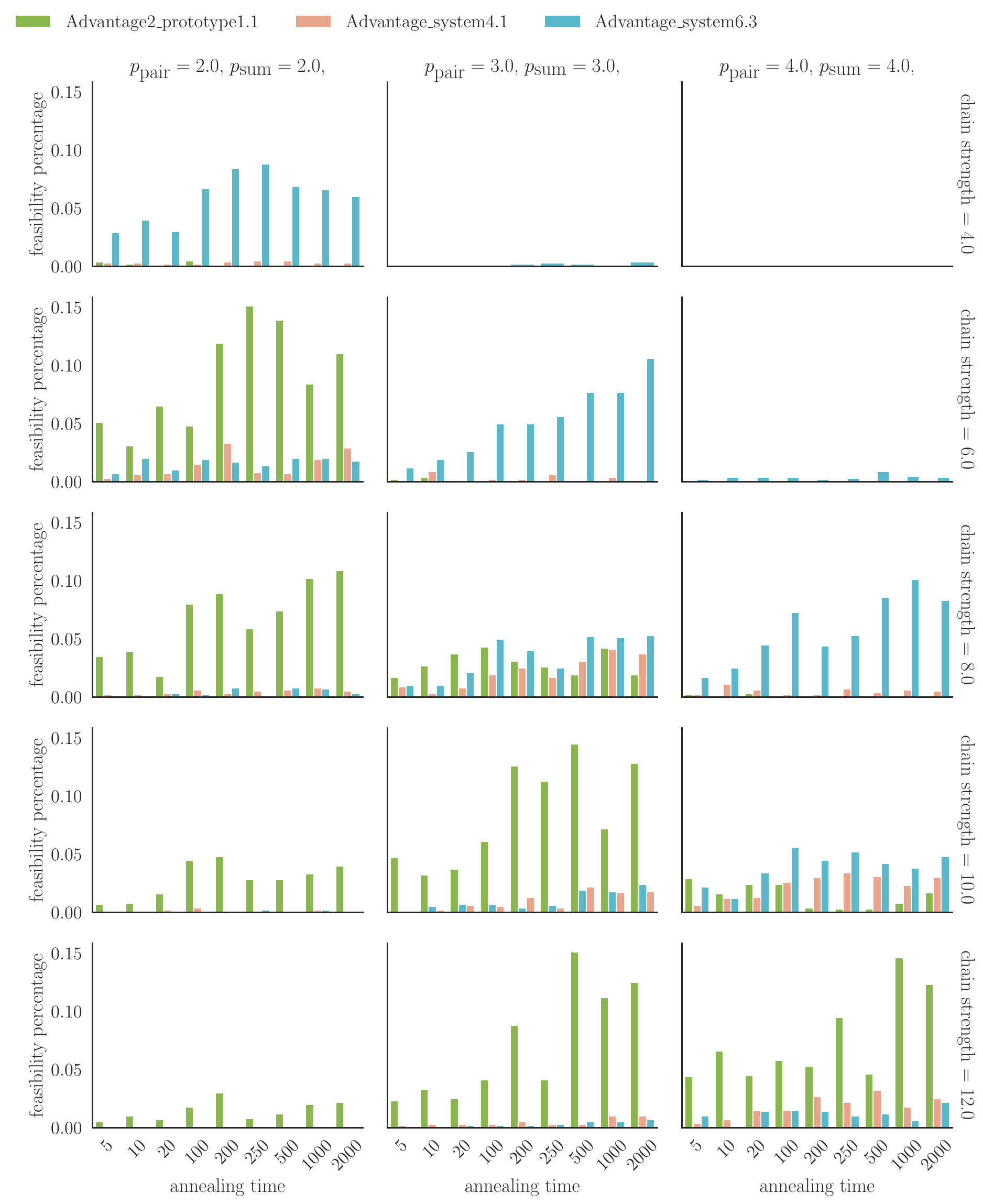}
  \caption{
    Frequency of finding a feasible solution for the problem defined for Line 216.
    Rows in the grid correspond to different values of chain strength and the
    columns correspond to different values of penalty scalings. In each cell, the
    X-axis depicts the annealing time $\tau$, while the $Y$-axis depicts the
    obtained fraction of the feasible solution (out of 1000 samples) }
  \label{fig:dwline216freq}
\end{figure}

All of the annealers were able to find a feasible solution to the problem for
at least some combination of parameters. However, their performance varied
highly depending on the parameter range. The frequency of finding a feasible
solution by the annealers is depicted in Fig. \ref{fig:dwline216freq}. As seen
there, the Advantage System6.3 and Advantage2 Prototype1.1 devices exhibited
much better performance than the older Advantage System4.1 device. As for the
quality of the solutions, all solvers managed to find an optimal solution,
although with different success rates. The summary of parameters for which a
ground state was obtained is presented in table \ref{tab:line216ground}.
Example ground states found are depicted in Fig. \ref{fig:dwline216grounds}.

\begin{table}
  \small
  \centering
  \begin{tabular}{|c|c|c|c|}
    \hline
    \rowcolor{theader} Solver & chain strength & annealing time & \# occurrences \\
    \hline
    Advantage System4.1       & 10             & 200            & 1              \\
    \hline
    Advantage System4.1       & 12             & 500            & 1              \\
    \hline
    \hline
    Advantage System6.3       & 10             & 500            & 1              \\
    \hline
    Advantage System6.3       & 12             & 100            & 1              \\
    \hline
    \hline
    Advantage2 Prototype1.1   & 12             & 5              & 1              \\
    \hline
    Advantage2 Prototype1.1   & 12             & 100            & 3              \\
    \hline
    Advantage2 Prototype1.1   & 12             & 1000           & 4              \\
    \hline
    Advantage2 Prototype1.1   & 12             & 2000           & 1              \\
    \hline
  \end{tabular}
  \caption{Parameters for which the D-Wave annealers managed to find the optimal solution to
    the problem defined on Line 216. All samples with ground states occurred at
    $p_{\text{pair}}=p_{\text{sum}}=4.0$. } \label{tab:line216ground}
\end{table}

One of the interesting observations one could make about the results presented
in Fig. \ref{fig:dwline216freq} is the performance difference between different
models of the annealers, which seem to be highly dependent on the regime of
parameters. Determining the sources of these differences requires further
research, but it is possible that they can be partially explained by
differences in the available range of quadratic coefficients between the
devices (see D-Wave QPU datasheets \cite{dwavedocs}), which in turn might
affect the DAC quantization effect (see discussion of error sources in Section
\ref{sec:parallel-in-time}).


%% file: summary.tex
\chapter*{Summary}

In this thesis, we focused on benchmarking quantum annealers and validating their
usefulness in practical settings. One of the most anticipated uses of quantum computers
is simulating physics, or, more precisely, simulating the dynamics of quantum systems.
Therefore, it seems there is no better benchmark for a quantum computer than to test
how far it is from achieving this long--awaited goal. To this end, we described in detail a
proof-of-concept algorithm for simulating the dynamics of a quantum (or in fact, any
dynamical) system using quantum annealers. Although the applicability of the algorithm
to current devices is limited by their small number of qubits and sparse connectivity,
our experiments indicated that already the present-day D-Wave annealers can capture
the dynamics of a very simple two-level system. We also contrasted the obtained results
with the ones produced by several classical solvers, concluding that they perform better
than the tested quantum devices. We also provided a possible explanation why the
particular optimization problems solved in our experiments are particularly hard for
D-Wave devices and checked our predictions with numerical experiments.

To assist in the process of benchmarking the annealers, we developed two distinctive
algorithms. The recent, tensor network-based algorithm allows one to solve Ising
spin--glass instances defined on Chimera graph and other similar layouts. The algorithm
is useful in itself as an optimization approach, but for other research conducted for
this thesis, provided a classical baseline for the results obtained by the D-Wave
annealer. The second of our algorithms, a massively parallelizable distributed brute-force
algorithm, allows for the exact computation of the low--energy spectrum of small,
but otherwise arbitrary, spin-glass instances. Importantly, this simple yet
efficient algorithm is exact and deterministic. We used the brute-force algorithm
to obtain low-energy spectra for some of the smaller instances used throughout
our experiments. This provided us not only with a means of assessing the quality of
solutions obtained from other solvers or annealers but also with valuable
insights into the structure of the solution space.

To benchmark another anticipated use of quantum annealers, i.e. solving hard
optimization problems stemming from real-life problems, we described an approach
for solving railway-dispatching problems by converting them to QUBO. We then
conducted experiments testing our approach on two generations of D-Wave quantum
annealers. Remarkably, for our tests, we used real railway timetables from two Polish
railway segments. In our experiments, D-Wave annealers were able to successfully
find an optimal solution to the small problem instances, although the performance
varied greatly depending on the parameters such as the annealing time and chain strength.


%% file: appendix.tex
\chapter{Asymptotic notation}

In order to characterize the complexity of algorithms, it is useful to use
asymptotic big-O notation. Consider two functions $f, g \colon \NN \to \RR$. We
say that $f$ is $O(g)$ if and only if there exists a constant $C > 0$ and a
natural number $n_{C}$ such that the inequality $0 \le f(n) \le C\cdot g(n)$
holds for all $n > n_{C}$ \cite{clrs}.

It is common to write $f=O(g)$ instead
of ``$f$ is $O(g)$'', slightly abusing the mathematical notation \cite{clrs}.
One should notice that big-O notation does not provide a tight bound. For
instance, we have $n + 1 = O(n)$ (since $n + 1 \le 2 \cdot n$) but also $n+1 =
  O(n^{10})$.

In the context of computational complexity, big-O notation is most commonly
used for expressing upper bound on number of (dominating) operations performed
by an algorithm as a function of its input size $N$. Since the number of
performed operations is roughly proportional to the algorithm's execution time,
it follows that algorithms with better bound can be considered as more
performant. However, care must be taken when applying this reasoning to judge
practical performance. In particular, one should be mindful of the constant
factor $C$ in the definition above, as well as any bottlenecks stemming from
the working of the underlying hardware. As a concrete example, Strassen's
algorithm for multiplying two $N \times N$ matrices requires $O(N^{\alpha})$
multiplications, where $2 < \alpha < 3$, and yet may perform worse than naive
algorithm peforming $N^{3}$ multiplications, even for $N$ of order of several
hundreds \cite{dalberto}.

We conclude this section by mentioning that there exist several other
asymptotic notations. For instance, $\Omega$, describing the asymptotic lower bound
of a function, and $\Theta$ combining big-O and $\Theta$. For more details, we
refer the reader to \cite{clrs}.

\chapter{Conditional probability on square lattice}
\chaptermark{Conditional probability}
\label{sec:probability}
\begin{figure}
  \centering
  \includegraphics[width=0.6\textwidth]{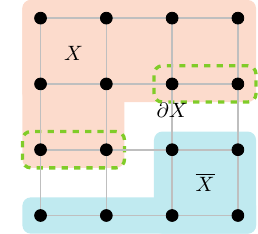}
  \caption{An example Ising spin-glass of 16 spins on a square lattice. The conditional
    probability for spins in the region $\overline{X}$ conditioned on the values of spins in the
    region $X$ depends only on the configuration on the border $\partial X$.}
  \label{fig:lattice}
\end{figure}

Consider a square lattice, such as the one depicted in Fig. \ref{fig:lattice}.
We will prove that the conditional probability for spins in the region $\overline{X}$
conditioned on the values of spins in the region $X$ depends only on the configurations of the spins on the
border $\partial X$.

Let denote by $H_X$ the usual Hamiltonian $H$ restricted to the graph
induced by vertices in $X$. Further, let $H_{X, \overline{X}} = H - H_X -
  H_{\overline{X}}$. Notice that $H_{X, \overline{X}}$ contains only quadratic
terms $J_{ij} s_i s_j$ such that $i \in X$ and $j \in \overline{X}$. Slightly
abusing the notation, one may thus write
\begin{equation}
  \small
  H(s_1, \ldots, s_N) = H_X(s_1, \ldots, s_k) + H_{\overline{X}}(s_{k+1}, \ldots, s_N) + H_{X, \overline{X}}(s_1, \ldots, s_N)
\end{equation}
Using definition of conditional probability applied to Boltzmann distribution,
one thus gets

\begin{align}
   & p(s_{k+1}|s_1, \ldots, s_k) = \frac{\sum\limits_{(z_{k+2}, \ldots, z_N)}e^{-\beta H(s_1, \ldots, s_{k+1}, z_{k+2},\ldots,z_N)}}{\sum\limits_{(z_{k+1}, \ldots, z_N)}e^{-\beta H(s_1, \ldots, s_k, z_{k+1},\ldots,z_N)}}                                                                                                                         \\
   & = \frac{\sum\limits_{(z_{k+2}, \ldots, z_N)}e^{-\beta (H_X(s_1, \ldots, s_k) + H_{\overline{X}}(s_{k+1}, z_{k+2},\ldots,z_N) + H_{X, \overline{X}}(s_1, \ldots, z_N))}}{\sum\limits_{(z_{k+1}, \ldots, z_N)}e^{-\beta (H_X(s_1, \ldots, s_k) + H_{\overline{X}}(z_{k+1}, \ldots,z_N) + H_{X, \overline{X}}(s_1, \ldots, z_N))}}                 \\
   & = \frac{e^{-\beta H_X(s_1, \ldots, s_k)}\sum\limits_{(z_{k+2}, \ldots, z_N)} e^{-\beta(H_{\overline{X}}(s_{k+1}, z_{k+2},\ldots,z_N) + H_{X, \overline{X}}(s_1, \ldots, z_N))}}{e^{-\beta H_X(s_1, \ldots, s_k)}\sum\limits_{(z_{k+1}, \ldots, z_N)}e^{ -\beta(H_{\overline{X}}(z_{k+1}, \ldots,z_N) + H_{X, \overline{X}}(s_1, \ldots, z_N))}} \\
   & = \frac{\sum\limits_{(z_{k+2}, \ldots, z_N)} e^{-\beta(H_{\overline{X}}(s_{k+1}, z_{k+2},\ldots,z_N) + H_{X, \overline{X}}(s_1, \ldots, z_N))}}{\sum\limits_{(z_{k+1}, \ldots, z_N)}e^{ -\beta(H_{\overline{X}}(z_{k+1}, \ldots,z_N) + H_{X, \overline{X}}(s_1, \ldots, z_N))}}
\end{align}
Note, in both numerator and denominator, spins with indices from $X$ appear
non-trivially only in $H_{X, \overline{X}}$ , i.e. the whole expression depends
only on those spins in $X$ that directly interact with spins in $\overline{X}$,
which was to be shown.

\chapter{Dispatching conditions}
\label{chapter:dispatching}
In the following appendix, we use the notation from Chapter \ref{chapter:trains}.
\section{The minimum passing time condition.}
Any train $j$ cannot travel through a block $b \in \Bj$ faster than the corresponding minimum
passing time:
\begin{equation}
  \label{eq:dc1}
  \tout(j, b) \ge \tin(j, b) + \pmin(j, b).
\end{equation}
Using \eqref{eq:djs} and \eqref{eq:pt} one can easily verify that inequality
\eqref{eq:dc1} is equivalent to the following inequality for station blocks:
\begin{equation}
  \label{eq:passingtime}
  d(j, s_{j,k+1}) \ge d(j, s_{j,k}) - \sum_{b}\alpha(j, b),
\end{equation}
where the sum runs over all blocks starting form the one succeeding $s_{j,k}$
and ending in $s_{j,k+1}$.
In binary variables, it means that if, for a fixed $j,s,m$, the $x_{j,s,m}=1$,
then delays $d(j,s)$ smaller than $m-\sum_{b}\alpha(j, b)$ are prohibited
and thus the corresponding variables have to zero out. Hence, we arrive at the
following condition:
\begin{equation}
  \label{eq:qubo:passingtime}
  \forall_{j} \forall_{s \in S_j \setminus \{s_{{j,\eend}}\}}
  \sum_{m \in A_{j,s}}
  \left(
  \sum_{ m' \in D(m) \cap A_{j, s_{j,k+1}}} x_{j, s, m}
  x_{j, s_{j,k+1}, m'} \right) = 0,
\end{equation}
where $D(m) = \{0, 1, \ldots, m - \sum_{b}\alpha(j, b) -1\}$.
\section{The single block occupation condition.}
Two trains cannot occupy the same line block. Consider two
trains, $j, j' \in \JJ_0$ leaving the same station $s_{j,k} \in \Sj$ in the direction of the
next station block $s_{j,k+1}$. Suppose further that the train $j$ leaves
first. i.e. $\tout(j', s) > \tout(j, s)$. Since two trains cannot occupy the
same block, some headway time has to pass after $\tout(j, s)$ before the
train $j'$ can leave. This headway is dependent on both $j$ and a
sequence of blocks, and hence we denote it by $\tauu(j, s_{j,k})$. Thus,
the condition becomes:
\begin{equation}
  \label{eq:single-block}
  \tout(j', s_{j,k}) \ge \tout(j, s_{j,k}) + \tauu(j, s_{j,k}).
\end{equation}
Substituting for $\tout$ in \eqref{eq:single-block} yields the following
inequality for delays:
\begin{equation}
\begin{split}
  \label{eq:single-block-delays}
  d(j', s_{j,k}) &\ge d(j, s_{j,k}) + \ttout(j, s_{j,k}) - \ttout(j', s_{j,k}) + \\
  &+\tauu(j, s_{j,k})
\end{split}
\end{equation}
or,
\begin{equation}
  d(j', s_{j,k}) \ge d(j, s_{j,k}) + \Delta(j, j',s_{j,k}) + \tauu(j, s_{j,k})
\end{equation}
where
\begin{equation}
  \label{eq:delta}
  \Delta(j, j', s_{j,k}) = \ttout(j, s) - \ttout(j', s)
\end{equation}
The precise form of the headway $\tauu$ depends on the dispatching detail of the problem.
In our approach, we propose the following form:
\begin{equation}
  \tauu(j, s_{j,k}) = \max_{b}\{\pt(j,b)\}
\end{equation}
where the maximum is taken over all blocks between stations $s_{j,k}$ and $s_{j,k+1}$.
For our decision variables, we use a similar scheme as with the previous
constraint and the condition becomes:
\begin{equation}
  \label{eq:qubo:singleblock}
  \forall_{i=0,1} \forall_{j, j' \in \JJ^{i}} \forall_{s \in S^{*}_{j} \cap S^{*}_{j'}} \sum_{m \in A_{j, s}} \left(
  \sum_{m' \in B(m) \cap A_{j', s}} x_{j,s,m}x_{j',s,m'}
  \right) = 0,
\end{equation}
where, as previously, $\Sjs = \Sj \setminus \{s_{j,\eend}\}$, and $B(m) = \{m + \Delta(j, j', s), m + \Delta(j, j', s)+ 1,\ldots, m +
  \Delta(j, j', s) + \tau_{(1)}(j,s)-1 \}$
is a set of delays violating condition \eqref{eq:single-block-delays}.

\section{The deadlock condition}
The deadlock condition is analogous to the single block occupation condition
but for trains going in opposite directions. Suppose trains $j$ and $j'$ are
heading in opposite directions on a route determined by two consecutive
stations $s_{j,k}$ and $s_{j,k+1}$. Note that for $j'$ the order is reversed, i.e. it
starts at $s_{j,k+1}$ and travels in the direction of $s_{j,k}$. In this case, if $j$
is supposed to to leave $s_{j,k}$ before $j'$ leaves $s_{j,k+1}$, the following
condition has to be satisfied:
\begin{equation}
  \label{eq:deadlock}
  \tout(j', s_{j,k+1}) \ge \tout(j,s_{j,k}) + \tauuu(j, s),
\end{equation}
where $\tauuu(j, s_{j,k})$ is the minimum time required for train $j$ to
get from station block $s_{j,k}$ to $s_{j,k+1}$. Rewritten in terms of delays, the
inequality \eqref{eq:deadlock} reads:
\begin{equation}
  \label{eq:deadlock2}
  d(j',s_{j,k+1}) \ge d(j, s_{j,k}) + \Delta(j,j',s) + \tauuu(j, s).
\end{equation}
In decision variables, the deadlock condition in its basic form looks as
follows:
\begin{equation}
  \label{eq:qubo:deadlock}
  \forall_{s \in S^{*}_{j} \cap S^{*}_{j'}} \sum_{m \in A_{j, s}} \left(
  \sum_{m' \in C(m) \cap A_{j', s}} x_{j,s,m}x_{j',s,m'}
  \right) = 0,
\end{equation}
and has to be applied for a limited number of trains $j \in \JJ^{0} (\JJ^{1})$
and $j' \in \JJ^{1}(\JJ^{0})$. Here, $C(m)$ is, similarly to $B(m)$, the set of delays
violating the condition for the given pair.
\section{The rolling stock circulation condition}
Our model assumes that some trains might be assigned the same train set. Naturally,
there exists some necessary \emph{turnover time}, before a train set can be
reused. Formally, if trains $j$ and $j'$ going in opposite directions are
assigned the same train set, then the following inequality has to hold:
\begin{equation}
  \tout(j', s_{j',1}) > \tout(j, s_{j,\eend}) + \Delta(j, j')
\end{equation}
where $\Delta(j, j')$ is the minimum turnover time. In the delay
representation, the inequality becomes:
\begin{equation}
  \label{eq:rolling}
  \begin{split}
    d(j',s_{j',1}) + \ttout(j',s_{j',1}) > & \; d(j, s_{j,\eend-1}) + \ttout(j, s_{j,\eend-1}) + \\
    & \; \tauuu(j, s_{j,\eend-1}) + \Delta(j,j').
  \end{split}
\end{equation}
Inequality \eqref{eq:rolling} can be simplified to:
\begin{equation}
  d(j',s_{j',1}) > d(j, s_{j,\eend-1}) - R(j,j'),
\end{equation}
by setting:
\begin{equation}
  \label{eq:rolling2}
  \begin{split}
    R(j, j') \coloneq &\ttout(j',s_{j',1}) - \ttout(j, s_{j,\eend-1}) \\
    &- \tauuu(j,s_{j,\eend-1}) - \Delta(j,j').
  \end{split}
\end{equation}
In decision variables, the rolling stock circulation condition for trains $j$
and $j'$ can be written as
\begin{equation}
  \label{eq:qubo:rollingstock}
  \sum_{m \in A_{j, s_{(j, \eend-1)}}} \sum_{m' \in E(d) \cap A_{j',s_{(j',1)}}} x_{j,s_{(j,\eend-1)},m}x_{j', s_{(j',1)},m'} = 0
\end{equation}
where $E(d) = \{0, 1, \ldots, m-R(j, j')\}$.

\section{The capacity condition}
Let $s$ be a station block with $b$ tracks and let $\{j_{1},\ldots,j_{b+1}\} \subset \JJ$ be any $b+1$-tuple of
trains. There should not exist time $t$ for which all the following conditions are simultaneously satisfied:
\begin{equation}
  \begin{split}
    \tin(j_{1}, s) \le &t \le \tout(j_{1}, s) \\
    &\ldots \\
    \tin(j_{b+1}, s_{j,k}) \le &t \le \tout(j_{b+1}, s).
  \end{split}
\end{equation}
In delay representation, the conditions read:
\begin{equation}
  \label{eq:buffer}
  \begin{split}
    d(j_{1}, s_{j_{1},k_{1}-1}) &+ \ttout(j_{1},s_{j_{1},k_{1}-1}) \le t \\
                            &\le d(j_{1},s_{j_{1},k_{1}}) + \ttout(j_{1},s_{j_{1},k_{1}})\\
    \ldots \\
    d(j_{b+1}, s_{j_{b+1},k_{b+1}-1}) &+ \ttout(j_{b+1},s_{j_{b+1},k_{b+1}-1}) \le t \\
                            &\le d(j_{b+1},s_{j_{b+1},k_{b+1}}) + \ttout(j_{b+1},s_{j_{b+1},k_{b+1}}),
  \end{split}
\end{equation}
where $k_{j_{i}}$ is the index of station $s$ in sequence $\Sj$.

The condition \eqref{eq:buffer} translated into binary variables can give a lot of additional terms.
In our problem instances, we ignore this condition, but verify the obtained solutions against it.